\pdfoutput=1
\documentclass[11pt,colorinlistoftodos]{article}
\usepackage{graphicx}
\usepackage[utf8]{inputenc}
\usepackage{amssymb}
\usepackage{bm}
\usepackage{booktabs}
\usepackage{array}
\usepackage{latexsym}
\usepackage{color}
\usepackage{verbatim}
\usepackage{enumerate}
\usepackage{chngpage}
\usepackage{mathrsfs}
\usepackage{euscript}
\usepackage{psfrag}
\usepackage{soul}
\usepackage{empheq}
\usepackage{graphicx}
\usepackage[subpreambles=true]{standalone}
\usepackage{tikz}
\usepackage{amsmath}
\usepackage{amsthm}
\usepackage{titlesec}
\usepackage{setspace}
\usepackage{tensor}
\usepackage{physics}
\usepackage{todonotes}  
\usepackage{mathtools}
\usepackage[nosort]{cite}
\usepackage{subcaption}
\usepackage{longtable}

\usepackage{pgfplots}
\pgfplotsset{compat=1.16}

\definecolor{cOne}{RGB}{198,161,210}    
\definecolor{cTwo}{rgb}{0.60546875,0.8203125,0.7265625}    
\definecolor{cThree}{rgb}{0.4921875,0.69921875,0.90234375} 

\definecolor{cOneDark}{RGB}{102,62,116}
\definecolor{cTwoDark}{RGB}{55,115,89}
\definecolor{cThreeDark}{RGB}{12,70,127}

\DeclareUnicodeCharacter{03A9}{\ensuremath{\Omega}}
\DeclareUnicodeCharacter{2126}{\ensuremath{\Omega}}

\titleformat{\paragraph}
{\normalfont\normalsize\bfseries}{\theparagraph}{1em}{}
\titlespacing*{\paragraph}
{0pt}{3.25ex plus 1ex minus .2ex}{1.5ex plus .2ex}

\usepackage[colorlinks=true,
            linkcolor=blue,
            urlcolor=blue,
            filecolor=blue,
            citecolor=blue,
            pdfstartview=FitH,
            pdfpagemode=UseNone,
            bookmarksopen=true]{hyperref}
\usepackage[all]{hypcap}
\usepackage{mciteplus}

\definecolor{cardinal}{rgb}{0.6,0,0}
\definecolor{darkgreen}{rgb}{0,0.4,0}
\definecolor{golden}{rgb}{0.92,0.7,0}
\definecolor{midnight}{rgb}{0,0,0.5}
\definecolor{darkblue}{rgb}{0,0,0.7}
\definecolor{purple}{rgb}{0.5,0,0.5}

\def\ZZ{\mathbb{Z}}

\def\cH{{\cal H}}\def\cJ{{\cal J}}
\def\cL{{\cal L}}\def\cM{{\cal M}}\def\cN{{\cal N}}

\newcommand{\Leff}{\Lambda_{\rm eff}}
\newcommand{\F}{\mathfrak F}

\numberwithin{equation}{section}

\begin{document}

\vspace*{1.2cm}

\begin{center}
{\LARGE\bfseries Rotating fuzzballs make Love real \dots\ and more}

\vspace{1.4cm}

{\large Massimo Bianchi$^{\,a,b}$, Ang\`ele Lochet$^{\,c}$ and Daniel Panella$^{\,a,b}$}

\vspace{0.8cm}

{\small\itshape
$^a$\,Dipartimento di Fisica, Universit\`a di Roma ``Tor Vergata'',\\
Via della Ricerca Scientifica 1, 00133 Roma, Italy\\[6pt]
$^b$\,INFN Section of Rome ``Tor Vergata'',\\
Via della Ricerca Scientifica 1, 00133 Roma, Italy\\[6pt]
$^c$\,Institut de Physique Th\'eorique, Universit\'e Paris-Saclay, CEA, CNRS,\\
Orme des Merisiers, 91191 Gif-sur-Yvette CEDEX, France\par}

\vspace{0.6cm}

{\footnotesize\ttfamily
\href{mailto:mbianchi@roma2.infn.it}{mbianchi@roma2.infn.it},\quad
\href{mailto:angele.lochet@ipht.fr}{angele.lochet@ipht.fr},\quad
\href{mailto:daniel.panella@outlook.it}{daniel.panella@outlook.it}\par}

\vspace{1.6cm}

{\large\scshape Abstract}
\end{center}

\vspace{2mm}

Rotating Top(ological) Stars (RTS) and Rotating Kerr-Taub Bolts (RKTB) are probably the simplest non-singular horizonless solution that may describe non-generic micro-states of rotating BHs from a 5-dimensional vantage point within the fuzzball proposal.  We study several observable properties and compare the ones to the other. Regularity of RTS requires a twisted identification of the azimuthal and the extra circle. In RKTBs the combined effect of NUT charge and fluxes break equatorial symmetry.  The quadrupole moment turns out to be always positive for RTS and may assume positive (yet small) values for RKTB, in some range of the parameters.
After showing separability of geodetic motion for RKTB, as already known for RTS we identify critical (null) geodesics. We then pass to consider massless scalar perturbations on these backgrounds. Relying on the connection with quantum Seiberg-Witten (SW) curves and on the Mano-Suzuki-Takasugi (MST) approach, we determine the expression of the spheroidal harmonics, including the RKTB without equatorial symmetry,  and their eigenvalue $\Lambda$ from the quantization of the quantum $a$-period to be identified with the `UN-renormalized angular momentum' of MST up to a shift of 1/2. Plugging the result in the radial equation. Quantization of the dual $a_D$-period yields the Quasi-Normal Modes, whose frequencies have negative imaginary part and are thus stable. Finally we study tidal deformations and compute tidal Love numbers, which turn out to be real at least for small non-zero frequency.

\thispagestyle{empty}
\newpage

\tableofcontents

\section*{Introduction}

Understanding the microscopic origin of Black Hole (BH) properties is a
very active endeavour \cite{BenaEtAl,Bena:2007kg,Mayerson:2020tpn}.
On the one hand, one may look for explicit solutions that describe
generic BH micro-states, on the other hand one may explore properties of
specific (typically simple) solutions that expose deviations from the BH
paradigm in General Relativity.
Along this line of investigation, static Top(ological) Stars and their
rotating versions play a prominent role \cite{BahHeid,Bah:2020pdz,Bah:2021irr,Bah:2022yji,
BDMR,pHpPjS,BL}. In some sense they represent
the simplest possible non-singular horizonless solution of gravity
coupled to Maxwell theory in five dimensions \cite{Jejjala:2005yu,Gibbons:2013tqa}. Their macroscopic
properties are easy to extract and the dynamics of linear perturbations
is integrable. One can then extract signatures that allow observations
to discriminate them from BHs in GR \cite{Heidmann:2022ehn,Heidmann:2023ojf,Bianchi:2023sfs,Cipriani:2024ygw,
Bena:2024hoh,Dima:2024cok,Bianchi-Bini-DiRusso,Bini:2025qyn,
Bini:2025ltr,Bah:2025vbr}. Within the fuzzball proposal \cite
{FuzzReview,Lunin:2001fv,Lunin:2001jy,Bena:2007kg,Skenderis:2008qn,
Bena:2016ypk,BenaEtAl}, (rotating) Top Stars represent extremely special
micro-states that however may capture properties shared by other --
hopefully a sizeable fraction of the whole lot -- that indicate the need
from a decisive departure from GR, like string theory. Quite remarkably
Rotating Top Stars (RTS) \cite{BDMR} and Rotating Kerr-Taub Bolts ({RKTB})
\cite{BL} both differ from their Kerr BH cousins in that there are
choices/ranges of their parameters for which the Mass Quadrupole Moment
${\cal M}_2$ is positive.

Aim of the present investigation is to study other observable properties
of Rotating Top Stars (RTS) \cite{BDMR} and Rotating Kerr-Taub Bolts ({RKTB})
\cite{BL} and compare the ones to the other\footnote{We will not address
the objects called ``Asymptotically Flat  Rotating Top Stars"
\cite{pHpPjS} because their metric and Maxwell field are rather
involved, even though the dynamics is separable. We hope to return to
this interesting case in the future. }

In particular after briefly reviewing the (known) properties of these
two classes of solutions, we will study geodetic motion in the {RKTB} case
and prove separability of the dynamics as in the RTS case. We then pass
to consider probe / inert (massless) scalar perturbations and, not
surprisingly, we find that polar angular dynamics separates in both
cases. The main novelty of the {RKTB} is the lack of equatorial symmetry,
caused by the presence of NUT charge in the background. The resulting
equatorially asymmetric spheroidal harmonics do not seem to be known in
the literature, while a lot is known in the equatorially symmetric case
\cite{BertiCardosoCasals}.

The relevant equation in the variable $\chi=\cos\theta$ can be put in
the form of a Confluent Heun Equation (CHE) that in turn can be viewed
as the quantum Seiberg-Witten curve for an $N=2$ Supersymmetric
Yang-Mills (SYM) theory with gauge group $SU(2)$  coupled to three
flavours of hyper-multiplets in the fundamental (doublet)
representation. As a result, the angular (momentum) eigenvalue $\Lambda
= \ell(\ell+1) + ...$, related to the classical Carter constant of separation, turns out to be given -- up to some simple shift -- by the
Coulomb branch parameter  $u = a^2 + ...$ evaluated at the
quantized value of the quantum SW period $a =\ell + {1\over 2}$ with
$\ell$ a positive integer. Moreover the corresponding eigen-functions,
i.e. the equatorially asymmetric (scalar) spheroidal harmonics, can be
expressed as ratio's of Nekrasov-Shatashvili (NS) partition functions.

  The eigenvalue $\Lambda$ can be plugged into the radial wave-equation
that -- not surprisingly -- can also be put in the form of a Confluent
Heun Equation (CHE) with a gauge/gravity dictionary which differs from
the one for the polar angular dynamics. This rewriting allows to extract
many informations: from the vanishing of the absorption rate (due to the
absence of a horizon, replaced by a smooth cap) to the tidal
deformability even in the static limit, from the Quasi Normal Modes
(QNMs) to the explicit expression of the radial wave-functions. We will
address the problem using the `qSW-improved' MST (Mano-Suzuki-Takasugi) approach. We find two branches
of QNMs. The ones located at the external unstable photon-ring have a
`largish' imaginary part $|{\rm Im}\omega|/{\rm Re}\omega \approx
n/\ell$ where $n$ is the `overtone' number \footnote{In our conventions
$\psi \sim e^{-i\omega t}$ linear stability of the solution requires
${\rm Im}\omega <0$. That's what we find.}. The ones located and the
internal photon-ring are meta-stable in that  $|{\rm Im}\omega|\sim
e^{-\Gamma} <<{\rm Re}\omega$.

  The plan of the presentation is as follows.

In Section {{1}} we will review the Rotating Kerr-Taub Bolt ({RKTB}) solution
\cite{BL}, including its static limits and its basic properties -- such
as the regime where ${\cal M}_2$ is positive, highlight the origin of
the lack of equatorial symmetry caused by the NUT charge, which in turn
allows to by-pass the problematic `twist'/entanglement of the azymuthal
angle with the extra-circle direction in the RTS solution of
\cite{BDMR}, and prove separability of the dynamics even for massive
probes with KK momentum.

In Section {{2}} we study linearized scalar perturbations, confirm
separability and explicitly derive the eigenvalue and eigenfunctions of
equatorially-asymmetric spheroidal (scalar) harmonics. For simplicity we
will not address the same problem for spin-weighted harmonics but
indicate the expected differences, based on the known equatorially
symmetric case \cite{BertiCardosoCasals}.

In Section {{3}} we pass to consider the consider the radial
wave-equation in the {RKTB} backround. After establishing the relevant
gauge/gravity dictionary, we restrict attention on the massless scalar
case with zero KK momentum\footnote{To uniformize the notation between RTS and RKTB we call $\tau$, rather than $y$, the coordinate along the circle.} $m_\tau =0$, show the vanishing of the
absorption rate and the non-vanishing of the tidal Love number even in
the static limit and its `reality' (imaginary parts are related to
dissipation that is impossible for a horizonless object at the
linearized level) and `compute' the QNMs by imposing the correct
quantization conditio, involving the `dual' quantum SW period $a_D$
rather than $a$ (as in the angular case).

In Section {{4}} we compare the results for {RKTB}'s to ones for RTS's,
that were only partially derived in \cite{BDMR}, where the radial
dynamics was analyzed only in the limit $a_b=a_s$ (that implies
$a_{RTS}=0$) whereby the angular dynamics becomes `trivial' so that
$\Lambda =\ell(\ell+1)$.

  In Section {{5}} we summarize our results, draw our conclusions and identify interesting
directions for future investigation.

We collect some useful formulae in the Appendices.

\section{Solutions and their static properties}\label{sec:solutions}

We start by recalling some known properties of RKTBs and RTSs and discuss their multipolar structure that allows to discriminate these micro-state geometries to standard BHs.

\subsection{Rotating Kerr-Taub-bolt solution}\label{sec:RKTB solution}

We consider five-dimensional solutions built on a four-dimensional Ricci-flat Euclidean base,
the Euclidean-Kerr-Taub-Bolt (EKTB) space \cite{Gibbons:1979nf}, whose line element reads
\begin{equation}
\label{EKTB base space}
    ds_{EKTB}^2 = {\Sigma} \left(\frac{dr^2}{\Delta} + d\theta^2\right)
    + \frac{\sin^2\theta}{{\Sigma}} \left(\alpha\, d\tau + U_r\, d\phi\right)^2
    + \frac{\Delta}{{\Sigma}}\left(d\tau + U_\theta\, d\phi\right)^2 ,
\end{equation}
with\footnote{Note that $(U_r,U_\theta)$ are usually denoted as $(P_r,P_\theta)$, but these notations will be used later.} :
\begin{align*}
        {\Sigma} &\equiv r^2- (N + \alpha \cos\theta)^2, & \Delta &\equiv r^2-2mr-\alpha^2+N^2,\\
        U_r &\equiv r^2-\alpha^2-\frac{N^4}{N^2-\alpha^2}\ , & U_\theta &\equiv -\alpha \sin^2\theta + 2 N \cos \theta -\frac{\alpha N^2}{N^2-\alpha^2}\ .
\end{align*}
The space depends on three parameters, ${M}$, $\alpha$ and $N$. Their physical meaning is
clearest from a Lorentzian viewpoint: adding a flat time direction to the EKTB space and
reducing the resulting $4+1$-dimensional geometry along $\tau$ gives a $3+1$-dimensional
solution, in which ${M}$, $\alpha$ and $N$ are related to the mass,  angular momentum
and NUT charge. For $N=0$ the EKTB metric
becomes the Euclidean Kerr metric, while for $\alpha=0$ and ${M}=N$ it reduces to the
Taub-NUT space.

Requiring the absence of conical singularities as well as non-removable Dirac strings, imposes two relations between the parameters. The first one is 
\begin{equation}
 \frac{ \alpha}{\sqrt{{M}^2 - N^2 +\alpha^2}}\equiv p \in \ZZ
 \label{periodicity of phi}
\end{equation}
and the second one is:
\begin{multline}
\label{same-sign}
    -16 N (N^2-\alpha^2)^2{M}^3 -4(5N^6-8\alpha^2 N^4+2\alpha^4N^2+\alpha^6){M}^2 \\
    + 16 N (N^2-\alpha^2)^3{M} + 20 N^8-52 \alpha^2 N^6 + 49 \alpha^4 N^4-16 \alpha^6 N^2 = 0 \, ,
\end{multline}
that applies to the cases $N>0$ and $U_r(r_+)<0$, we choose to work with for simplicity. Indeed an another class of solutions exist for $N>0$ and $U_r(r_+)>0$, which gives \eqref{same-sign} up to the transformation $N\rightarrow-N$, but we will restrict to only one branch of solutions. Combining these two equations, we have for a fixed integer $p$, three unknowns for two equations. Fortunately equations \eqref{periodicity of phi} and \eqref{same-sign} are homogeneous of degree respectively 0 and 8 in $N,M,\alpha$. We could define new variables $\widetilde\alpha = {\alpha}/{N}, {\widetilde{M}}= {M}/{N}$ which reduces the problem to two unknowns for two equations, or simply fix $N=1$ in numerical applications, which we will do in the following. Note that regular solutions need $M<N$ from \eqref{periodicity of phi}.

Running-Bolt solutions are obtained by threading the Euclidean-Kerr-Taub-Bolt base with
magnetic fluxes that are either self-dual or anti-self-dual with respect to the
four-dimensional metric \eqref{EKTB base space} \cite{Bena:2009qv,Bobev:2009kn,FloatingBrane}.
 Our supergravity conventions, together with the brane configuration underlying
the solution, are collected in Appendix~\ref{Appendix Supergravity}.

Through the Chern-Simons interaction, these fluxes source both warp factors and a rotation
one-form, and the full eleven-dimensional geometry takes the form\footnote{For simplicity we set $Z_1=Z_2=Z_3={Z}$ in the present investigation. The general formulae can be found in \cite{BL}. We also denote the one-form $k$ by $W$, $\mu$ as $U_\phi$, and $\nu$ as $U_\tau$.}
\begin{equation}
\label{eq:11d metric}
ds^2 = -{Z}^{-2} (dt +{W})^2 + {Z} \, ds^2_{EKTB}
+ds^2_{T^6} \: ,
\end{equation}
where ${Z}$ is a warp factor  and ${W}$ is a one-form on the base.

Dimensionally reducing the eleven dimensional solution along the six-torus produces a five dimensional solution, whose resulting metric is 
\begin{equation}
\label{five dim metric unboosted}
\begin{split}
    ds_5^2 &=-Z^{-2}(dt+{W})^2+ Z ds_{EKTB}^2 \\
    &=-Z^{-2}(dt+{W})^2+Z V^{-1}(d\tau +\widetilde{U}_\theta d\phi)^2 + Z V ds^2_3
\end{split}
\end{equation}
with
\begin{align*}
        ds^2_3 &= \dfrac{\Delta_\theta}{\Delta}dr^2+ \Delta_\theta d\theta^2+\Delta \sin^2\theta d\phi^2 , & \Delta_\theta &\equiv \Delta + \alpha^2\sin^2\theta\ ,\\
        \widetilde{U}_\theta &= U_\theta + \alpha \dfrac{{\Sigma}}{\Delta_\theta}\sin^2\theta\ , &
        V &\equiv \dfrac{{\Sigma}}{\Delta_\theta}\ .
\end{align*}
As anticipated, we will restrict our analysis to the minimal supergravity case, taking ${Q}_1={Q}_2={Q}_3 \equiv {Q}$. The general case can be found in \cite{BL}.

For the self-dual case, warp-factors and rotation vector of the Running-Kerr-Taub-Bolt (RKTB) solution are \cite{Bena:2009qv}:
\begin{equation}
\label{selfdualsol}
     \left\{
    \begin{array}{ll}
        {Z} &= 1 +\dfrac{{Q}^2}{(N- M)\big(r- (N+\alpha \cos \theta)\big)} \\
        {W} &= {U_\phi} (d\tau + U_\theta d\phi) +{U_\tau} d\phi
    \end{array}
\right. \, .
\end{equation}
The scalars ${U_\phi} $ and ${U_\tau}$, entering the angular momentum vector, are given by: 
\begin{equation}
     \left\{
    \begin{array}{ll}
        {U_\phi} \equiv &\gamma \Big(1-\dfrac{2N}{r+N+\alpha \cos \theta}\Big)-3 {Q}\, \dfrac{r}{{\Sigma}}\\[1em]
        &+~ \dfrac{{Q}^3}{2(N-{M})^2}\Big[\dfrac{{M}-N-2\alpha \cos\theta}{{\Sigma}}-\dfrac{2(N-{M})}{(r-N-\alpha \cos \theta)^2}\Big]\\[1em]
        {U_\tau}\equiv &\gamma \alpha \sin^2 \theta -\dfrac{\alpha {Q}^3 \sin^2\theta}{(N-{M})^2(r-N-\alpha \cos \theta)}\\[1em]
        \gamma\equiv& -\dfrac{3 {Q}}{2(N-{M})}+\dfrac{{Q}^3}{4(N-{M})^3}\Big(2-\dfrac{{M}+N}{r_{+}}\Big)\\
    \end{array}
\right. \, .
\label{def mu nu gamma}
\end{equation}
where $r_{\pm}$ are the two roots of $\Delta(r)$. By Kaluza-Klein (KK) reduction (detailed in \cite{BL}), one obtains the four-dimensional metric. It corresponds to an asymptotically flat space with ADM mass, angular momentum, and spin-induced quadrupole moment:
\begin{align}
    G_4 {\cal M}_0 &= G_4 {\cal M} = \frac{1}{4(1-\gamma^2)}\Big(2{M} +\frac{3 {Q}^2}{N-{M}}+ 6 \gamma {Q} +4 \gamma^2 (N-{M})\Big)\, , \label{M self dual}\\[1em]
    G_4 {\cal S}_1 &= G_4 \mathcal{J}=-\frac{1}{(1-\gamma^2)^{1/2}}\frac{\alpha {Q}^3 ({M}+N)}{4(N-{M})^2r_{+}}\ ,
\end{align}

\begin{equation}    
\begin{split}\label{M J M2 self dual}
    G_4 {\cal M}_2 &= \frac{\alpha ^2}{4 \left(1-\gamma ^2\right)(N-{M})^3} \Big[(N-{M})^2 \big[3 {Q}^2+2 (N-{M}) \left(2 \gamma ^2 (N-{M})+{M}+ 3 \gamma {Q} \right)\big]\\[0.8 em]
    &\qquad\qquad\qquad\qquad\qquad-\frac{\left[(N-{M}) \left(-3{Q}^2+2 N (N-{M})\right)-2 \gamma  {Q}^3\right]^2}{3{Q}^2+2 (N-m) \left(2 \gamma ^2 (N-{M})+{M}+ 3 \gamma q \right)}\Big]\\[0.8 em]
    &=\alpha^2 G_4 {\cal M} - \alpha^2\dfrac{\left((N-{M}) \left(2 N (N-{M})-3 {Q}^2\right)-2 \gamma  {Q}^3\right)^2}{4(N-{M})^3(1-\gamma^2)D} \\
    &= G_4 {\cal M} \left(\alpha^2-{d}^2 \right)
\end{split}
\end{equation}
where 
\begin{equation}
\begin{split}
    D &\equiv 3 {Q}^2+2 {M}(N-{M})+6(N-{M}) {Q} \gamma +4(N-{M})^2\gamma^2 \\
    {d} &\equiv \frac{ \alpha \left( (N-{M}) \left(2 N (N-{M})- 3 {Q}^2\right)-2 \gamma  {Q}^3\right)}{(N-{M}) \left(2 (N-{M}) \left(2 \gamma ^2 (N-{M})+{M}+ 3\gamma  {Q}\right)+3 {Q}^2\right)} \, .
\end{split}
\end{equation}
$D$ is positive for positive ${Q}$, and ${d}$ corresponds to $b$ defined in equation $(3.3)$ of \cite{BL}. Physically, it is the shift that mass-centers the coordinate system. Thus the quadrupole moment is given by reversing the sign of the quadrupole moment of Kerr (since $\alpha=ia_{\text{Kerr}}$) and then subtracting a positive term. 

For a given solution $({M},\alpha)$, the four-dimensional metric obtained by
KK reduction of the five-dimensional metric \eqref{five dim metric unboosted},
as written in \cite{BL}, is free of closed time-like curves only for a
certain range of the parameter ${Q}$. This range is essentially determined by requiring
the boost parameter to satisfy $|\gamma|\le 1$. For instance, for
$({M},\alpha)=(0.766,0.742)$ one needs $0.76\le {Q}\le 0.92$. Within this range, ${\cal M}_2$ can be either positive or negative.

\subsection{Rotating Topological Star}
A rotating version of the Top(ological) Stars originally found in \cite{BahHeid,Bah:2020pdz} was found in \cite{BDMR} exploiting double Wick rotation on a solution with mass parameter $\mathtt{m}$ and then performing a boost with parameter $\delta$ \cite{Compere:2009zh}. The general solution can be expressed in terms of four parameters\footnote{For easier comparison with RKTB we changed the notation used in \cite{BDMR} and put $-a^2=\alpha^2$, which is positive in the solitonic regime. } 
\begin{equation}
r_b\,=\,2 \mathtt{m}\sinh^2\delta\,, \hspace{1cm} r_s\,=\, 2 \mathtt{m} \cosh^2\delta, \, \hspace{1cm}  a_{RTS}^2=a_s^2-a_b^2 =-\alpha^2\, ,
\end{equation}
which are subject to the following constraint
\begin{equation}
\label{constraint}
r_s\,a_b^2 \,=\, r_b\,a_s^2\, .
\end{equation}
In terms of these parameters and the variables $\chi=\cos \theta$, $r=r_{_\text{CdBJV}}+r_b$ and $\tau=y_{RTS}$ the solution of \cite{BDMR}, based on \cite{Compere:2009zh}, can be written as:
\begin{equation}
\label{RTS_solution}
\hspace{-7mm}
\begin{aligned}
{d} s^2\,=\,&-\left(1-\frac{r_s r}{\Sigma}+r_sr_b\,a_s^2\frac{\chi^2}{\Sigma^2}\right) {d} t^2+\left(1-\frac{r_b r}{\Sigma}-r_sr_b\,a_b^2\frac{\chi^2}{\Sigma^2}\right) {d} \tau^2 + \Sigma \left(\frac{{d} r^2}{\Delta}+\frac{{d} \chi^2}{1-\chi^2}\right) \\[1mm]
+&\frac{1-\chi^2}{\Sigma}\left[\left(r^2-\alpha^2\right)^2+\alpha^2\left(1-\chi^2\right)\Delta-r_s r_b\, \alpha^2\frac{ r^2 \left(1-\chi^2\right)}{\Sigma}\right]{d}\phi^2 +2 \,r_sr_b\,a_sa_b\frac{\chi^2}{\Sigma^2}\, {d} t \, {d} \tau\\[1mm]
-&2\,\frac{1-\chi^2}{\Sigma} \left[ r_s\,a_s \left(r+r_b\, \alpha^2\frac{\chi^2}{\Sigma}\right){d} t \,{d} \phi-  r_b\,a_b \left(r+r_s \,\alpha^2\frac{\chi^2}{\Sigma}\right){d} \tau \,{d} \phi\right]\,, \\[1mm]
A\,=\,& \sqrt{3 r_s r_b}\, \frac{\chi}{\Sigma}\left(a_s \,{d} t-\left(r^2-\alpha^2\right)\,{d} \phi-a_b\,{d} \tau\right)\, ,
\end{aligned}
\end{equation}
where 
\begin{equation}
\Sigma\,=\,r^2-\alpha^2\,\chi^2\,, \hspace{1cm} \Delta=\left(r-r_s\right)(r-r_b)-\alpha^2=(r-r_+)(r-r_-)   \,.
\end{equation}
In the `solitonic' RTS regime, corresponding to $r_b>r_s$ and $\alpha^2>0$ the solution is non-singular and horizonless. The geometry is capped at $r=r_+$ and has neither ergo-region nor closed time-like curves. Unlike RKTBs RTS enjoy equatorial symmetry and require a twisted identifications \cite[(3.33)]{BDMR}, under which $t$, $\varphi$ and $\tau$ (called $y$ there) are mixed. These are untwisted by the boost \cite[(3.34)]{BDMR}, and the frequency and momentum conjugate to the asymptotic coordinates $(\hat t,\hat y)$ are
\begin{equation}
\hat\omega=\frac{\omega-{\xi} {m_\tau}}{\sqrt{1-\xi^{2}}},\qquad
{\hat{m}_\tau}=\frac{{m_\tau}-{\xi}\omega}{\sqrt{1-\xi^{2}}} .
\end{equation}
The single-valuedness condition of \cite[(5.3)]{BDMR} then reads
\begin{equation}
K\equiv\left({m_\tau}-{\xi}\omega\right)R_\tau-\frac{m_\phi {h_1}}{{h_2}}
=\sqrt{1-\xi^{2}}\,R_\tau\,{\hat{m}_\tau}-\frac{m_\phi {h_1}}{{h_2}}\in\mathbb Z .
\label{eq:RTSquant}
\end{equation}
with $R_\tau$ the radius of the KK circle. In analogy with the sector $m_\tau=0$ of the {RKTB} we will restrict attention onto KK-neutral modes, ${\hat{m}_\tau}=0$, i.e.\ ${m}_\tau={\xi}\omega$. Since ${h_1}$ and ${h_2}$ are coprime, eq.~\eqref{eq:RTSquant} then implies, for ${h_1}\neq0$,
\begin{equation}
m_\phi \in {h_2}\,\mathbb Z ,
\label{eq:RTSmq}
\end{equation}
which is precisely the set of modes captured by the reduction to four dimensions along $\hat\tau$ (called $\hat y$ there) \cite[Sec. 3.6]{BDMR}.
The ADM mass and angular momentum can be computed from the asymptotics of the metric, and are given by\footnote{Note that the expression of the mass in \cite{BDMR} has a wrong sign in front of the $\xi^2$ term in the numerator. Notwithstanding the mass of RTS is positive in the soliton regime.}:
\begin{align}
    G_4 {\cal M}_0 &= G_4 {\cal M} = \frac{2r_s+r_b-\xi^2(r_s+2r_b)}{4(1-\xi^2)}\, , \label{eq:M RTS}\\[1em]
    G_4 {\cal S}_1 &= G_4 \mathcal{J}=\frac{a_sr_s-a_br_b\xi}{2\sqrt{1-\xi^2}}\ .
\end{align}
To compute the spin-induced quadrupole moment, we can follow the procedure of \cite{RevModPhys.52.299}. We first need to find an ``Asymptotically Cartesian and Mass Centered to order N'' (ACMC-N) system of coordinates, and following \cite{BL} we introduce the set of ACMC coordinates:
\begin{equation}
    \tilde{r}\sin\tilde{\theta} = \sqrt{r^2-\alpha^2} \sin\theta \, ,\qquad \tilde{r}\cos\tilde{\theta} =r\cos\theta
\end{equation}
Note that here, thanks to equatorial symmetry, we have $d=0$: the system is already mass-centered, there is no need to shift it. As a consequence, all odd mass multipoles and even current multipoles vanish. Then, doing the expansion of the metric, one finds:
\begin{equation}
     \mathcal{M}_2 = \alpha^2 {\cal M} >0 \,.
\end{equation}
Thus the spin-induced quadrupole moment is always positive. Furthermore we can also compute the multipole moments at all orders:
\begin{equation}
    \mathcal{M}_{2n} = \alpha^{2n} {\cal M}\, \qquad \mathcal{S}_{2n+1}=\mathcal{J} \alpha^{2n}
\end{equation}
These formulae are the same as the ones obtained with the Kerr solution, but with $\alpha=ia_{Kerr}$.
\section{(Critical) geodesics of RKTB}
We will here derive the geodesics equation of the RKTB solution, and show that despite appearances, the equation is separable.

\noindent The geometry admits the three commuting Killing vectors
$\partial_t,\partial_\tau,\partial_\phi$, which can be associated to the conserved momenta:
\begin{equation}\label{eq:momenta}
P_t\;=\;-E\,,\qquad P_\tau\;=\;{m_\tau} \,,\qquad P_\phi\;=\;J \,
\end{equation}
%
We study the Hamilton--Jacobi (HJ) equation
$\cH \equiv g^{\mu\nu}P_\mu P_\nu=-{\mu}^2$ for a test particle of mass
${\mu}$. It is convenient to introduce the two effective moments:
\begin{align}
\widetilde{P}_\tau&\;\equiv\; P_\tau-{U_\phi}\,P_t\,,\label{eq:Meff}\\
\widetilde{P}_\phi&\;\equiv\; P_\phi-V\,P_\tau-{U_\tau}\,P_t\label{eq:Leff}
\end{align}
as well as the variables:
\begin{align}\label{Y and X}
    Y\;&\equiv\; r- N-\alpha \cos\theta \\
    \Sigma\;&\equiv\; r^2- \left(N+\alpha \cos\theta\right)^2 
\end{align}
which absorb the frame-dragging generated by the shift functions ${U_\phi},{U_\tau}$ and the base function $V$, and the co-dependance in $r$ and $\theta$. In these
variables, the contravariant metric organizes the HJ equation, after
multiplication by the global conformal factor $Z\Sigma$, into the form:
\begin{equation}\label{eq:HJ}
Z\,\Sigma\big(\mathcal H+{\mu}^2\big)\;=\;\Delta\,P_r^2+P_\theta^2-Z^3\Sigma\,P_t^2
+\frac{{\Sigma}^2}{\Delta_\theta}\widetilde{P}_\tau^2
+\frac{\Delta_\theta}{\Delta\sin^2\theta}\bigg(\widetilde{P}_\phi-\alpha\frac{\Sigma}{\Delta_\theta}\sin^2\theta\,\widetilde{P}_\tau\bigg)^2\;=\;0\,.
\end{equation}
where $P_r$ and $P_\theta$ denote the canonical radial and polar momenta, and are equal respectively to $\dfrac{{\Sigma}}{\Delta}\,\dot r$ and ${\Sigma} \, \dot \theta$.

\noindent The last two terms of \eqref{eq:HJ} can be rewritten in a form free of the
mixed denominator $\Delta_\theta$. Using the definition of $\Delta_\theta$ we obtain:
\begin{equation}\label{eq:HJred}
\Delta\,P_r^2+P_\theta^2-Z^3\Sigma\,P_t^2+\frac{\widetilde{P}_\phi^2}{\sin^2\theta}
+\frac{(\Sigma\widetilde{P}_\tau -\alpha \widetilde{P}_\phi)^2}{\Delta}+Z\,\Sigma\,{\mu}^2\;=\;0
\end{equation}
Equation \eqref{eq:HJred} is the starting point for the separability analysis.
The only obstacle to additive separation into a radial part $R(r)$ and an
angular part $\Theta(\theta)$ is the factor $Y=r-N-\alpha\cos\theta$, which
appears in the denominators of $Z,{U_\phi},{U_\tau}$ and couples $r$ and $\theta$. Using the identity:
\begin{equation}\label{eq:bridge}
\alpha^2\sin^2\theta\;=\;\alpha^2-(r-N-Y)^2\,,
\end{equation}
we trade the dependence of \eqref{eq:HJred} in $(r,\theta)$ to a dependence on $(r,Y)$. The equation is then separable, if there are no poles in $Y^n$. At first sight, it seems that there are terms of order $Y^{-2}$ and $Y^{-1}$. However after algebraic manipulations, each of these poles vanish identically, as well as terms in $Y^{3}$ and $Y^{4}$. Finally, by replacing $Y$ by its expression (eq.\eqref{Y and X}), one obtains a cancellation of the cross-terms in $r$ and $\theta$, and we can separate the equation \eqref{eq:HJred} as:
\begin{equation}\label{eq:radial_angular_ode}
\begin{split}
\Delta(r)\,P_r^2+\mathcal{R}(r)+{\mu}^2\,\mathcal{R}_Z(r)&\;=\;-{\Lambda}\,,\\
P_\theta^2+\Theta(\theta)+{\mu}^2\,\Theta_Z(\theta)&\;=\;+{\Lambda}\,.
\end{split}
\end{equation}
where ${\Lambda}$ is the separation constant, often called in this context the Carter constant. Henceforth we will focus on massless geodesics and we will not need the explicit expressions of $\mathcal{R}_Z(r)$ and $\Theta_Z(\theta)$ ... that we display in Appendix \ref{app:massive} for the interested reader. 

We will need instead $\mathcal{R}$, given by
\begin{equation}
\begin{split}
	\mathcal{R}(r) =& \left(-\frac{{\cal P}_{2}(r)}{4(m-N)^{4} \Delta} + \frac{{\cal P}_{1}(r)(N-r)}{(m-N)^{2}}\right)P_{t}^{2}+2\alpha \left(\frac{f(r)}{\Delta}-\gamma\right) P_{t} P_{\phi} + \frac{\alpha^{2}}{\Delta} P_{\phi}^{2} \\
    &+ \frac{g(r)^2}{\Delta} P_\tau^2 \,+\, 2 \alpha \frac{g(r)}{\Delta} P_{\tau} P_\phi+2\left(\frac{(r+N){Q}^3}{(N-m)^3}+\frac{g(r)f(r)}{\Delta}\right) P_\tau P_t\, ,
\end{split}
\end{equation}
with ${\cal P}_{1},{\cal P}_{2},f$ and $g$ four polynomials in $r$, of degree, respectively, $1$, $3$ and $2$ for the last two, given explicitly by:
\begin{align}
	{\cal P}_{1}(r) &\equiv {M}^2 \left(\gamma ^2 N+N-\gamma ^2 r+r\right)+{M} \left(-2 \left(\gamma ^2+1\right) N^2+2 \left(\gamma ^2-1\right) N r+3 {Q}^2\right)\label{P1}\\
	&+\left(\gamma ^2+1\right) N^3+N^2 \left(r-\gamma ^2 r\right)-3 N {Q}^2-2 \gamma  {Q}^3 \nonumber\\
	{\cal P}_{2}(r) &\equiv -8 \gamma ^2 r ({M}-N)^5 \left(-\alpha ^2+N^2-2 N r+r^2\right) \nonumber\\
	&+ 24 \gamma  r ({M}-N)^4 \left(-\alpha ^2+N^2-2 N r+r^2\right) {Q} \nonumber\\
	&+12 r ({M}-N)^3 \left(2 \alpha ^2+{M} r-2 N^2+3 N r-2 r^2\right) {Q}^{2}\label{P2}\\
	&+ 4 \gamma  ({M}-N)^2 \left({M} \left(-\alpha ^2+N^2+6 N r-7 r^2\right)-3 N^3+3 \alpha ^2 N+N r^2+2 r^3-2 \alpha ^2 r\right) {Q}^{3} \nonumber\\
	&+12 ({M}-N)^2 \left(\alpha ^2+{M} r-N^2-N r+r^2\right) {Q}^{4} \nonumber\\
	&+\left(-4 \alpha ^2-{M}^2-2 {M} (N+2 r)+3 N^2+4 N \right){Q}^{6}\nonumber\\
    f(r)&\equiv\frac{{Q}^3 (-{M}-N+2 r)}{2 (N-{M})^2}+\gamma  \left((r-N)^2-\alpha ^2\right)-3 {Q} r \label{f}\\
    g(r) &\equiv N^2-r^2+\alpha^2\left(1+\frac{N^2}{N^2-\alpha^2} \right)
\end{align}

The polar angle effective potential $\Theta$ is given by:
\begin{equation}
    \begin{split}
        \Theta(\theta) \equiv& \big(\cos^2\theta\alpha^2(1-\gamma^2)+\tilde\Omega_1 \cos\theta \big)P_t^2+2\left(\gamma \alpha U_\theta+\frac{\alpha {Q}^3 \cos\theta}{(N-m)^3}\right) P_t P_\tau\\
        &+\frac{U_\theta^2}{\sin^2\theta} P_\tau^2-\frac{2U_\theta}{\sin^2\theta}P_\tau P_\phi+\frac{1}{\sin^2\theta}P_\phi^2
    \end{split}
\end{equation}
with 
$$
\Omega_1 \equiv \alpha E^2 \left(2N  - \dfrac{3{Q}^2}{(N-{M})}-\dfrac{2 \gamma {Q}^3}{(N-{M})^2}\right) \quad {\rm and} \quad \tilde{\Omega}_1 = \dfrac{\Omega_1}{E^2}\: . $$
It will be convenient for the following analysis to use the degree 4 polynomial
\begin{equation}
    \cN_4(r) \equiv \Delta(r) \big(\mathcal{R}(r)+{\Lambda}\big) = \sum_{i=0}^4\lambda_ir^{i}
\end{equation}
whose coefficients $\lambda_i$ are, for $P_\tau=0$:
\begin{equation}
    \begin{split}
        \lambda_0=&E^2\big(N^2-\alpha^2\big)\left[N^2\big(1+\gamma^2\big)+\frac{Q^3\gamma-3NQ^2}{N-M}+\frac{3Q^4}{(N-M)^2}-\frac{Q^6}{(N-M)^4}\right]\\[0.7em]
        &+\left(\alpha\, J+\frac{E\, Q^3\,(N+M)}{2\,(N-M)^2}\right)^2+ \Lambda \big(N^2-\alpha^2\big)\\[0.7em]
        \lambda_1=&-2M\left(\Lambda+E^2\Big[N^2\big(1+\gamma^2\big)+\frac{Q^3\gamma-3NQ^2}{N-M}+\frac{3Q^4}{(N-M)^2}-\frac{Q^6}{(N-M)^4}\Big]\right)\\[0.7em]
        &+E\left(2\gamma\,(N-M)+3Q-\frac{Q^3}{(N-M)^2}\right)\left[2\alpha\, J+\frac{E\, Q^3\,(N+M)}{(N-M)^2}\right]\\[0.7em]
        &-E^2\big(N^2-\alpha^2\big)\left[2\,(2N-M)\,\gamma^2+6Q\gamma+\frac{3Q^2}{N-M}\right]\, ,\\[0.7em]
        \lambda_2=&\Lambda+ E^2\Big(\alpha^2\big(1-\gamma^2\big)+6\,\big(N\gamma+Q\big)^2+\frac{3Q^2}{N-M}\left[M-Q\gamma-\frac{Q^2}{N-M}\right]\Big)\\[0.7em]
        \lambda_3=&E^2\Big(-\frac{3 Q^2}{N-M}+2 M-4 \gamma ^2 N-6 \gamma  Q\Big)\,,\\[0.7em]
        \lambda_4=&-E^2(1-\gamma^2)\,.
    \end{split}
\end{equation}
Note that $\mathcal{N}_4$ behaves like $-r^4$ at infinity. The radial potential is $E^2(1-\gamma^2)+\dfrac{\mathcal{N}_4}{\Delta^2}$, and it asymptotes to zero at infinity which is what we expected. Near the bolt, $\mathcal{N}_4$ is finite and positive as $\mathcal{N}_4(r_+)=\alpha^2 J^2$, and therefore the radial potential diverges and creates a potential barrier, which is what is expected for smooth horizonless geometries. 
Then for massless probes and for $m_\tau=0$, \eqref{eq:radial_angular_ode} becomes:
\begin{figure}
    \centering
    \includegraphics[width=0.4\linewidth]{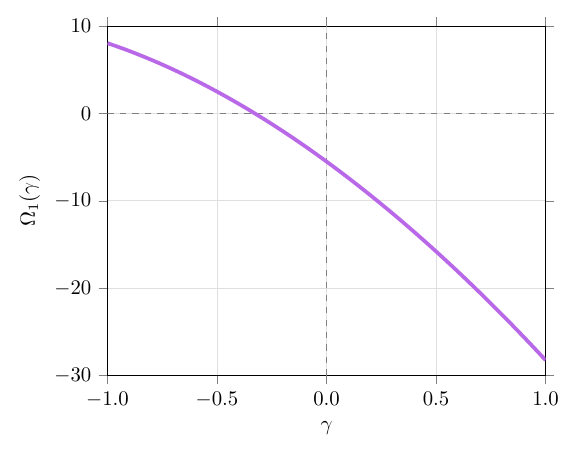}
    \caption{$\tilde\Omega_1$ as a function of $\gamma$, for $p=2$ and $N=1$, leading to $({M},\alpha)=(0.766,0.742)$}
    \label{fig:omega}
\end{figure}
\begin{equation}
     \left\{
    \begin{array}{ll}
        \Delta(r) P_r^2+ \dfrac{ \cN_4(r)}{\Delta(r)}=  0\\[1em]
        \left(1-\chi^2\right) P_\chi^2+\dfrac{J^2}{1-\chi^2}+\alpha^2 E^2\left(1-\gamma^2\right)\chi^2+\Omega_1\,\chi =  {\Lambda} \\
    \end{array}
\right. \, .
\label{EoM}
\end{equation}
where we performed the change of variables $\chi=\cos\theta$. The term in $\Omega_1$ breaks the equatorial symmetry $\chi\rightarrow-\chi$ as a consequence of the presence of the NUT charge $N$, as well as of the fluxes (see Figure \ref{fig:omega}). In similar geometries it does not appear: neither in Kerr, nor in the RTS geometry.

\begin{figure}[!htbp]
  \centering
  \includegraphics[width=0.9\textwidth]{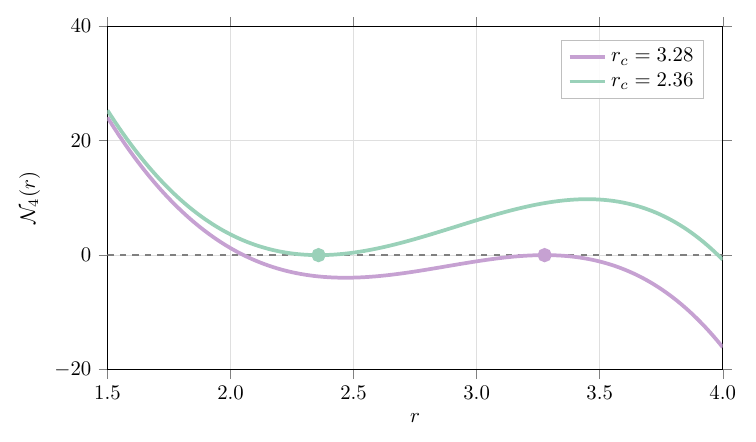}
  \caption{$\mathcal{N}_4(r)$ for several values of $r_c$, for the solution $(p,{M},\alpha,{Q})=(2,0.766,0.742,0.82)$ and for counter-rotating photon sphere. The greatest $r_c$ (purple curve) corresponds to an unstable light ring, as $r_c$ corresponds to a local maximum. For $r_c$ lower (the green curve), $r_c$ corresponds to a local minimum: it is a stable light ring. As expected, the stable light ring is closer to the bolt than the unstable one.}
  \label{fig:N4}
\end{figure}

Critical null geodesics are those that remain at a fixed
radius $r=r_c$, which is not necessarily equivalent to having a (non-)planar spherical geodesics, as $r$ is not the Cartesian coordinates radius but the radius in a Boyer{-}Lindquist kind coordinates. A geodesic stays at constant
$r=r_c$ if and only if $\dot r=0$ and $\ddot r=0$ at $r_c$. Assuming $\Delta(r_c)\neq 0$, 
the condition $\dot r(r_c)=0$, i.e.\ $P_r(r_c)=0$, gives
\begin{equation}\label{eq:N4=0}
  \mathcal{N}_4(r_c)=0\,,
\end{equation}
while differentiating the radial equation
with respect to $t$, and using \eqref{eq:N4=0} and $\dot r(r_c)=\ddot r (r_c)=0$ gives
\begin{equation}\label{eq:N4'=0}
  \mathcal{N}_4'(r_c)=0\,.
\end{equation}
However these two conditions are still not enough to fully determine the geodesics. Indeed we have two equations for three unknowns, $\tilde J, r_c, {\Lambda}$, where $\tilde J\equiv{J}/{E}$. The angular equation tells us that ${\Lambda}$ is only constrained by a bound. In general, critical geodesics in rotating BHs and fuzzballs are non-planar and the polar angle $\theta$ can vary for non-zero $P_\theta$. The condition on ${\Lambda}$ thus reduces to:
\begin{equation}
 {\Lambda}\geq \Theta(\theta)\,.
\end{equation}
\begin{figure}[!htbp]
  \centering
  \includegraphics[width=0.9\textwidth]{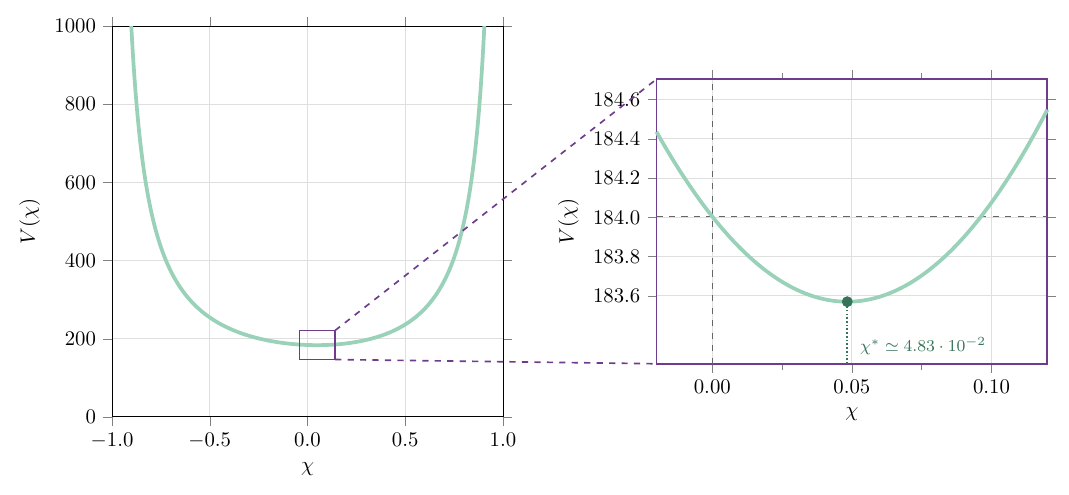}
  \caption{Angular potential $V(\chi)= \dfrac{\tilde{J}^2}{1-\chi^2}+\alpha^2 \left(1-\gamma^2\right)\chi^2+\tilde{\Omega}_1\,\chi$ for the solution $(p,M,\alpha,Q)=(2,0.766,0.742,0.9)$ and taking $\tilde{J}=13.56$, which corresponds to the stable geodesic for this set of parameters.}
  \label{fig:V(chi)}
\end{figure}
Equivalently, ${\Lambda}$ must be greater than the angular potential $V(\chi)$ plotted for a given solution on Figure \ref{fig:V(chi)}. When ${\Lambda}$ is equal to the minimum of this potential, $V(\chi_0)$, equation \eqref{EoM} tells us that $P_\chi$ vanishes: the geodesics keeps a fixed angle $\theta\neq \dfrac{\pi}{2}$, because of the breaking of equatorial symmetry. 
To compute explicit geodesics, we will evaluate it at this polar
turning point $\theta_0$, where $P_\theta(\theta_0)=0$. It fixes the Carter constant as a function of the impact parameter,
\begin{equation}
  {\Lambda}=\Theta(\theta_0(\tilde J))\,.
\end{equation} 
Then using this Carter constant in \eqref{eq:N4=0}-\eqref{eq:N4'=0} one finds an expression of $r_c,\tilde{J}$ in terms of the parameters. For each set of parameters $({M},\alpha,{Q})$ we tried, we found a counter-rotating (meaning $\tilde{J}$ has the opposite sign as $\mathcal{J}$, which is negative for $q$ positive) stable and an unstable geodesics, as pictured in Figure \ref{fig:N4}. There is no co-rotating (ie $\tilde{J}<0$) geodesics, as their critical radius lies inside the bolt. Co-rotating geodesics have a smaller $r_c$ than counter-rotating ones, and this is the expected usual behavior, because of the frame-dragging effect.

%
\section{Linearized scalar perturbations}\label{angular}

We now pass to the study of linearized scalar perturbations around RKTBs and RTSs. In order to compute the spectrum of QNMs and other dynamical properties, we need Carter separation constant $\Lambda$.  As it appears both in the angular and radial equation, the idea is to compute it from the angular equation and then substitute it into the radial equation, solving them in a coupled manner. Both angular and radial equations can be brought to the form of Confluent Heun Equations (CHEs). CHEs can be solved by expansions in (confluent) hyper-geometric functions, following and generalizing the Mano-Suzuki-Takasugi (MST) approach \cite{MST1,MST2,MST3,MST4}. Alternatively they can be related to quantum Seiberg-Witten (qSW) curves for ${\cal N}=2$ supersymmetric Yang-Mills (SYM) theories with $SU(2)$ gauge group coupled to $N_f=3$ hypermultiplets  \cite{Aminov:2020yma,Bianchi:2021xpr,BCGM,BILT-Kerr,
Bonelli-Iossa-Lichtig-Tanzini,CFMP,Bianchi:2023sfs,
DiRusso:2024hmd,Bianchi-Bini-DiRusso,Cipriani:2024ygw}.

The massless scalar field, $\Box\Phi=0$, on the RKTB background separates similarly to the null geodesics. The Killing vectors $\partial_t$, $\partial_\tau$ and $\partial_\phi$ allow us to decompose it into modes
\begin{equation}\label{eq:mode}
\Phi=e^{-i\omega t+i m_\tau\tau+i m_\phi\phi}\,R(r)\,S(\theta),\qquad m_\tau,m_\phi\in\mathbb Z,
\end{equation}
where $\omega$ is the frequency, $m_\phi$ the azimuthal number and $m_\tau$ the KK momentum. These are the wave counterparts of the conserved momenta \eqref{eq:momenta}. 
In the KK-neutral sector $m_\tau = 0$ the angular problem for RKTB is a prolate spheroidal equation deformed by a term $\Omega_1 \chi$, which is absent in the RTS case. The wave equation separates exactly as the HJ equation, and its radial and angular parts are obtained from \eqref{EoM} through $(E,J,m_\tau)\rightarrow (\omega,m_\phi,m_\tau)$ and
\begin{equation}\label{eq:quantisation}
\Delta\,P_r^2\;\to\;-\frac{d}{dr}\Big(\Delta\,\frac{d}{dr}\Big),\qquad
(1-\chi^2)\,P_\chi^2\;\to\;-\frac{d}{d\chi}\Big((1-\chi^2)\,\frac{d}{d\chi}\Big).
\end{equation}
In particular, the radial equation reads
\begin{equation}\label{eq:radial-wave}
\Delta\,\frac{d}{dr}\Big(\Delta\,\frac{dR}{dr}\Big)=\cN_4^{\rm w}(r)\,R,\qquad
\cN_4^{\rm w}\equiv\cN_4\big|_{E\to\omega,\,J\to m_\phi,\,p\to m_\tau,}\,.
\end{equation}
From now on we will thus use previous notations with $(E,J,m_\tau)\rightarrow (\omega,m_\phi,m_\tau)$, and $\cN_4^{\rm w}$ will be identified to $\cN_4$.

\subsection{The angular equation for both RKTB and RTS}

We start with the angular dynamics and initially focus on RKTB because in general and contrary to RTS they break equatorial symmetry. Relying on the relation with qSW and on the MST approach, we solve the Sturm-Liouville problem and find a perturbative expansion for the eigenvalue $\Lambda$ that replaces Carter's separation constant in this context. Moreover we determine the explicit form of the prolate spheroidal harmonics, that to the best of our knowledge has never appeared in the literature. Eventually, we will discuss the modifications needed to describe the angular dynamics of scalar perturbations around RTs. For simplicity, we will not consider vector, tensor or spinor perturbations.     

Introducing
\begin{equation}\label{omega_1_2}
    \Omega_2 \equiv \alpha \omega \sqrt{1- \gamma^2}, \hspace{5mm}\Omega_1 \equiv \alpha \omega^2\left(2N - \frac{3{Q}^2}{(N-{M})}-\frac{2 \gamma {Q}^3}{(N-{M})^2} \right),
\end{equation}
the angular equation becomes
\begin{equation}\label{angular_equation}
    \frac{d}{d \chi}\left[ (1- \chi^2)\frac{dS}{d\chi}\right] + \left[\Lambda - \Omega_1 \chi - \Omega_2^2\chi^2 - \frac{m_\phi^2}{1-\chi^2} \right]S = 0.
\end{equation}


that has singularities at $\chi=+1$ and $\chi=-1$ (the two regular points, i.e. the poles of the sphere) and at $\chi=\infty$ (the irregular singular point). As it has the form of a confluent Heun equation (CHE), it can be related to the qSW curve of $\mathcal{N} = 2$ SYM with group $SU(2)$ coupled to $N_f = 3$ hypermultiplets \cite{BDMR}, whose canonical Schr\"odinger-like form is
\begin{equation}\label{Q_SW}
    \Psi(z)'' + {\cal Q}_{\rm SW}(z)\Psi(z) = 0,
\end{equation}
where ${\cal Q}_{\rm SW}(z)$ is given by
\begin{equation}
\label{eq:QSW}
{\cal Q}_{\rm SW}(z)=-\frac{{q}^2}{4z^4}+\frac{{q}\,m_3}{z^3}
+\frac{1-(m_1-m_2)^2}{4(z-1)z}
+\frac{1-(m_1+m_2)^2}{4(z-1)^2z}
+\frac{u-\tfrac14+\tfrac{q}{2}(m_1+m_2-1)}{(z-1)z^2}.
\end{equation}
The angular equation \eqref{angular_equation} can also be reduced to a canonical, Schr\"odinger-like form. To this aim, one has to use the following dictionary
\begin{equation}\begin{split}\label{dictionary}
    &\chi = \frac{2}{z} - 1, \qquad S(\chi) = \frac{\Psi(z)}{\sqrt{2(1-z)}}, \qquad {q} =4i\Omega_2,\qquad m_1 = m_\phi,\\
    &m_2 = 0, \qquad m_3 = \frac{2\Omega_1}{q} = -\frac{i\Omega_1}{2\Omega_2}, \qquad u = \Lambda - \Omega_2^2 - \Omega_1 + \tfrac{1}{4} - 2i \Omega_2 (m_\phi - 1).
    \end{split}
\end{equation}
For real $\omega$, ${q}$ and $m_3$ are purely imaginary and ${q}m_3 = 2 \Omega_1$ is real.

For RTS, the angular dictionary is ${q} = 4a \sqrt{\omega^2 - m_\tau^2}= \pm 4i\alpha \sqrt{\omega^2 - m_\tau^2}$, $m_1 = {m_\phi}$ and $m_2$.  
Notice that $\Omega_2^2$ can have either sign: positive in the solitonic regime $-a_{RTS}^2=\alpha^2\ge 0$ ($r_b\ge r_s$), negative $-a_{RTS}^2=\alpha^2\le 0$ ($r_b\le r_s$) in the rotating black string regime. Here, \eqref{dictionary} gives ${q}^2 = -16 \alpha^2\omega^2(1 - \gamma^2)$. This is the RTS expression under $a \to i \alpha$ and ${m_\tau} \to \gamma \omega$. Even at $m_\tau = 0$, whereby $m_2=0$ as in the RTS, we have a new mass parameter $m_3 \ne 0$ generated by the term odd in $\cos \theta$. This is a major difference with respect to the RTS case.


Following the Mano-Suzuki-Takasugi (MST) approach \cite{MST1,MST2,MST3,MST4} in the form adapted to non-rotating Top Stars in \cite{Bianchi-Bini-DiRusso}, we expand the solution in Legendre (hypergeometric) functions multiplied by $e^{2i\Omega_2y}$. We prove that the expansion coefficients obey an exact three-term recurrence and show that for integer ``UN-renormalised angular momentum''\footnote{Contrary to the radial case, $\nu$ will turn out be identified with $\ell$.} $\nu = \ell$ the chain terminates at the lowest admissible degree, so that no second regularity condition is needed. We solve the associated continued fraction perturbatively in $(\Omega_1,\Omega_2)$, determine the eigen-value $\Lambda$ and construct the corresponding eigen-functions $S_\Lambda$. For RTS, that are symmetric under $\chi \to -\chi$, one simply has to put $\Omega_1=0$ in the formulae that we will find.

Let us preliminarily observe some properties of  $\Lambda$. For real $\Omega_1,\Omega_2^2$ the problem is a regular Sturm--Liouville problem, hence $\Lambda$ is real. The map $\chi\to-\chi$ leaves Eq.~\eqref{angular_equation} invariant provided $\Omega_1\to-\Omega_1$; therefore
\begin{equation}
\Lambda(\Omega_1,\Omega_2^2)=\Lambda(-\Omega_1,\Omega_2^2).
\label{eq:parity}
\end{equation}
For $m_2=0$ the eigenvalue $\Lambda$ is therefore blind to the sign of the odd term; the breaking of equatorial ($\chi\to-\chi$) symmetry only shows up in the eigen-functions not in the spectrum. Even though the eigenvalue $\Lambda$ is even in $\Omega_1$, $\langle \chi \rangle$ is odd and generically non-zero. The statement can be proven as follows: write equation \eqref{angular_equation} as $\cL S = \Lambda S$ with
\begin{equation}
    \cL = -\frac{d}{d\chi} (1-\chi^2)  \frac{d}{d\chi} +\frac{m_\phi^2}{1-\chi^2} + \Omega_1\chi + \Omega_2^2\chi^2
\end{equation}
Differentiating $\cL S_\Lambda = \Lambda S_\Lambda$ with respect to $\Omega_1$, multiplying by $S_\Lambda$ and integrating over $\chi$, the terms containing $\partial_{\Omega_1}S_\Lambda$ cancel after integrating by parts twice and one gets
\begin{equation}\label{Hellmann-Feynman}
    \langle \chi \rangle \equiv \frac{\int_{-1}^1 \chi S_\Lambda^2\,d\chi}{\int_{-1}^{1} S_\Lambda^{2}\,d\chi} = \frac{\partial\Lambda}{\partial \Omega_1},
\end{equation}
whenever $\int S_\Lambda^2\, d\chi \ne 0 $. This is always the case for real parameters, for which $S_\Lambda$ can be chosen real. Since $\Lambda$ is even in $\Omega_1$, $\langle \chi \rangle$ is odd and generically non-zero. Next, we analyze the local exponents. In particular near $\chi = 1$ one put ${\delta} = 1 - \chi$, then $1-\chi^2 = {\delta}(2 - {\delta}) \simeq 2 {\delta}$ and $d/d\chi =- d/d {\delta}$. For $S \sim {\delta}^{\beta}$ the leading terms of eq. \eqref{angular_equation} give
\begin{equation}
    2{\beta}^2 {\delta}^{{\beta}-1} - \frac{m_\phi^2}{2}{\delta}^{{\beta}-1} = 0 \longrightarrow {\beta} = \pm \frac{|m_\phi|}{2},
\end{equation}
and the same holds at $\chi = -1$. Regularity of $S$ selects ${\beta} = |m_\phi|/2$ at both poles. For $m = 0$ the two exponents coincide, and the second solution is logarithmic. As $\chi \to \infty$ the dominant terms are $-\chi^2 S'' - \Omega_2^2 \chi^2 S \simeq 0$, so that $S \sim e^{\pm i \Omega_2\chi}$: the point at infinity is an irregular singular point of rank one.

\subsection{Equatorially asymmetric spheroidal harmonics}

\subsubsection{Reduction to hypergeometric form}
In order to reduce the angular equation to a hypergeometric-like equation, we introduce the variable $y$ 
\begin{equation}
  y = \frac{1 + \chi}{2} = \cos^2{\theta\over 2} \longrightarrow \chi = 2y - 1
\end{equation}
Then $1 - \chi^2 = 4 y (1-y)$ and $d/d\chi = \frac{1}{2}d/dy$, so that
\begin{equation}
\frac{d}{d\chi}\Big[(1-\chi^2)\frac{d S}{d\chi}\Big]
=\frac12\frac{d}{d y}\Big[4y(1-y)\cdot\frac12 S'\Big]=\frac{d}{d y}\big[y(1-y)S'\big],
\end{equation}
and eq.~\eqref{angular_equation} becomes
\begin{equation}
\big[y(1-y)S'\big]'+\Big[\Lambda-\Omega_1(2y-1)-\Omega_2^2(2y-1)^2-\frac{m_\phi^2}{4y(1-y)}\Big]S=0 .
\label{eq:ang-y}
\end{equation}
Assuming $m_\phi\ge0$, for simplicity, and setting
\begin{equation}
    S = e^{2i\Omega_2y} \rho F,
\end{equation}
with
$$
\rho(y)=\left({y\over 1-y}\right)^{m_\phi/2}
$$  
so that
$$
{\rho'\over \rho}=\frac{m_\phi}{2}\big(\frac1y+\frac1{1-y}\big)=\frac{m_\phi}{2y(1-y)}
$$
we obtain
\begin{equation}
y(1-y)F''+(1+m_\phi-2y)F'+\Big[\Lambda-\Omega_2^2+2i\Omega_2 m_\phi+(2i \Omega_2+\Omega_1)(1-2y)\Big]F+4i \Omega_2\,y(1-y)F'=0 .
\label{eq:F-full}
\end{equation}
Moving the terms in $y(1-y)F'$ and $yF$ to the right-hand side, thus treating them as a ``source'', and defining
\begin{equation}
\tilde\Lambda\equiv\Lambda-\Omega_2^2+\Omega_1+2i \Omega_2(1+m_\phi) \:,
\label{eq:Ltilde}
\end{equation}
we get
\begin{equation}
y(1-y)F''+(1+m_\phi-2y)F'+\tilde\Lambda\,F=-4i \Omega_2\,y(1-y)F'+2(2i \Omega_2+\Omega_1)\,y\,F .
\label{eq:F-source}
\end{equation}
The left-hand side has the form of Gauss hypergeometric equation
\begin{equation}
    y(1-y)F''+[c-(a+b+1)y]F'-abF = 0
\end{equation}
with
\begin{equation}
c=1+m_\phi,\qquad a+b=1,\qquad ab=-\tilde\Lambda\equiv-{k}({k}+1),\quad a=-{k},\ b={k}+1 .
\end{equation}
Notice that $c=1+m_\phi$ and $a+b=1$ hold for all $(\Omega_1,\Omega_2)$. All the dependence on
$\Omega_{1,2}$ resides in $\tilde\Lambda$ (i.e.\ in ${k}$) and in the source. For $\Omega_1=\Omega_2=0$ one has $\tilde\Lambda=\Lambda=\ell(\ell+1)$, $a=-\ell$, $b=\ell+1$,
$c=1+m_\phi$ (the standard spherical harmonics).

\subsubsection{The Legendre basis and its contiguous relations}\label{legendre_basis}
In order to lighten the notation, in the following we will use $m_\phi=m$ and define
\begin{equation}
F_k(y)\equiv{}_2F_1(-k,k+1;1+m;y),\qquad \ P_k(y)\equiv\rho(y)F_k(y).
\label{eq:Fk}
\end{equation}

At $y=1$ the hypergeometric function satisfies Chu-Vandermonde formula 
\begin{equation}\label{23}
F_k(1)=\frac{(m-k)_k}{(1+m)_k},\qquad (m-k)_k=(m-k)(m-k+1)\cdots(m-1).
\end{equation}
For integer $k\ge m\ge1$ the product contains a zero factor, so $F_k(1)=0$, more precisely $F_k=O((1-y)^m)$.
For integer $0\le k<m$ all factors are positive, $F_k(1)\neq0$ and $S$ is singular (for $m=0$ every integer $k\ge0$ is regular).  Moreover, for integer $k\ge0$, $F_k$ is a polynomial of degree $k$, 
hence the $F_k$, are linearly independent for $k\ge m$. They can be related to Ferrers functions or, equivalently, to Legendre associated functions through $\rho F_k (y) =P_k (y)= \Gamma(m+1) P^{-m}_k(2y-1) $ .

Moreover, since ${}_2F_1$ is symmetric in its first two parameters, $F_{-k-1}=F_k$: for integer index, negative indices generate no new functions, while $0\le k<m$ gives only singular ones. For any $k$ and ${m}$, one has  
\begin{align}
(2k+1)(1-2y)F_k&=(k+m+1)F_{k+1}+(k-m)F_{k-1},\label{eq:L1}\\[2pt]
y(1-y)F_k'&=-\frac m2F_k+\frac{k(k+m+1)F_{k+1}-(k-m)(k+1)F_{k-1}}{2(2k+1)}.\label{eq:L2}
\end{align}

Combining \eqref{eq:L1} with $y=\frac12-\frac12(1-2y)$ yields
\begin{equation}
yF_k=\frac12F_k-\frac{(k+m+1)F_{k+1}+(k-m)F_{k-1}}{2(2k+1)} .
\label{eq:yF}
\end{equation}

\subsubsection{Exact three-term recurrence}\label{Exact_three_term}

Using Gauss hypergeometric equation 
$$
y(1-y)F''_k+(1+m-2y)F'_k+k(k+1)F_k=0
$$
with the standard redefinition\footnote{In analogy with the MST approach \cite{Cipriani:2025ikx} we introduce an `UN-renormalized angular momentum' $\nu$ that will turn out to be simply $\nu=\ell=a-{1\over 2}$.}
$$k=\nu +n$$
and inserting the expansion $F=\sum_nC_nF_{\nu+n}$ into Eq.~\eqref{eq:F-source}, one finds 
\begin{align}
4{i}\Omega_2 y(1-y)F_k'&=-2{i}\Omega_2 mF_k+\frac{2{i}\Omega_2}{2k+1}\Big[k(k+m+1)F_{k+1}-(k-m)(k+1)F_{k-1}\Big],\\
-2(2{i}\Omega_2+\Omega_1)yF_k&=-(2{i}\Omega_2+\Omega_1)F_k+\frac{2{i}\Omega_2+\Omega_1}{2k+1}\Big[(k+m+1)F_{k+1}+(k-m)F_{k-1}\Big].
\end{align}
thanks to Eqs.~\eqref{eq:L2}--\eqref{eq:yF}.
Adding up the three contributions:
\begin{itemize}
\item \emph{Diagonal:} $\tilde\Lambda-k(k+1)-2{i}\Omega_2 m-(2{i}\Omega_2+\Omega_1)=\Lambda-\Omega_2^2-k(k+1)$ from \eqref{eq:Ltilde}. The shift $2{i}\Omega_2(1+m)+\Omega_1$ of $\tilde\Lambda$ is exactly canceled by the diagonal pieces of the source. This is the explicit link with the $\tilde\beta$-trick of Sec.~V.A in \cite{Bianchi-Bini-DiRusso}. 
\item \emph{$F_{k+1}$:} $\dfrac{(k+m+1)}{2k+1}\big[2{i}\Omega_2 k+2{i}\Omega_2+\Omega_1\big]=\dfrac{(k+m+1)\big[\Omega_1+2{i}\Omega_2(k+1)\big]}{2k+1}$;
\item \emph{$F_{k-1}$:} $\dfrac{(k-m)}{2k+1}\big[-2{i}\Omega_2(k+1)+2{i}\Omega_2+\Omega_1\big]=\dfrac{(k-m)\big[\Omega_1-2{i}\Omega_2 k\big]}{2k+1}$.
\end{itemize}
Collecting the coefficient of each $F_{\nu+n}$: $C_n$ from the diagonal with $k=\nu+n$, $C_{n-1}$ from the ``up'' term with $k=\nu+n-1$, and $C_{n+1}$ from the ``down'' term with $k=\nu+n+1$ and imposing that $S$ solves Eq.~\eqref{angular_equation} one finds
\begin{equation}
\alpha_nC_{n+1}+\beta_nC_n+\gamma_nC_{n-1}=0,
\label{eq:rec}
\end{equation}
where
\begin{align}
\beta_n&=\Lambda-\Omega_2^2-(\nu+n)(\nu+n+1),\\
\alpha_n&=\frac{(\nu+n+1-m)\,\big[\Omega_1-2i\Omega_2(\nu+n+1)\big]}{2\nu+2n+3}, \label{alpha}\\
\gamma_n&=\frac{(\nu+n+m)\,\big[\Omega_1+2i\Omega_2(\nu+n)\big]}{2\nu+2n-1}.
\end{align}
The structure is that of the (radial) MST recursion: the perturbations only appear off-diagonally if $\Lambda-\Omega_2^2$ is assigned the role of the unperturbed eigenvalue usually played by $\Lambda_0=\ell(\ell+1)$.

Only the products $t_n\equiv\alpha_n\gamma_{n+1}$ enter the continued fraction
\begin{equation}
    t_n=\frac{(\nu +n+1-m)(\nu +n +1+m)\,\big[\Omega_1-2i\Omega_2(\nu+n+1)\big]\big[\Omega_1+2i\Omega_2(\nu+n+1)\big]}{(2\nu+2n +3)(2\nu+ 2n+1)},
    \label{eq:tn}
\end{equation}
Henceforth we find it convenient to redefine $k$  by a unit shift $${k} \equiv \nu + n +1$$
and define
\begin{equation}\label{tkappa}
t_n \equiv T ({k}) = \frac{({k}-{m})({k}+{m})\,\big[\Omega_1-2i\Omega_2{k}\big]\big[\Omega_1+2i\Omega_2{k}\big]}{(2{k}+1)(2{k}-1)}
=\frac{({k}^2-{m}^2)\,\big(\Omega_1^2+4\Omega_2^2{k}^2\big)}{4{k}^2-1}.
\end{equation}
Each eigenvalue is an even function of $\Omega_1$ and of $\Omega_2$ separately, consistent with \eqref{eq:parity} and with the fact that Eq. \eqref{angular_equation} only involves $\Omega_2$ through $\Omega_2^2$, and it is
analytic in a neighborhood of $\Omega_1=\Omega_2=0$. The radius of convergence of the
Taylor expansion is finite, being limited by branch points in the complex domain where
two eigenvalues with the same $m=m_\phi$ coalesce, as for spheroidal eigenvalues.

With regard to the asymptotic behavior and convergence of the MST series one should study the large $n$ behaviour of the coefficients. For $|n|\to\infty$, $\alpha_n\simeq-{i}\Omega_2 n$, $\gamma_n\simeq{i}\Omega_2 n$, $\beta_n\simeq-n^2$. Setting $R_n\equiv C_n/C_{n-1}$ the ratio of the coefficients of the minimal solution (with $C_0=1$)  at $n\to+\infty$ from \eqref{eq:rec} reads
\begin{equation}
R_n=-\frac{\gamma_n}{\beta_n+\alpha_nR_{n+1}}\;\Rightarrow\;R_n\simeq\frac{{i}\Omega_2}{n},\qquad C_n\simeq\frac{({i}\Omega_2)^n}{n!}\ (\text{up to a constant}).
\end{equation}
Since the $F_k$, $k\ge m$, are bounded on $[0,1]$, the series built on the minimal solution converges absolutely and uniformly, together with its derivatives, for any (not necessarily small) $\Omega_{1,2}$.

\subsubsection{Closure at integer \texorpdfstring{$\nu=\ell$}{nu=ell}}\label{sec:closure}

Let us finally consider the case $\nu=\ell\in\mathbb{N}$, $\ell\ge m$. Throughout this subsection we keep the two labels distinct: $k$ denotes the degree of a basis function, $F_k$ , exactly as in the previously section \ref{legendre_basis}; $n$ labels the equations of the recurrence, that is, eq.~\eqref{eq:rec} with index $n$ is the coefficient of $F_{\ell+n}$ obtained after inserting the MST ansatz, and $C_{n-1},C_n,C_{n+1}$ are the coefficients of $F_{\ell+n-1},F_{\ell+n},F_{\ell+n+1}$. The two labels are related by $k=\ell+n$. We denote by
\begin{equation}
    c_k\equiv C_{k-\ell},
    \label{eq:ck}
\end{equation}
the coefficient of $F_k$ in the expansion. Consider the equation that projects onto $F_{m-1}$, i.e.\ eq. \eqref{eq:rec} with
\begin{equation}
n_*\equiv m-\ell-1\le -1 .
\end{equation}
Its coefficient $\alpha_{n_*}$ multiplies $C_{n_*+1}=c_m$. From eq. \eqref{alpha}, $\alpha_{n}$ contains the factor
$(\ell+n+1-m)$, so that
\begin{equation}
\alpha_{n_*}=0 .
\label{eq:alphazero}
\end{equation}
In the language of the contributions listed in Section \ref{Exact_three_term}, this is the statement that the
``down'' term generated by $F_k$, proportional to $(k-m)F_{k-1}$, vanishes for $k=m$: acting on $F_m$, the operator in eq. \eqref{eq:F-source} produces only $F_m$ and $F_{m+1}$, and never $F_{m-1}$. As a consequence, the subspace spanned by $\{F_k\}_{k\ge m}$ is invariant under that operator, and starting from degrees $k\ge m$ one never generates degrees $k<m$.

If one sets $C_n=0$ for $n<m-\ell$, i.e.\ $c_k=0$ for $k<m$ one then finds
\begin{itemize}
\item the equations with $n\le n_*-1$ (projections onto $F_k$, $k\le m-2$) involve only
      vanishing coefficients and are identically satisfied;
\item the equation with $n=n_*$ (projection onto $F_{m-1}$) reduces to
      $\alpha_{n_*}C_{n_*+1}=0$. This would force $c_m=0$, and hence the trivial
      solution, were it not for eq.~\eqref{eq:alphazero}, which makes it an identity;
\item the equation with $n=m-\ell$ (projection onto $F_m$) reduces to the two-term relation
      $\alpha_{m-\ell}C_{m-\ell+1}+\beta_{m-\ell}C_{m-\ell}=0$, since $\gamma_{m-\ell}$ multiplies $C_{m-\ell-1}=0$.
\end{itemize}
Equivalently, the product entering the continued fraction, eq.~\eqref{eq:tn}, vanishes at
$n=n_*$, $t_{n_*}=\alpha_{n_*}\gamma_{n_*+1}=0$, so the continued fraction towards negative $n$
terminates.

By the regularity analysis of Section~\ref{legendre_basis}, the functions $\rho F_k$ with $0\le k<m$ are precisely the singular ones, cf.\ eq. \eqref{23}. Therefore, we can state
\begin{quote}
    For $\nu=\ell\in\mathbb{N}$ the regular solution is 
\begin{equation}
S=e^{2i\Omega_2 y}\,\rho(y)\sum_{k\ge m}c_k\,F_k(y),
\label{eq:Sregular}
\end{equation}
and no additional regularity condition is needed.
\end{quote}

Three remarks complete the argument. (i) By the completeness and redundancy properties discussed in
Section \ref{legendre_basis}, every regular eigenfunction admits an expansion
$e^{-2i\Omega_2 y}S=\rho \sum_{k\ge m}c_kF_k$. An integer $\nu$ is therefore not an ansatz but a
consequence of regularity, and the only unknown is $\Lambda$. A two-sided series with integer
$\nu$ would in any case be redundant, since $F_{-k-1}=F_k$.

(ii) Written in terms of $c_k$, the chain $\{F_k\}_{k\ge m}$ and its coefficients are the same
for every $\ell\ge m$: $\nu=\ell$ is only a label, specifying which root of the eigenvalue condition is selected, namely the one that reduces to $\Lambda_0 =\ell(\ell+1)$ at $\Omega_1=\Omega_2=0$.

(iii) This is the crucial difference with the radial problem, where $\Lambda$ is given,
$\nu\notin\mathbb{Z}$ is the unknown and a genuinely two-sided series is required. In the
radial MST chain the vanishing of $\beta_n$ at $n=-2\ell-1$ (for $\nu\simeq\ell$) produces
$L_{-2\ell-1}\equiv C_{-2\ell-1}/C_{-2\ell}\sim 1/\epsilon$. Here that point is never
reached, since $n_{\min}=m-\ell>-2\ell-1$. The only remaining condition is that $C_n$ be the
minimal solution as $n\to+\infty$.

\subsubsection{Continued-fraction condition}\label{cf}

Define $L_n\equiv C_n/C_{n+1}$. From \eqref{eq:rec}, $R_n=-\gamma_n/(\beta_n+\alpha_nR_{n+1})$ and $L_n=-\alpha_n/(\beta_n+\gamma_nL_{n-1})$. The equation at $n=0$,
$\beta_0+\alpha_0R_1+\gamma_0L_{-1}=0$, becomes
\begin{equation}
\beta_0-\frac{\alpha_0\gamma_1}{\beta_1-\dfrac{\alpha_1\gamma_2}{\beta_2-\cdots}}
-\frac{\alpha_{-1}\gamma_0}{\beta_{-1}-\dfrac{\alpha_{-2}\gamma_{-1}}{\beta_{-2}-\cdots-\dfrac{\alpha_{m-\ell}\gamma_{m-\ell+1}}{\beta_{m-\ell}}}}=0,
\label{eq:CF-ang}
\end{equation}
where the lower chain terminates at $k=m$ ($n_{\min}=m-\ell$) automatically, since
$t_{n_{\min}-1}=T(m)=0$, $T({k})$ being proportional to ${k}^2-m^2$; this is the closure
property of Section~\ref{sec:closure}, $\alpha_{n_*}=0$ with $n_*=n_{\min}-1$. The upper fraction is
infinite, while the lower one is finite and contains $\ell-m$ levels. Following MST,
the existence of a minimal solution as $n\to+\infty$ is equivalent to \eqref{eq:CF-ang}.
Writing
\begin{equation}
\Lambda=\ell(\ell+1)+\Omega_2^2+\Delta,\qquad b_n\equiv n(n+2\ell+1),
\label{eq:Delta-def}
\end{equation}
one has $\beta_n=\Delta-b_n$ and \eqref{eq:CF-ang} takes the fixed-point form
\begin{equation}
\Delta=\cfrac{t_0}{\Delta-b_1-\cfrac{t_1}{\Delta-b_2-\cdots}}
+\cfrac{t_{-1}}{\Delta-b_{-1}-\cfrac{t_{-2}}{\Delta-b_{-2}-\cdots-\cfrac{t_{m-\ell}}{\Delta-b_{m-\ell}}}},\qquad t_n=T(\ell+n+1)\equiv T(k).
\label{eq:Delta-fp}
\end{equation}
The second fraction is finite and contains $\ell-m$ levels; for $\ell=m+1$ it reduces to
$t_{-1}/(\Delta-b_{-1})$, and for $\ell=m$ it is absent. Equation \eqref{eq:Delta-fp} has infinitely many roots;
inverting the continued fraction at $n=0$ makes the $\ell$-th eigenvalue the root
continuously connected to $\Delta=0$ at $\Omega_1=\Omega_2=0$ (Leaver's inversion).

\subsubsection{Eigenvalues $\Lambda$}\label{sec:Lambda}

We restrict to the regime of small $\Omega_{1,2}$. Since every $t_n$ is $O(\Omega^2)$,
eq.~\eqref{eq:Delta-fp} can be solved by iteration. Rescaling $\Omega_{1,2}\to\epsilon\,\Omega_{1,2}$,
\begin{equation}
\Lambda=\ell(\ell+1)+\Omega_2^2+\sum_{j\ge1}\Delta^{(2j)},\qquad \Delta^{(2j)}=O(\epsilon^{2j}),
\label{eq:Lambda-series}
\end{equation}
where each iteration fixes two further orders. $\Delta^{(2j)}$ involves only
$t_{-j},\dots,t_{j-1}$ and $b_{\pm1},\dots,b_{\pm j}$. Because the $t_n$ depend on
$\Omega_1^2$, $\Omega_2^2$ and $m^2$ only, $\Lambda$ is separately even in $\Omega_1$ and
$\Omega_2$, invariant under $m\to-m$, and real for real $\Omega_1$ and $\Omega_2^2$.
The first two orders read
\begin{align}
\Delta^{(2)}&=-\frac{t_0}{b_1}-\frac{t_{-1}}{b_{-1}},\label{eq:D2}\\
\Delta^{(4)}&=\frac{t_0^2}{b_1^3}+\frac{t_{-1}^2}{b_{-1}^3}
+t_0t_{-1}\Big[\frac1{b_1b_{-1}^2}+\frac1{b_1^2b_{-1}}\Big]
-\frac{t_0t_1}{b_1^2b_2}-\frac{t_{-1}t_{-2}}{b_{-1}^2b_{-2}},\label{eq:D4exp}
\end{align}
with $b_{1}=2(\ell+1)$, $b_2=2(2\ell+3)$, $b_{-1}=-2\ell$, $b_{-2}=-2(2\ell-1)$ and
$t_n=T(\ell+n+1)$. Using
$T({k})$, eq.~\eqref{eq:D2} gives
\begin{equation}
\Lambda=\ell(\ell+1)
+\Omega_2^2\,\frac{2\ell(\ell+1)-2m^2-1}{(2\ell-1)(2\ell+3)}
+\Omega_1^2\,\frac{\ell(\ell+1)-3m^2}{2\ell(\ell+1)(2\ell-1)(2\ell+3)}
+O(\Omega^4).
\label{eq:Lambda2}
\end{equation}
The denominators $b_{-j}=-j(2\ell+1-j)$ vanish for $j=2\ell+1$. For $\ell=0$ this affects the order $\epsilon^2$, and for $\ell=1$ the order $\epsilon^6$.
The corresponding terms are absent, since they multiply $t_{m-\ell-1}=T(m)=0$.
Equivalently, the general-$\ell$ expressions, such as the last term of eq.~\eqref{eq:Lambda2},
are to be understood as limits at fixed integer ${M}$. Denoting by $\Lambda_{\ell m}$ the eigenvalue of the mode $(\ell,m)$, explicitly, up to $O(\Omega^8)$,
\begin{align}
\Lambda_{00}&=0+\Omega_2^2-\tfrac{1}{6}\big(\Omega_1^2+4\Omega_2^2\big)
+\tfrac{1}{1080}\big(\Omega_1^2+4\Omega_2^2\big)\big(11\Omega_1^2-4\Omega_2^2\big)\nonumber\\
&\quad-\tfrac{1}{34020}\big(\Omega_1^2+\Omega_2^2\big)\big(\Omega_1^2+4\Omega_2^2\big)\big(47\Omega_1^2-4\Omega_2^2\big),\\
\Lambda_{10}&=2+\Omega_2^2+\tfrac{1}{10}\big(\Omega_1^2-4\Omega_2^2\big)
-\tfrac{1}{7000}\big(73\Omega_1^4+296\Omega_1^2\Omega_2^2+48\Omega_2^4\big)\nonumber\\
&\quad+\tfrac{1}{2362500}\big(\Omega_1^2+\Omega_2^2\big)\big(3281\Omega_1^4+13112\Omega_1^2\Omega_2^2-144\Omega_2^4\big),\\
\Lambda_{1,\pm1}&=2+\Omega_2^2-\tfrac{1}{20}\big(\Omega_1^2+16\Omega_2^2\big)
+\tfrac{1}{56000}\big(\Omega_1^2+16\Omega_2^2\big)\big(19\Omega_1^2-16\Omega_2^2\big)\nonumber\\
&\quad-\tfrac{1}{302400000}\big(\Omega_1^2+16\Omega_2^2\big)\big(1661\Omega_1^4+7232\Omega_1^2\Omega_2^2-2304\Omega_2^4\big).
\end{align}
These values have been explicitly displayed because, in addition to being the lowest-order multipoles, they are the only cases where the formulas for a generic $\ell$ cannot be used via direct substitution, as this would result in an indeterminate form, but must instead be understood as limits. For $\ell \ge 2$, direct substitution suffices.

Two checks should be mentioned. (i) For $\Omega_1=0$, eq. \eqref{angular_equation} is the prolate spheroidal equation. In the conventions of
ref. \cite{BertiCardosoCasals} (spin weight $s=0$) one has $A_{\ell m}=\Lambda$ and $c^2=-\Omega_2^2$.
The $\Omega_2^2$ and $\Omega_2^4$ coefficients of eq.~\eqref{eq:Lambda-series} reproduce the known small-$c$
expansion of the spheroidal eigenvalue, e.g.\ $\Lambda_{00}=-\tfrac13c^2-\tfrac{2}{135}c^4+O(c^6)$.

(ii) For $\Omega_2=0$, eq. \eqref{angular_equation} reduces to the eigenvalue problem of a quantum rigid
rotor in a uniform field, $-\nabla^2_{S^2}+\Omega_1\cos\theta$. The first-order shift vanishes by
parity, and the $\Omega_1^2$ term of eq. \eqref{eq:Lambda2} coincides with the standard
second-order Stark shift of a linear molecule \cite{Autler:1955zz},
$\propto[\ell(\ell+1)-3m^2]/[\ell(\ell+1)(2\ell-1)(2\ell+3)]$.

For generic $\Omega_1$, $\Omega_2$, eq. \eqref{angular_equation} is the angular generalized (Coulomb) spheroidal
equation (DLMF \S30.12). In black-hole perturbation theory, by contrast, a linear term in the angular equation arises only for non-zero spin weight $s$, with
coefficient $-2cs$ fixed by the spheroidal parameter~\cite{BertiCardosoCasals}. In the present
background it appears already for a scalar field, with $\Omega_1$ independent of $\Omega_2$.

\subsubsection{Relation to the quantum Seiberg--Witten curve}\label{sec:qsw-bridge}

The MST label $\nu$ of the angular problem is the quantum $a$-period. In the annulus $0<|z|<1$, i.e.\
$|y|>1$, which separates the irregular point $z=0$ ($y=\infty$) from the two regular ones, the series $\sum_nC_nP_{\nu+n}$ carries the exponents $y^{\nu}$ and $y^{-\nu-1}$, since $F_k\sim y^{k},y^{-k-1}$. The factors $e^{2i\Omega_2y}$, $\rho(y)$ and $(z-1)^{1/2}$ have trivial monodromy around this cycle.  Comparing with the exponents $\tfrac12\pm a$ at $z=0$ of the solution $\Psi(z)$ of eq \eqref{Q_SW}, related to $S$ by \eqref{dictionary}, and normalised so that $u=a^2$ at ${q}=0$, gives
\begin{equation}
a=\nu+\frac{1}{2} ,
\label{eq:a-nu}
\end{equation}
exactly in ${q}$. Regularity (Section \ref{sec:closure}) then implies $a=\ell+\tfrac12$, i.e. $a-m_1=n+\tfrac12$ with $n=\ell-|m_\phi|$. The angular problem is solved by quantising the $a$-period, whereas the radial one is solved by quantising the dual $a_D$-period.

With the dictionary in eq. \eqref{dictionary} on the branch ${q}=4i\Omega_2$, the continued fraction in eq. \eqref{eq:Delta-fp} read
\begin{equation}
T({k})=-\frac{{q}^2}{4}\,\frac{({k}^2-m_1^2)({k}^2-m_3^2)}{4{k}^2-1},\qquad
b_n=n(n+2a),\qquad \Delta=\Lambda-\Omega_2^2+\tfrac14-a^2 .
\label{eq:gauge-rec}
\end{equation}
They depend on ${q}^2$, $m_1^2$ and $m_3^2$ only. Expanding \eqref{eq:Delta-fp} at generic $a$, with both chains infinite, and using \eqref{dictionary} yields
\begin{equation}
u=a^2+\frac q2\,(1-m_1-m_3)
+\frac{{q}^2}{32(a^2-1)}\Big[a^2-\tfrac54+m_1^2+m_3^2-\frac{12\,m_1^2m_3^2}{4a^2-1}\Big]+O(x^4).
\label{eq:u-series}
\end{equation}
This coincides with the result of the `quantum' Matone relation, $u=-q\,\partial_q{\cal F}_{\rm NS}$, through $O({q}^{10})$ (the comparison also fixes the normalisation of the instanton parameter to ${q}=4i\Omega_2$). The only odd term in \eqref{eq:u-series} cancels exactly the shift in \eqref{dictionary}, so that
\begin{equation}
\Lambda=a^2-\tfrac14+\Omega_2^2+\sum_{j\ge1}\Delta^{(2j)}(a),\qquad
\Delta^{(2j)}(a)=O({q}^{2j}).
\label{eq:Lambda-gauge}
\end{equation}
Since each $t_n$ carries a single factor ${q}^2({k}^2-m_3^2)$, $\Delta^{(2j)}(a)$ is a polynomial of degree at most $j$ in $m_3^2$. By \eqref{dictionary} it is therefore homogeneous of degree $2j$ in $(\Omega_1,\Omega_2)$, and at $a=\ell+\tfrac12$, $m_1=m$ it coincides with $\Delta^{(2j)}$ of \eqref{eq:Lambda-series}. The NS series and the continued-fraction expansion are thus the same expansion. In particular $j=1$ reproduces \eqref{eq:Lambda2}, and the parity \eqref{eq:parity} is manifest, since $\Omega_1\to-\Omega_1$ amounts to $m_3\to-m_3$.

The coefficients $\Delta^{(2j)}(a)$ have poles at $a=\ell+\tfrac12$, inherited from $b_{-(2\ell+1)}=0$, which first appear at order ${q}^{4\ell+2}$. They are removable. Setting $\delta=a-\ell-\tfrac12$ the first occurrence is $1/b_{-(2\ell+1)}=O(\delta^{-1})$ that multiplies $t_{-1}\cdots t_{-(2\ell+1)}$, which contains $T(m+\delta)\,T(-m+\delta)$, or $T(\delta)$ for $m=0$, and this factor is $O(\delta^2)$. Accordingly, $\Lambda$ is obtained by setting $m_1=m$ before taking $a\to\ell+\tfrac12$. The limit reproduces the eigenvalues of Section \ref{sec:Lambda} through $O({q}^{10})$. Validity to all orders is
expected from the general correspondence between confluent Heun equations and the
qSW curve \cite{BCGM,BDMR}.


\subsubsection{Eigenfunctions}

Since $y={1\over 2} (1+\chi) = \cos^2(\theta/2)$, eq. \eqref{eq:Sregular} already gives the spheroidal harmonics in closed form.
Normalised to unity, they read
\begin{equation}
S_{\ell m_\phi}(\theta)=\mathcal N_\ell\;e^{2i\Omega_2\cos^2\frac\theta2}
\sum_{k\ge m_\phi}c_k\,P_k\big(\cos^2\tfrac\theta2\big)
\end{equation}
where ${\cal N}_\ell$ a normalization constant, that we will write below, such that
\begin{equation}
\int_{-1}^{1}S_{\ell m_\phi}S_{\ell' m'_\phi} d\chi d\phi=\delta_{\ell\ell'} \delta_{m_\phi, m'_\phi},
\label{eq:Sl}
\end{equation}
and $P_k$ are Ferrers functions of order $-m_\phi$, related to associated Legendre functions by 
\begin{equation}
 P_k(y)=\Gamma(1+m)\,\mathsf P_k^{-m}(1-2y)   
\end{equation} 
The coefficients $c_k=C_{k-\ell}$ are given by the minimal solution of \eqref{eq:rec} with $C_0=1$, i.e.\
$C_{n}=\prod_{j=1}^{n}R_j$ and $C_{-n}=\prod_{j=1}^{n}L_{-j}$ in terms of the ratios of Sec. \ref{cf}. The constant ${\cal N}_\ell$ is
\begin{equation}
\mathcal N_\ell=\sqrt{\frac{\tilde\sigma}{\sigma\,\sum_k h_k\,c_k\tilde c_k}},\qquad
\sigma\equiv\sum_{k\ge m_\phi}c_k,\qquad
h_k\equiv\int_{-1}^1P_k^2\,\mathrm d\chi=\frac{2\,(m_\phi!)^2\,(k-m_\phi)!}{(2k+1)\,(k+m_\phi)!},
\label{eq:Nl}
\end{equation}
where $\tilde c_k$ and $\tilde\sigma$ are obtained from $c_k$ and $\sigma$ by $\Omega_2\to-\Omega_2$, and the root is continued from $\mathcal N_\ell=(-1)^\ell h_\ell^{-1/2}$ at $\Omega_{1,2}=0$, where $S_{\ell m_\phi}$ reduces to the normalised associated Legendre function
$$\bar{\mathsf P}^{m_\phi}_\ell\equiv[(2\ell+1)(\ell-m_\phi)!/(2(\ell+m_\phi)!)]^{1/2}\mathsf P^{m_\phi}_\ell. $$ Eq.~\eqref{eq:Nl} follows from comparing \eqref{eq:Sl} with the equivalent representation with
$\Omega_2\to-\Omega_2$ at the pole $y=0$, where $y^{-m_\phi/2}P_k=1$ for every $k$.
For real $\Omega_{1,2}$ one has $\tilde c_k=c_k^*$, so $\mathcal N_\ell^2=\sigma^*/(\sigma\sum_kh_k|c_k|^2)$:
$\mathcal N_\ell$ removes the constant phase of the series, and $S_{\ell m_\phi}$ is real, as required
by the real Sturm--Liouville problem \eqref{angular_equation}. For the complex frequencies of
quasi-normal modes (QNMs) the same expressions define $S_{\ell m_\phi}$ analytically in $\omega$, with the bilinear normalisation of \eqref{eq:Sl}.
For later purposes let us stress that for fixed $\ell$ and $m_\phi$
\begin{equation}
S_{\ell m_\phi}(\theta)e^{im_\phi \phi} = \sum_{\ell'\ge|m_\phi|} \hat{C}_{\ell'}^{(\ell,m_\phi)} Y_{\ell' m_\phi}(\theta, \phi)
\end{equation}
so that only terms with $\ell'\ge |m_\phi|$ can contribute.

\textbf{Small frequency:} For $\Omega_{1,2}=\mathcal O(\varepsilon)$ the weight of $P_{\ell\pm n}$ is $\mathcal O(\varepsilon^n)$,
\begin{equation}
C_{n}=\prod_{j=1}^{n}\frac{\gamma_j}{b_j}\,\big[1+\mathcal O(\varepsilon^2)\big],\qquad
C_{-n}=\prod_{j=1}^{n}\frac{\alpha_{-j}}{b_{-j}}\,\big[1+\mathcal O(\varepsilon^2)\big],
\label{eq:Clead}
\end{equation}
with $\alpha_n,\gamma_n$ of \eqref{alpha} at $\nu=\ell$ and $b_n$ of \eqref{eq:Delta-def}; the
second product stops at $n=\ell-m_\phi$. 

\begin{figure}[!htbp]
    \centering
    \includegraphics[width=0.42\textwidth]{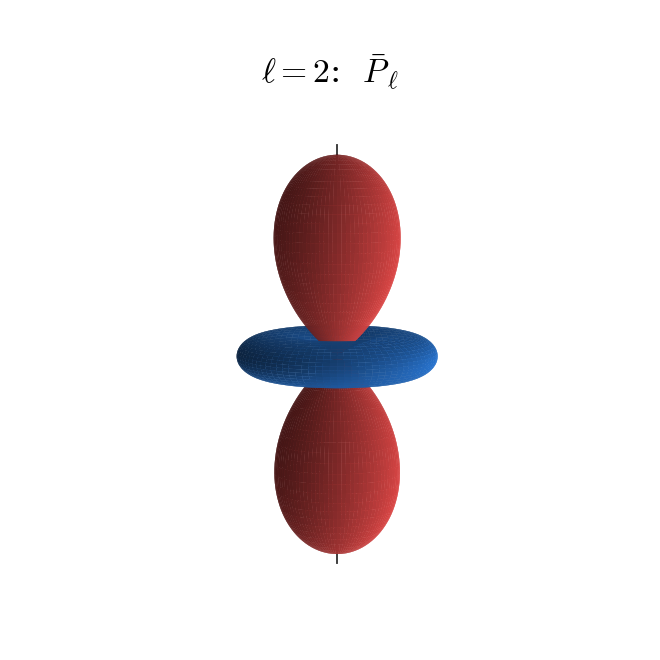}
    \hfill
    \includegraphics[width=0.42\textwidth]{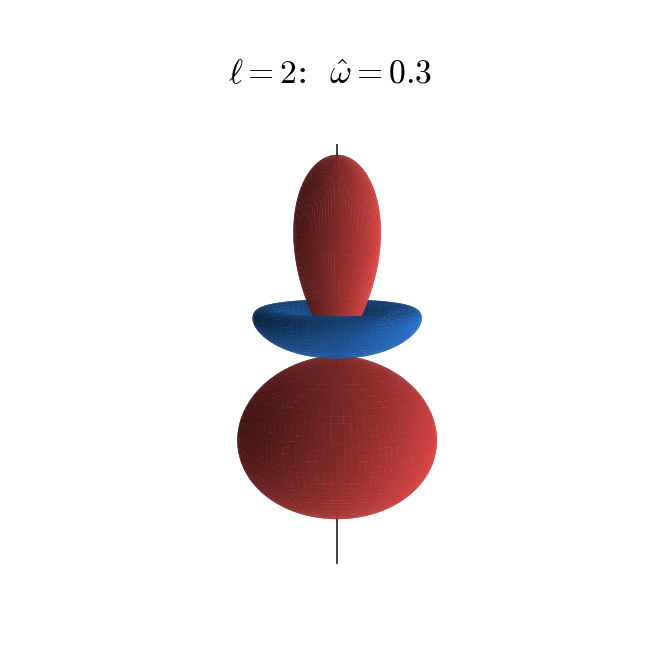}
    \vspace{0.5cm}
    \includegraphics[width=0.42\textwidth]{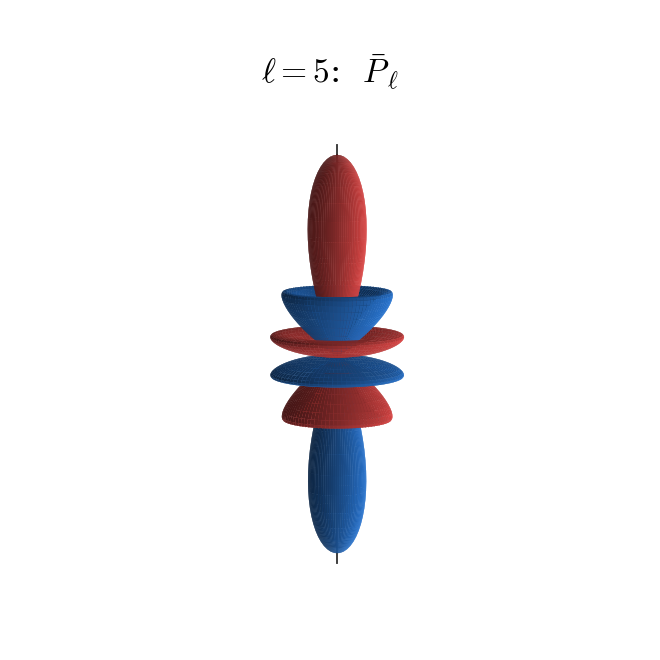}
    \hfill
    \includegraphics[width=0.42\textwidth]{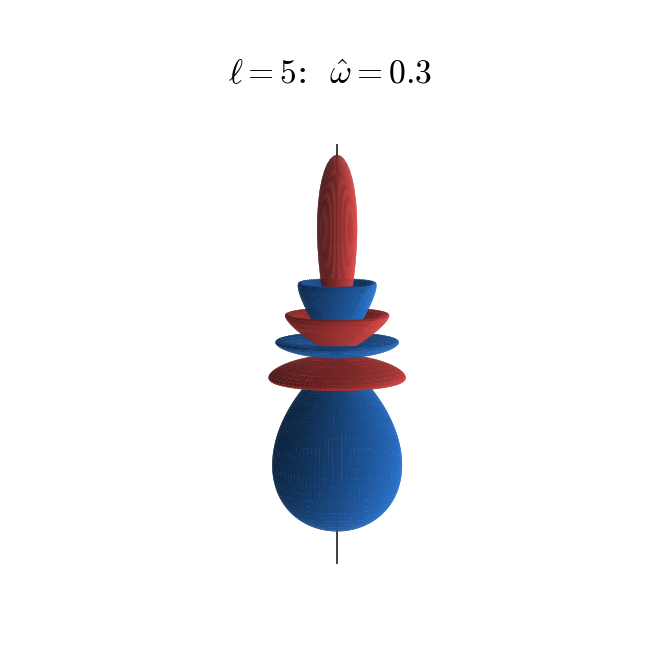}
    \caption{Polar (``lobe'') representation of the equatorially asymmetric
  harmonics $S_{\ell 0}(\theta)$ of eq. \eqref{eq:Sl}, for the $p=2$ background
  of Table \ref{tab:bg} ($q=0.88$, $N=1$), where eq. \eqref{omega_1_2} gives
  $\Omega_1=-11.51\,\omega^2$ and $\Omega_2=0.707\,\omega$.
  Each surface is the surface of revolution $r(\theta)=|S_{\ell0}(\theta)|/\max_\theta|S_{\ell0}|$
  about the polar axis (vertical segment, $\theta=0$ on top), normalised to its own
  maximum. Red (blue) marks $S_{\ell0}>0$ ($<0$), with the convention $S_{\ell0}(0)>0$,
  so that the $\ell+1$ lobes are separated by $\ell$ nodal cones.
  First and third panels: the spherical limit $\Omega_{1,2}=0$, $S_{\ell0}=\bar P_\ell(\cos\theta)$.
  Second and fourth panels: real frequency $\omega=(\ell+\tfrac12)\,\hat\omega$ with
  $\hat\omega=0.3$, i.e.\ $\omega=0.75$ ($\Omega_1=-6.48$, $\Omega_2=0.53$, $\Lambda=7.19$)
  for $\ell=2$ and $\omega=1.65$ ($\Omega_1=-31.3$, $\Omega_2=1.17$, $\Lambda=35.65$)
  for $\ell=5$. The constant $\hat\omega$ is illustrative: it stands for the light-ring
  frequency $\Omega_{\rm LR}$ in the eikonal scaling ${\rm Re}\,\omega\simeq(\ell+\tfrac12)\Omega_{\rm LR}$,
  under which the equatorial asymmetry does not decay with $\ell$, whereas at fixed $\omega$ it vanishes as $\ell^{-2}$,
  cf.\ eq. \eqref{eq:asym}.
  The Stark term $\Omega_1\cos\theta$ shifts the nodal cones towards $\theta=0$
  (from $54.7^\circ,125.3^\circ$ to $37.6^\circ,93.4^\circ$ for $\ell=2$). The southern
  lobes become broader but lower, giving $\langle\cos\theta\rangle_{\ell0}=-0.27$ ($\ell=2$)
  and $-0.34$ ($\ell=5$); axial symmetry is unaffected.
  $\ell=2$ is the lowest multipole with an equatorial lobe between the two polar ones.
  $\ell=5$ is already close to the eikonal plateau of $\langle\cos\theta\rangle_{\ell0}$ while its lobes remain resolvable.
  All curves are obtained from the exact recursion~\eqref{eq:rec}, since these
  frequencies lie beyond the radius of convergence of the small-$\omega$ expansion.}
    \label{fig:harmonic_lobes}
\end{figure}

At first order the $\Omega_2$ terms of
\eqref{eq:Sl} cancel, and
\begin{equation}
S_{\ell m_\phi}=\bar{\mathsf P}^{m_\phi}_\ell-\frac{\Omega_1}{2}\Big[\frac{{\zeta}^{(m_\phi)}_{\ell+1}}{\ell+1}\,\bar{\mathsf P}^{m_\phi}_{\ell+1}
-\frac{{\zeta}^{(m_\phi)}_\ell}{\ell}\,\bar{\mathsf P}^{m_\phi}_{\ell-1}\Big]+\mathcal O(\varepsilon^2),\qquad
{\zeta}^{(m_\phi)}_k\equiv\sqrt{\frac{k^2-m_\phi^2}{4k^2-1}},
\label{eq:S1}
\end{equation}
where $k$ is the degree of the basis function, $k=\ell+n$ as in \eqref{eq:Sregular}, and
${\zeta}^{(m_\phi)}_k=\int_{-1}^1\chi\,\bar{\mathsf P}^{m_\phi}_k\bar{\mathsf P}^{m_\phi}_{k-1}\,\mathrm d\chi$, so that
$T({k})={\zeta}_{k}^2\,(\Omega_1^2+4\Omega_2^2{k}^2)$ in \eqref{tkappa}. For $\ell=m_\phi$ the term in $\bar{\mathsf P}^{m_\phi}_{\ell-1}$ is absent, since
the chain starts at ${k}=m_\phi$ (Sec. \ref{sec:closure}) and ${\zeta}^{(m_\phi)}_{m_\phi}=0$. The deformation is the Stark mixing produced by $\Omega_1\cos\theta$. It is odd in $\Omega_1$, whereas $\Lambda$ is even: the harmonics lose equatorial symmetry while the spectrum only depends on $\Omega_1^2$. The asymmetry is measured by
\eqref{Hellmann-Feynman} and \eqref{eq:Lambda2},
\begin{equation}
\langle\cos\theta\rangle_{\ell,m_\phi}=\Omega_1\,\frac{\ell(\ell+1)-3m_\phi^2}{\ell(\ell+1)(2\ell-1)(2\ell+3)}
+\mathcal O(\varepsilon^3),
\label{eq:asym}
\end{equation}
equal to $-\Omega_1/3$ for $\ell=m_\phi=0$. By \eqref{omega_1_2}, $\langle\cos\theta\rangle=\mathcal O(\omega^2)$. A representation of these spheroidal harmonics, for some choice of the parameters, is shown in  Fig. \ref{fig:harmonic_lobes}.
\section{The radial dynamics for RKTB and RTS}\label{sec:{RKTB}}

We are now ready to address and solve the radial wave equation for scalar perturbations of RKTB in the KK-neutral sector $m_\tau=0$ with $\gamma\neq0$.  We adapt the Mano--Suzuki--Takasugi (MST) construction \cite{MST1,MST2,MST3,MST4} following ~\cite{Bianchi-Bini-DiRusso} to the bolt and identify the quantum Seiberg--Witten (qSW) period whose quantization yields the quasi-normal modes (QNMs). This allows us to compute the spectrum for the two regular, CTC-free geometries with $p=2,3$. The radial equation for scalar perturbations of RTS in \cite{BDMR} has exactly the same structure. In the simple case $r_s=r_b$, corresponding to extremal rotating strings, it was addressed in \cite{BDMR}. We therefore present the method for RKTB and give, in Section~\ref{sec:RTS}, the map between the parameters of the two backgrounds. For the RTS we then quote only the results, at the end of each subsection\footnote{While completing this project, Alba Grassi informed us that in collaboration with Michele Giorgi she had proven linear stability of RTS with respect to massless scalar perturbations by a rigorous analysis of the spectrum of QNMs. The results will appear soon.} .

\subsection{Confluent Heun form and gauge dictionary}\label{sec:radial-CHE}

Let $\hat\omega\equiv\sqrt{1-\gamma^2}\,\omega$ denote the frequency measured by an asymptotic static observer. In the rescaled and shifted radial variable
\begin{equation}\label{eq:z}
z=\frac{r_+-r}{r_+-r_-}\qquad(\text{cap }\mapsto z=0,\ \ r_-\mapsto z=1,\ \ \infty\mapsto z=-\infty)
\end{equation}
making the ansatz
\begin{equation}\label{eq:ansatz}
R(r)=e^{i\hat\omega r}\,(r-r_+)^{A}\,(r-r_-)^{B}\,H(z),
\end{equation}
with 
\begin{equation} \label{eq:ABdef}
A=\frac{p|m_\phi|}{2},\qquad B=\frac{p\,m_\phi}{2}- \kappa
\end{equation}
where
\begin{equation}\label{eq:C}
\kappa \equiv \omega \left( 2\gamma(N-{M})+3{Q}-\frac{{Q}^3}{(N-{M})^2}\right),
\end{equation}
the function $H$ satisfies a confluent Heun equation (CHE),
\begin{equation}\label{eq:CHE}
H''+\Big(\frac{\gamma_{c}}{z}+\frac{\delta_{c}}{z-1}+\eta_{c}\Big)H'+\frac{\alpha_{c}z-\beta_{c}}{z(z-1)}H=0,
\end{equation}
with parameters 
\begin{align}
\gamma_{c}&=1+2A=1+p|m_\phi|,\qquad \delta_{c}=1+2B,\qquad \eta_{c}=2i\epsilon,\\
\alpha_{c}&=2i\epsilon\,(1+A+B)+2\epsilon\tau,\qquad 
\beta_{c}=\Lambda_{\rm eff}-(A+B)(A+B+1)+i\epsilon(1+2A),
\end{align}
where
\begin{equation}\label{eq:epsLeff}
\epsilon\equiv-\hat\omega\,(r_+-r_-)\,,\qquad {\tau}\equiv2G_4{\cal M}\,\hat\omega\,\quad
\Lambda_{\rm eff}\equiv\lambda_2+\lambda_3(r_-+2r_+)+\lambda_4(r_-^2+2r_-r_++3r_+^2)=\Lambda-g_{\rm eff}\,\omega^2.
\end{equation}
with $\lambda$'s the coefficients of the polynomial $\cN_4(r)$.

The exponent $A$ at the cap does not depend on $\omega$: the radial wavefunction should be regular at the bolt $r=r_+$, as befits a smooth cap, rather than being ingoing, as befits to an event horizon.
For $A=0$ and $B=-\kappa$ one recovers exactly the parameters of \cite[eq. (4.11)]{Bianchi-Bini-DiRusso},
$\gamma_{c}=1$, $\delta_{c}=1-2\kappa$, $\eta_{c}=2i\epsilon$, $\alpha_{c}=2i\epsilon+2\epsilon(\tau-i\kappa)$, with the same sign convention for $\epsilon$. The structural difference with respect to the non-rotating Top Stars lies entirely in $\gamma_{c}=1+2A=1+p|m_\phi|\neq1$, except for $m_\phi=0$.

The radial equation can also be brought to the canonical form of the qSW curve
for $N_c=2$  and $N_f=3$, eq.~\eqref{eq:QSW}, in the variable
$z_{\rm SW}=(r_+-r_-)/(r-r_-)$, with $R(r)=\Psi(z_{\rm SW})/\sqrt{2-2z_{\rm SW}}$. Since ${\cal Q}_{\rm SW}$
depends on $q$ and on $m_1\pm m_2$ partly through their squares, the matching determines the gauge
parameters only up to discrete sign choices. In its most general form the radial dictionary reads 
\begin{equation}\label{eq:dict-signed}
\begin{aligned}
{q}&=2i\,s_q\sqrt{-\lambda_4}\,(r_--r_+)\,,\qquad
m_3=-\,\frac{i\,s_q\big[\lambda_3+2\lambda_4(r_++r_-)\big]}{2\sqrt{-\lambda_4}}\,,\\
m_1+m_2&=\frac{2\,s_+\sqrt{\cN_4(r_+)}}{r_+-r_-}\,,\qquad
m_1-m_2=\frac{2\,s_-\sqrt{\cN_4(r_-)}}{r_+-r_-}\,,\\
u&=\tfrac14+\Lambda_{\rm eff}+i\,s_x\sqrt{-\lambda_4}\,\Big[2\,s_+\sqrt{\cN_4(r_+)}-(r_+-r_-)\Big],
\end{aligned}
\end{equation}
with $s_{q},s_\pm\in\{+1,-1\}$ chosen independently; we have verified that the matching holds for all eight choices. The three signs reflect invariances of  the system under exchanges of the masses, compensated (when needed) by an action on ${q}$ and $u$.
For definiteness, we take $s_{q}=s_\pm=+1$  and starting from  $\cN_4(r_+)=\alpha^2m_\phi^2$ and $\cN_4(r_-)=(r_+-r_-)^2\big(\kappa-\tfrac{p\,m_\phi}{2}\big)^2$, define the square roots as
\begin{equation}\label{eq:branches}
\sqrt{\cN_4(r_+)}\equiv\alpha|m_\phi|,\qquad
\sqrt{\cN_4(r_-)}\equiv(r_+-r_-)\Big(\kappa-\frac{p\,m_\phi}{2}\Big),
\end{equation}
the first being the one selected by regularity at the bolt. With $\sqrt{-\lambda_4}=\hat\omega$,
$\alpha/(r_+-r_-)=p/2$ and~\eqref{eq:ansatz}--\eqref{eq:epsLeff}, the dictionary~\eqref{eq:dict-signed}
then takes the compact form
\begin{equation}\label{eq:diz}
{q}=2i\epsilon,\qquad m_1=A-B,\qquad m_2=A+B,\qquad m_3=i\tau,\qquad u=\tfrac14+\Lambda_{\rm eff}-i\epsilon(2A-1),
\end{equation}
valid for either sign of $m_\phi$.

\subsubsection{The RTS background and the parameter map}\label{sec:RTS}\label{sec:RTSCHE}

We now turn to the RTS of \cite{BDMR}. In \cite{BDMR} the radial dynamics was analysed only along the extremal line $r_s=r_b$, where $a_{\rm RTS}^2=-\alpha^2=0$ and the angular problem reduces to spherical harmonics: $S^{(0)}_{\ell,m} = Y_{\ell,m}$ with $\Lambda^{(0)}=\ell(\ell+1)$. Here we consider instead the generic smooth soliton, with  $r_+>r_b>r_s$, and $\alpha^2 = - a_{\rm RTS}^2>0$ fixed by the angular-momentum quantisation of \cite[(3.41)]{BDMR}.

The massless scalar wave equation separates as for RKTB. The radial equation reads
\begin{equation}
\frac{d}{dr}\left[\Delta(r)\,\frac{dR}{dr}\right]-\left[\Lambda+\frac{\mathcal{\widetilde{N}}_4(r)}{\Delta(r)}\right]R=0,
\qquad
\mathcal{\widetilde{N}}_4(r)=\sum_{k=0}^{4}\lambda_k\,r^k ,
\label{eq:RTSradial}
\end{equation}
where $\Lambda$ is the separation constant (denoted $A$ in \cite{BDMR}), $\Delta(r) =(r-r_+)(r-r_-)$, with
\begin{equation}\label{eq:RTSLove-rpm}
r_\pm=\begin{cases}r_b+\lambda\,(r_b-r_s)\\ r_s-\lambda\,(r_b-r_s)\end{cases},\qquad
\lambda=\frac{{h_2}}{2\sqrt{{h_2}^2-{h_1}^2}}-\frac12,
\end{equation}
Hence
\begin{equation}
r_++r_- =r_s+r_b \quad , \quad r_+-r_-=(1+2\lambda)(r_b-r_s) .
\end{equation}
In the soliton regime $r_-<r_s<r_b<r_+$: the regular singular point $r_-$ lies outside the spacetime, exactly as the inner root $r_-$ of the {RKTB}.
\footnote{In this subsection only, $p$ denotes the momentum along $y$, as in \cite{BDMR}, and not the integer $p$ of the bolt.} From \eqref{eq:RTSmq}, we know that non-axisymmetric modes start at $\ell\geq {h_2}$. Note that a frequency-independent choice ${m}_\tau=0$ would be incompatible with \eqref{eq:RTSquant} for complex $\omega$. In this sector 
\begin{equation}
\lambda_4=-\hat\omega^2,\qquad
\lambda_3=\frac{r_b-\xi^{2}r_s}{1-\xi^{2}}\,\hat\omega^2,\qquad
\lambda_2={\alpha}^2\hat\omega^2 ,
\end{equation}
so that $\hat\omega$ is the frequency measured by an asymptotic static observer, as in the {RKTB}.

For $\hat{m}_\tau=0$ the polar equation of \cite[(5.5)]{BDMR} coincides with eq. \eqref{angular_equation} upon setting
\begin{equation}
\Omega_1=0,\qquad \Omega_2^2=+{\alpha}^2\,\hat\omega^2>0 ,
\end{equation}
i.e.\ it is the equatorially symmetric, prolate case of Section \ref{angular}: the recursion \eqref{eq:rec}, the eigenvalues \eqref{eq:Lambda-series} and the quantisation $a=\ell+\tfrac12$ apply verbatim. A peculiarity of the RTS is that, by (3.9) and (3.41) of \cite{BDMR},
\begin{equation}
+{\alpha}^2=\left(\frac{{h_1}}{2{h_2}}\right)^{2}(r_+-r_-)^2
\qquad\Longrightarrow\qquad
\Omega_2=\frac{{h_1}}{2{h_2}}\,(r_+-r_-)\,\hat\omega ,
\label{eq:RTSOmega2}
\end{equation}
so that the spheroidal parameter is fixed by the orbifold data and by the same length $r_+-r_-$ that controls the radial problem (see eq.~\eqref{eq:RTSxang} below).

\textbf{Exponents at $r_\pm$:} the value of the radial potential at the cap is fixed by the integer~\eqref{eq:RTSquant}: one finds 
\begin{equation}
\mathcal{\widetilde{N}}_4(r_+)=\frac{(r_+-r_-)^2}{4}\,{K}^2
\qquad\Longrightarrow\qquad
\mathsf A=\frac{|K|}{2}\ \xrightarrow{\ \hat{m}_\tau=0\ }\ \mathsf A=\frac{{h_1}\,|m_\phi|}{2{h_2}} .
\label{eq:RTSA}
\end{equation}

The geometric origin is transparent. As for the bolt, the exponent at the cap does not depend on $\omega$: the boundary condition at $r=r_+$ is regularity rather than an ingoing condition, and $2\mathsf A={h_1}|m|/{h_2}\in\mathbb N_{0}$ by \eqref{eq:RTSmq}.
$\mathcal{\widetilde{N}}_4(r_-)$ is a perfect square as well. 

In the sector $\hat{m}_\tau=0$ we choose the root analytic in $\omega$,
\begin{equation}
\mathsf B=\frac{{h_1}\,m}{2{h_2}}-\kappa,\qquad
\kappa\equiv\omega\,\frac{r_s^{3/2}\sqrt{r_b-r_s}}{r_+-r_s} .
\label{eq:RTSB}
\end{equation}

With the ansatz \eqref{eq:ansatz} for $(A,B)\to(\mathsf A,\mathsf B)$, i.e.\ $\mathcal{\widetilde{N}}_4(r_+)=(r_+-r_-)^2\mathsf A^2$ and $\mathcal{\widetilde{N}}_4(r_-)=(r_+-r_-)^2\mathsf B^2$, the function $H$ satisfies the CHE \eqref{eq:CHE} with parameters of exactly the same form as in the {RKTB} case, where now
\begin{equation}
\begin{aligned}
&\epsilon\equiv-\hat\omega\,(r_+-r_-),\qquad
\tau\equiv-\frac{\lambda_3+2\lambda_4(r_++r_-)}{2\hat\omega},\\
&\Lambda_{\rm eff}\equiv \Lambda+\Big[\lambda_2+\lambda_3(r_-+2r_+)+\lambda_4\big(r_-^2+2r_-r_++3r_+^2\big)\Big].
\end{aligned}
\label{eq:RTSeps}
\end{equation}
In the sector $\hat{m}_\tau=0$ the whole explicit frequency dependence of $\Lambda_{\rm eff}$ is quadratic,
\begin{equation}
\Lambda_{\rm eff}=\Lambda(\hat\omega)-g_{\rm RTS}\,\hat\omega^2,\qquad
g_{\rm RTS}=-{\alpha}^2+\frac{\xi^{2}r_s-r_b}{1-\xi^{2}}\,(r_-+2r_+)+(r_s+r_b)^2+2r_+^2 ,
\label{eq:RTSgeff}
\end{equation}
the RTS counterpart of \eqref{eq:epsLeff}, and $\tau$ is the Coulomb parameter of the reduced four-dimensional geometry,
\begin{equation}
\tau=2G_4{\cal M}\,\hat\omega,\qquad
4G_4{\cal M}=\frac{2r_s+r_b-\xi^{2}(r_s+2r_b)}{1-\xi^{2}} ,
\label{eq:RTStau}
\end{equation}
with ${\cal M}$ the ADM mass read off from the asymptotic expansion (3.59) of~\cite{BDMR} after reduction along $\hat\tau$.\footnote{The corresponding expression~(3.67) of \cite{BDMR} contains a misprint in the sign of the term proportional to $\xi^{2}$; eq.~\eqref{eq:RTStau} is the one consistent both with~(3.59) and with the large-$r$ behaviour of the wave equation.}

\textbf{Parameter map:} the RTS is therefore obtained from the {RKTB} by the replacements collected in Table~\ref{tab:map}. In particular, the integer $p$ of the bolt is replaced by the rational number ${h_1}/{h_2}$, and $\kappa_{RKTB}$ by $\kappa_{RTS}$.
\begin{table}[h]
\centering\small
\renewcommand{\arraystretch}{1.3}
\begin{tabular}{@{}lll@{}}
\toprule
 & {RKTB} ($m_\tau=0$) & RTS ($\hat p=0$)\\
\midrule
radial potential & $\mathcal N_4(r)$ & $\mathcal{\widetilde{N}}_4(r)$, eq.~\eqref{eq:RTSradial}\\
asymptotic frequency & $\hat\omega=\sqrt{1-\gamma^2}\,\omega$ & $\hat\omega=\sqrt{1-\xi^{2}}\,\omega$\\
$r_+-r_-$ & $\dfrac{2\alpha}{p}$ & $\dfrac{2\alpha{h_2}}{{h_1}}$, eq.~\eqref{eq:RTSOmega2}\\
integer at the cap & $p$ & ${h_1}/{h_2}$\\
azimuthal number & $m_\phi\in\mathbb Z$ & $m\in {h_2}\mathbb Z$, eq.~\eqref{eq:RTSmq}\\
exponent at $r_+$ & $A=p|m_\phi|/2$ & $\mathsf A={h_1}|m|/(2{h_2})$, eq.~\eqref{eq:RTSA}\\
exponent at $r_-$ & $B=p\,m_\phi/2-\kappa$ & $\mathsf B={h_1}\,m/(2{h_2})-\kappa$, eq.~\eqref{eq:RTSB}\\
$O(\omega)$ shift & $\kappa$, eq.~\eqref{eq:C} & $\kappa$ , eq.~\eqref{eq:epsRTS}\\
$\epsilon$ & $-\hat\omega\,(r_+-r_-)$ & $-\hat\omega\,(r_+-r_-)$\\
$\tau$ & $2G_4{\cal M}\hat\omega$ & $2G_4{\cal M}\hat\omega$, eq.~\eqref{eq:RTStau}\\
$\Lambda_{\rm eff}$ & eq.~\eqref{eq:epsLeff} & eq.~\eqref{eq:RTSeps}, \eqref{eq:RTSgeff}\\
angular problem $(\Omega_1,\Omega_2)$ & eq.~\eqref{angular_equation} & $\big(0,\alpha\hat\omega\big)$, eq.~\eqref{eq:RTSOmega2}\\
co-rotating modes & $m_\phi<0$ \ ($G_4J<0$) & $m>0$ \ ($G_4J>0$)\\
\bottomrule
\end{tabular}
\caption{Map between the parameters of the {RKTB} and of the RTS. Under this map the radial equations of the two backgrounds take the same form, and all the results of this section carry over from the {RKTB} to the RTS.}
\label{tab:map}
\end{table}

Under this map all the equations of this section hold verbatim for the RTS. In particular, in the variable $z_{\rm SW}=(r_+-r_-)/(r-r_-)=1/(1-z)$, with $R(r)=\Psi(z_{\rm SW})/\sqrt{2-2z_{\rm SW}}$, the radial equation takes identically the canonical form~\eqref{eq:QSW}, with the same dictionary \eqref{eq:diz} after replacing $(A,B)\to(\mathsf A,\mathsf B)$. The discrete sign ambiguities discussed below \eqref{eq:dict-signed} apply unchanged.\footnote{This dictionary is equivalent to (5.14) of \cite{BDMR} once two misprints are corrected there: the normalisation of the masses should read $m_{1,2}=\tfrac{1}{r_+-r_-}\big(\sqrt{\mathcal{\widetilde{N}}_4(r_+)}\mp\sqrt{\mathcal{\widetilde{N}}_4(r_-)}\big)$, as required by the characteristic exponents at $r_\pm$, and the last term in the bracket multiplying $\lambda_4$ in $u$ should read $r_-^2+2r_+r_-+3r_+^2$.}
Combining~\eqref{eq:RTSOmega2} and~\eqref{eq:RTSeps}, the instanton parameters of the angular and radial curves are rigidly related,
\begin{equation}
{q}_{\rm ang}\equiv4i\,\Omega_2=-\frac{{h_1}}{{h_2}}\,{q} ,
\label{eq:RTSxang}
\end{equation}
up to the irrelevant sign of $\Omega_2$. In the RTS the two quantum curves are thus controlled by a single expansion parameter, with a relative weight fixed by the orbifold data.
For ${h_1}\to0$ one has $a_{\rm RTS}\to0$, $r_+\to r_b$, $r_-\to r_s$, ${\xi}\to0$, and \eqref{eq:RTSA}--\eqref{eq:RTSB} give $\mathsf A=0$, $\mathsf B=-\kappa$ with $\kappa$ as in \cite[(4.6)]{Bianchi-Bini-DiRusso}. Moreover $\epsilon\to-(r_b-r_s)\omega$, $\tau\to\omega\,(r_s+r_b/2)$ and $\Lambda_{\rm eff}\to\ell(\ell+1)-\omega^2(r_s^2+r_sr_b+r_b^2)$, so that the CHE parameters reproduce exactly those of \cite[(4.11)]{Bianchi-Bini-DiRusso}, $\gamma_{c}=1$, $\delta_{c}=1-2\kappa$, $\alpha_{c}=2i\epsilon+2\epsilon(\tau-i\kappa)$.

\subsection{Generalized MST recursion}\label{sec:MST}

Following \cite{Bianchi-Bini-DiRusso} Sec. V-A, we write the CHE as
\begin{equation}\label{Gauss}
z(1-z)H''+[\gamma_{c}-(\gamma_{c}+\delta_{c})z]H'+\beta_{c}H=-\eta_{c}\, z(1-z)H'+\alpha_{c}\,z\,H
\end{equation}
and we expand $H$ in terms of a family of hyper-geometric functions -- each an eigenfunction of the Gauss operator -- whose indices are shifted by integers relative to the `renormalized angular momentum' $\nu$\footnote{Please not that $\nu$ here (radial dynamics) is different from the $\nu$ in the polar angular dynamics. The latter is `UN-renormalized' $\nu_{\rm ang} =\ell$, while the former is `renormalized' $\nu_{\rm rad} = \ell+ ...$. 
Both are exactly given by the corresponding quantum $a$-period up to a sifht by 1/2: $\nu_{\rm ang} =a_{\rm ang} -{1\over2}\neq \nu_{\rm rad} =a_{\rm rad} -{1\over2}$.}
\begin{equation}\label{eq:base}
H(z)=\sum_{n=-\infty}^{+\infty}C_n\,F_{\nu+n}(z),\qquad
F_k(z)\equiv{}_2F_1\big(k+1+A+B,\ A+B-k;\ 1+2A;\ z\big).
\end{equation}
The symmetric choice of the first two indices renders the basis invariant under $k\to-k-1$. For $A=0$ and $B=-\kappa$, one recovers the basis (5.10) from \cite{Bianchi-Bini-DiRusso}. Each $F_k$ is regular at $z=0$; therefore, the series \eqref{eq:base} is automatically the in-solution, which is regular at that point.

For $\gamma_{c}=1+2A$ (the third parameter of the ${}_2F_1$ in~\eqref{eq:base}) and arbitrary $k$, the following hold
\begin{align}
z(1-z)F_k'&=\mathcal A_+F_{k+1}+\mathcal A_0F_k+\mathcal A_-F_{k-1} \label{4.14},\\
zF_k&=\mathcal B_+F_{k+1}+\mathcal B_0F_k+\mathcal B_-F_{k-1},
\end{align}
with
\begin{align}
\mathcal A_+&=\frac{(k-A-B)(k+1+A-B)(k+1+A+B)}{2(k+1)(2k+1)}, &
\mathcal B_+&=-\frac{(k+1+A-B)(k+1+A+B)}{2(k+1)(2k+1)},\\
\mathcal A_0&=-\frac{(A-B)(k-A-B)(k+1+A+B)}{2k(k+1)}, &
\mathcal B_0&=\frac{A^2-B^2+k(k+1)}{2k(k+1)},\\
\mathcal A_-&=-\frac{(k-A-B)(k-A+B)(k+1+A+B)}{2k(2k+1)}, &
\mathcal B_-&=-\frac{(k-A-B)(k-A+B)}{2k(2k+1)}\label{4.18}.
\end{align}
We plug \eqref{eq:base} into the CHE. The Gauss operator acts diagonally, $F_k\mapsto(k+1+A+B)(A+B-k)F_k$; the ``source'' terms are redistributed among $F_{k\pm1}$ via the above formulae. Collecting the coefficient of $F_{\nu+n}$ yields the central result: the coefficients satisfy the three-term recursion relation
\begin{equation}
\alpha_nC_{n+1}+\beta_nC_n+\gamma_nC_{n-1}=0   
\end{equation}
with $k\equiv\nu+n$ and
\begin{align}
\alpha_n&=-\frac{i\epsilon\,(k+1+i\tau)(k+1-m_2)(k+1-m_1)}{(k+1)(2k+3)},\label{eq:alpha}\\
\gamma_n&=\frac{i\epsilon\,(k-i\tau)(k+m_1)(k+m_2)}{k(2k-1)},\label{eq:gamma}\\
\beta_n&=\Leff-\epsilon\tau-k(k+1)-\frac{\epsilon\tau\,m_1m_2}{k(k+1)}.\label{eq:beta}
\end{align}
where $m_1=A-B$, $m_2=A+B$, $m_1m_2=A^2-B^2$. The three masses enter the continued fraction only through
\begin{equation}\label{product}
    \alpha_n\gamma_{n+1}=-\frac{{q}^2}{4}\,T(k+1),\qquad T(k)=\frac{(k^2-m_1^2)(k^2-m_2^2)(k^2-m_3^2)}{k^2(4k^2-1)}.
\end{equation}
Notice that $T(k)$ looks different from \eqref{tkappa}. There one of the masses was set to zero $m_2=0$ by choosing $m_\tau=0$ zero KK momentum.
The renormalised angular momentum $\nu$ is fixed by requiring a minimal solution in both directions:
\begin{equation}\label{eq:CF}
\beta_0+\alpha_0R_1+\gamma_0L_{-1}=0,\qquad R_n=-\frac{\gamma_n}{\beta_n+\alpha_nR_{n+1}},\quad L_n=-\frac{\alpha_n}{\beta_n+\gamma_nL_{n-1}}.
\end{equation}
At small frequency it gives $\nu=\ell+\big[\Leff-\ell(\ell+1)-\epsilon\tau\big]/(2\ell+1)+O(\epsilon^2)$, which reduces to eq. (5.19) of \cite{Bianchi-Bini-DiRusso} at $A = 0$.

\textbf{Small-frequency structure:} under $(\omega,m_\phi)\to(-\omega,-m_\phi)$ one has $A\to A$, $B\to-B$, hence $m_1\leftrightarrow m_2$, while $(\epsilon,\tau)\to(-\epsilon,-\tau)$ and $\Lambda_{\rm eff}$ is invariant; $\beta_n$ and $\alpha_n\gamma_{n+1}$ are invariant, and therefore
\begin{equation}
\nu(\omega,m_\phi)=\nu(-\omega,-m_\phi).
\end{equation}
At fixed $m_\phi\neq0$, however, $\nu$ is not even in $\omega$, in contrast with the static Top Star \cite{Bianchi-Bini-DiRusso}. The odd part originates from
\begin{equation}
m_1m_2=A^2-B^2=p\,m_\phi\,\kappa-\kappa^2=O(\omega),
\end{equation}
and solving \eqref{eq:CF} by iteration gives
\begin{equation}
\begin{aligned}
\nu=\ell+\frac{1}{2\ell+1}\bigg[&\Lambda_{\rm eff}-\ell(\ell+1)-\epsilon\tau\left(1+\frac{m_1m_2}{\ell(\ell+1)}\right)+\epsilon^2\left(\frac{T(\ell+1)}{2(\ell+1)}-\frac{T(\ell)}{2\ell}\right)\bigg]+O(\omega^4),
\end{aligned}
\label{eq:nusmall}
\end{equation}
whose leading odd term is $O(\omega^3)$, proportional to $m_\phi$. Finally, as $\omega\to0$ the masses tend to integers, $\{m_1,m_2\}\to\{0,2A\}$, so that
\begin{equation}
T(k)\big|_{\omega=0}=\frac{k^2\,(k^2-4A^2)}{4k^2-1} .
\end{equation}
The additional zero at $k=2A$, absent in the static case, modifies the pattern of suppression of the coefficients $C_n$ at small frequency with respect to (5.43) of \cite{Bianchi-Bini-DiRusso}.

A comparison with the polar-angular problem is enlightening.
\begin{itemize}
\item \textbf{Angular problem:} Both boundary conditions sit at regular singular points, $\chi = \cos\theta=\pm1$. The coefficient of the descending term vanishes at integer $k=|m_\phi|$, so the chain closes in itself and regularity enforces $\nu_{\rm ang}=\ell$; the unknown is $\Lambda$.
\item \textbf{Radial problem:} One boundary condition sits at a regular point (the bolt) and the other at the irregular point at infinity. The zeros of $\alpha_n$ occur at $k+1=m_{1,2}$, which are not integers for $\omega\neq0$, so the chain does not close. Here $\Lambda$ is a datum and $\nu_{\rm rad}$ is a computable, generically complex, quantity.
\end{itemize}
The quantum $a$-period of the qSW curve with dictionary~\eqref{eq:diz} satisfies $a=\nu+\tfrac12$, as found perturbatively for Top Star \cite[App. A]{Bianchi-Bini-DiRusso} and rigorously shown in \cite{HPoghosyan}.

\subsubsection{RTS.}\label{sec:RTSMST}
With the map of Table~\ref{tab:map}, eqs.~\eqref{eq:base}--\eqref{eq:CF} hold verbatim, with $m_{1,2}=\mathsf A\mp\mathsf B$ and $(\epsilon,\tau,\Lambda_{\rm eff})$ given by \eqref{eq:RTSeps}. The third parameter $\gamma_{c}=1+2\mathsf A=1+{h_1}|m|/{h_2}$ is a positive integer: every $F_k$ is regular at $z=0$, while the second Frobenius solution, $\propto z^{-2\mathsf A}$ (logarithmic for $m=0$), is excluded by regularity at the cap. For $\mathsf A=0$, $\mathsf B=-\kappa$, eqs.~\eqref{eq:alpha}--\eqref{eq:beta} reduce exactly to the In-recursion (5.16) of \cite{Bianchi-Bini-DiRusso}, with $\Lambda_{\rm eff}=\ell(\ell+1)-\tfrac13(4\tau^2+\tau\epsilon+\epsilon^2)+\epsilon\tau$.
The odd part of $\nu$ now originates from $m_1m_2=\frac{{h_1}\,m}{{h_2}}\,\kappa-\kappa^2$, and is proportional to $m$. We have verified \eqref{eq:nusmall} against the exact root of \eqref{eq:CF}, the residual scaling as $\omega^4$. In the static limit the $O(\omega^2)$ coefficient of~\eqref{eq:nusmall} reproduces exactly eq. (5.19) of \cite{Bianchi-Bini-DiRusso}. As $\omega\to0$, $\{m_1,m_2\}\to\{0,N\}$ with $N=2\mathsf A={h_1}|m|/{{h_2}}<|m|\leq\ell$.

\subsection{Quantisation of the dual period $a_D$}\label{sec:radial-aD}
Three coordinates adapted to the different steps of the construction appear in the following:
the MST variable $z=(r_+-r)/(r_+-r_-)$ of~\eqref{eq:z}, which places the cap at $z=0$;
the variable
\begin{equation}
z_B\equiv\frac{r-r_-}{r_+-r_-}=1-z ,
\end{equation}
which places the two regular singular points at $z_B=0$ ($r_-$) and $z_B=1$ (cap) and the
irregular one at $z_B=\infty$, as required by the connection formulae of \cite{Bonelli-Iossa-Lichtig-Tanzini}; and $z_{\rm SW}=1/z_B$, in which the radial equation takes the canonical qSW form~\eqref{eq:QSW}. They are related by M\"obius transformations that map singular points to singular points and preserve the characteristic exponents, so the
boundary conditions and the quantum periods $a$, $a_D$ (defined by monodromies around cycles
mapped into themselves, up to the orientation reversal $a\to-a$, under which the
quantisation condition is invariant) do not depend on this choice. The only
convention-dependent quantity is the accessory parameter: the parameter $u_{\rm BPZ}$ of the
BPZ equation in $z_B$ is related to the Coulomb-branch parameter $u$ of~\eqref{eq:diz} by
\begin{equation}
u_{\rm BPZ}=\tfrac14-u-\tfrac{q}{2}\,(m_1+m_2-1)=-\Lambda_{\rm eff}.
\end{equation}
In the variable $z_B$ the normal form of the radial equation for $\Psi=\Delta^{1/2}R$
coincides with the semiclassical BPZ equation (3.2.40) of~\cite{Bonelli-Iossa-Lichtig-Tanzini}, after the following identifications: $a_0^2=B^2$, $a_1^2=A^2$, $L={q}=2i\epsilon$, $m=m_3=i\tau$.

Taking the In-solution regular at the bolt, $a_1=|A|$, and requiring the absence of the incoming wave at infinity, the connection formulae (3.2.47)--(3.2.48) \cite{Bonelli-Iossa-Lichtig-Tanzini} then give the quantization condition\footnote{Notice the different choice of convention on the signs of $m_3=-\tilde{m}_3$ and $q=-L$ with respect to \cite{Bonelli-Iossa-Lichtig-Tanzini}.}
\begin{equation}\label{eq:Xi}
 {\varrho}(\omega) \equiv {{q}}^{2a}\,\frac{\Gamma^2(1-2a)}{\Gamma^2(1+2a)}\prod_{i=1,2,3}\frac{\Gamma(\frac12+m_i+a)}{\Gamma(\frac12+m_i-a)}\;e^{-\partial_a {\cal F}^{\rm NS}(a,{{q}})}=1,\qquad a=\nu(\omega)+\tfrac12 .
\end{equation}
Comparison with the Nekrasov--Shatashvili (NS) prepotential ${\cal F}^{\rm NS}(a,{{q}})$ for $N_f=(1,2)$ \cite{BCGM} shows that the left-hand side equals $\exp(-\partial_a{\cal F}^{\rm NS})$, including the tree-level and one-loop terms, at $q=-L$ and masses $(m_1,m_2,-m_3)$. With $2\pi ia_D=-\partial_a{\cal F}^{\rm NS}$, the QNM spectrum is therefore
\begin{equation}
\label{eq:aDquant}
a_D(\omega)\in\mathbb Z .
\end{equation}
while $a=\nu+\tfrac12$ (exactly) is not quantized. In this convention the quantisation coincides with that of~\cite{BCGM}; it is equivalent to $\Pi_B=2\pi(n+\frac12)$ in the conventions of~\cite{Aminov:2020yma}. The angular and the radial problems are solved by quantising two different cycles of quantum SW curves of the same kind (three flavours):  $a$ in the former case,  $a_D$ in the latter.

The instanton part of the prepotential follows from the MST recursion itself. Since $u_{\rm BPZ}=-\Leff$, quantum Matone's relation reads
\begin{equation}\label{eq:Matone}
{{q}}\,\partial_{{q}}{\cal F}^{\rm NS}(a,{{q}})=-\big[\Leff(\nu;\epsilon)-\nu(\nu+1)\big]_{m_i\ \text{fixed}}
\quad\Longrightarrow\quad
{\cal F}^{\rm NS}(a,{{q}})=-\int_0^1\frac{dt}{t}\,{\cal L}(a;t{{q}}),
\end{equation}
where ${\cal L} \equiv\Leff-\nu(\nu+1)$ is obtained from~\eqref{eq:CF} solved for $\Leff$ at fixed $\nu$ and fixed masses. The expansion
\begin{equation}\label{F_frk}
{\cal F}^{\rm NS}_{\rm inst} (a,{{q}}) =\frac{{{q}}\,m_3(4a^2-1+4m_1m_2)}{2(4a^2-1)}+\frac{{{q}}^2}{8}\Big[\frac{T(a-\frac12)}{2a-1}-\frac{T(a+\frac12)}{2a+1}\Big]+O({{q}}^3),
\end{equation}
agrees with ${\cal F}^{\rm NS}_{\rm inst}$ at one- \cite{BCGM} and at two-instantons \cite{Bianchi-Bini-DiRusso}, up to $a$-independent terms. The integral representation in~\eqref{eq:Matone} is the one we use numerically, since it does not rely on the convergence of the instanton series.

\subsubsection{RTS.}\label{sec:RTSaD}
In $z_B=(r-r_-)/(r_+-r_-)$ the normal form $\Psi=\Delta^{1/2}R$ of \eqref{eq:RTSradial} coincides identically with the semiclassical BPZ equation of \cite{Bonelli-Iossa-Lichtig-Tanzini}, with $a_0^2=\mathsf B^2$, $a_1^2=\mathsf A^2$, $L={q}$, $m=m_3$ and $u_{\rm BPZ}=-\Lambda_{\rm eff}$. Regularity at the cap selects $a_1=\mathsf A$, which, unlike for the static Top Star, is non-vanishing for $m\neq0$. The QNMs are then given by \eqref{eq:Xi} with the RTS masses $m_{1,2}=\mathsf A\mp\mathsf B$, $m_3=i\tau$, i.e.\ by \eqref{eq:aDquant}, which reduces to the static Top Star condition for ${h_1}\to0$. The representative of $a$ is fixed by continuity from $a=\ell+\tfrac12$ at $\hat\omega=0$, using~\eqref{eq:nusmall} as a seed.

A feature specific to the RTS is that the angular and radial curves share very similar instanton counting parameters. The spectrum is fixed by the simultaneous quantisation
\begin{equation}
a\big(u_{\rm ang};{q}_{\rm ang},m\big)=\ell+\tfrac12,\qquad
a_D\big(u;{q},m_1,m_2,m_3\big)\in\mathbb Z,\qquad
{q}_{\rm ang}=-\frac{{h_1}}{{h_2}}\,{q} ,
\label{eq:RTScoupled}
\end{equation}
as underlined in \eqref{eq:RTSxang}, with $u_{\rm ang}$ and $u$ linked by the separation constant ${\Lambda}$. In the {RKTB}, instead, the angular curve carries the independent mass $m_3\propto\Omega_1$. Finally, the connection coefficients of~\cite{Bonelli-Iossa-Lichtig-Tanzini} are derived as expansions in $L={q}$, and for the RTS $|{q}|=2|\hat\omega|(r_+-r_-)$ or $|{q}|=4|\hat\omega|\frac{{h_2} \alpha}{{h_1}}$ with $(r_+-r_-)={h_2}(r_b-r_s)/\sqrt{{h_2}^2-{h_1}^2}$.

\subsection{The QNMs spectrum}\label{sec:QNMspectrum}
We solve \eqref{eq:Xi} for the regular, CTC-free geometries with $p=2,3$ at $q=0.88$, $N=1$ (background data in Table \ref{tab:bg}). The accessory parameter \eqref{eq:epsLeff} depends on the frequency as $\Lambda_{\rm eff}=\Lambda(\omega)-g_{\rm eff}\,\omega^2$, with $g_{\rm eff}$ fixed by the background and $\Lambda(\omega)$ obtained from \eqref{eq:rec}; $\partial_a{\cal F}^{\rm NS}(a,{{q}})$ is computed from \eqref{eq:Matone}. Every frequency has also been determined independently from a Leaver continued fraction built directly on the radial equation, which uses neither the MST recursion nor the connection formula (Appendix \ref{app:numerics}). The two agree with a median difference of $4\times10^{-12}$, which confirms directly that the QNMs are the zeros of $1-e^{2\pi ia_D}$. We determined all modes with $\ell\le2$ in the region $0.01\le{\rm Re}\,\omega\le0.46$, $|{\rm Im}\,\omega|\le0.26$ ($p=2$) and $0.005\le{\rm Re}\,\omega\le0.30$, $|{\rm Im}\,\omega|\le0.13$ ($p=3$), in units $N=1$. We quote them in units of the ADM mass, ${\cal M}\hat\omega\equiv G_4{\cal M}\sqrt{1-\gamma^2}\,\omega$; the fundamental modes are listed in Table~\ref{tab:{RKTB}-fund}, the complete list in Table~\ref{tab:RRB-full}.

\begin{table}[t]
\centering\scriptsize
\renewcommand{\arraystretch}{1.15}\setlength{\tabcolsep}{1.5pt}
\begin{tabular}{@{}cc|lll|lll@{}}
\toprule
 & & \multicolumn{3}{c|}{$p=2$} & \multicolumn{3}{c}{$p=3$}\\
$\ell$ & $m_\phi$ & $M\hat\omega_0$ & $\frac{|{\rm Im}\,\hat\omega_0|}{{\rm Re}\,\hat\omega_0}$ & $\nu$ & $M\hat\omega_0$ & $\frac{|{\rm Im}\,\hat\omega_0|}{{\rm Re}\,\hat\omega_0}$ & $\nu$\\[2pt]
\midrule
0 & $0$ & $0.240949-i\,0.04432$ & $0.184$ & $-0.1862+i\,0.0956$ & $0.229587-i\,0.01193$ & $0.0520$ & $-0.1037+i\,0.0121$\\
\midrule
1 & $-1$ & $0.797354-i\,0.2098$ & $0.263$ & $0.4345+i\,0.7118$ & $0.986882-i\,0.1216$ & $0.123$ & $0.2425+i\,0.5928$\\
1 & $0$ & $0.520565-i\,0.01375$ & $0.0264$ & $0.5256+i\,0.1697$ & $0.489363-i\,1.686\times10^{-4}$ & $3.45\times10^{-4}$ & $0.7774+i\,1.9\times10^{-4}$\\
1 & $+1$ & $0.496718-i\,0.04046$ & $0.0815$ & $0.5882+i\,0.2955$ & $0.480514-i\,2.104\times10^{-3}$ & $4.38\times10^{-3}$ & $0.7454+i\,0.0033$\\
\midrule
2 & $-2$ & $1.331531-i\,0.3967$ & $0.298$ & $1.0136+i\,1.2500$ & $1.715789-i\,0.2696$ & $0.157$ & $0.6290+i\,1.1124$\\
2 & $-1$ & $1.107439-i\,0.1213$ & $0.110$ & $0.8807+i\,0.6449$ & $1.297160-i\,0.02814$ & $0.0217$ & $0.8698+i\,0.1517$\\
2 & $0$ & $0.813673-i\,3.493\times10^{-3}$ & $4.29\times10^{-3}$ & $1.2851+i\,0.0115$ & $0.744510-i\,7.140\times10^{-7}$ & $9.59\times10^{-7}$ & $1.7038+i\,6.5\times10^{-7}$\\
2 & $+1$ & $0.785246-i\,0.01782$ & $0.0227$ & $1.2403+i\,0.0879$ & $0.739479-i\,2.956\times10^{-5}$ & $4.00\times10^{-5}$ & $1.6855+i\,3.1\times10^{-5}$\\
2 & $+2$ & $0.762025-i\,0.03518$ & $0.0462$ & $1.2611+i\,0.1861$ & $0.733898-i\,1.889\times10^{-4}$ & $2.57\times10^{-4}$ & $1.6648+i\,2.4\times10^{-4}$\\
\bottomrule
\end{tabular}
\caption{Fundamental QNM of each $(\ell,m_\phi)$ tower and the value of $\nu=a-\frac12$ at the quasinormal frequency; in the angular problem the same quantity is fixed to $\nu=\ell$. Since $G_4J<0$, $m_\phi<0$ is co-rotating.}
\label{tab:{RKTB}-fund}
\end{table}
\textbf{Structure of the spectrum:} all the modes have ${\rm Im}\,\hat\omega<0$: there is no sign of a linear instability in this sector. For fixed $(\ell,m_\phi)$ the modes form a single tower, labelled by $n$ in order of increasing ${\rm Re}\,\hat\omega$, along which the damping grows monotonically. For $\ell\ge1$ the fundamental axisymmetric and counter-rotating modes are long-lived, with $|{\rm Im}\,\hat\omega_0|/{\rm Re}\,\hat\omega_0$ between $8\times10^{-2}$ and $10^{-6}$, while co-rotating modes ($m_\phi<0$, since $G_4{\cal J}<0$) and the monopole are strongly damped. At $\ell\le2$ the spectrum does not split into two separate branches, and $|{\rm Im}\,\hat\omega|/{\rm Re}\,\hat\omega$ does not follow $n/\ell$.

\textbf{Long-lived modes:} for $m_\phi\ge0$, $m_1=\kappa$ by~\eqref{eq:diz}, and the ratio 
\begin{equation}\label{eq:epsp}
\varepsilon_p\equiv\frac{|{q}|}{|m_1|}=\frac{2\sqrt{1-\gamma^2}\,(r_+-r_-)}{|C|}\simeq
\begin{cases}0.15 & p=2\\ 0.025 & p=3\end{cases}
\end{equation}
is independent of the frequency, while $|m_3/m_1|=O(1)$. The spectrum is therefore organised by an expansion in ${q}$ at fixed masses, in which the radial equation reduces to its near-zone (hyper-geometric) part. At ${q}\to0$, having ${\varrho}=1$ requires a pole of $\Gamma(\frac12+m_1+a)$. With the $a$-period at first order in ${q}$, given by $\beta_0=0$ in~\eqref{eq:CF},
\begin{equation}\label{eq:a0}
{v}^2-\big(\Lambda_{\rm eff}-\epsilon\tau\big){v}+\epsilon\tau\,m_1m_2=0,\qquad {v}=a_0^2-\tfrac14 ,
\end{equation}
the real part and the width of the modes follow from
\begin{equation}\label{eq:ReQNM}
\tfrac12+m_1(\omega_n^{(0)})+a_0(\omega_n^{(0)})=-n,\qquad
\omega_n\simeq\omega_n^{(0)}+\frac{(-1)^n}{n!}\,\frac{\hat{\varrho}}{\partial_\omega(m_1+a_0)}\bigg|_{\omega_n^{(0)}},
\end{equation}
where $\hat{\varrho}={\varrho}/\Gamma(\frac12+m_1+a)$, so that $|{\rm Im}\,\omega_n|\propto|{q}|^{2\,{\rm Re}(a)}$. For the fundamental long-lived modes of $p=3$ (except $\ell=m_\phi=1$) and for $p=2$, $\ell=2$, $m_\phi=0$, eqs.~\eqref{eq:a0}--\eqref{eq:ReQNM} reproduce ${\rm Re}\,\omega$ to better than $1\%$ and, including the two-instanton term of~\eqref{F_frk}, ${\rm Im}\,\omega$ within $2\%$. The width is thus suppressed as a power of $|{q}|=2|\hat\omega|(r_+-r_-)$. This accounts for the narrowness of the $p=3$ modes, whose $|{q}|$ is about five times smaller, for the growth of the lifetime with $\ell$, and for the short-lived monopole (${\rm Re}\,a\simeq0.3$--$0.4$). For co-rotating modes the shift $p|m_\phi|$ in $m_1$ moves the pole to higher frequency, where $|{q}|$ is no longer small. In the MST language, \eqref{eq:ReQNM} is the vanishing of $\gamma_{n+1}\propto\nu+n+1+m_1$ in~\eqref{eq:gamma}: the chain truncates as in the angular closure of Section~\ref{sec:closure}, but at the frequency-dependent point $k=-m_1(\omega)$ and only up to $O({q}^{2a})$.

\underline{\textbf{$a$ versus $a_D$:}} at every quasi-normal frequency $a=\nu+\frac12$ is not a half-integer (Table~\ref{tab:{RKTB}-fund}): $\nu$ is genuinely complex for the damped modes, and nearly real but far from an integer for the long-lived ones, e.g.\ $\nu=1.7038$ for $p=3$, $\ell=2$, $m_\phi=0$, where the angular problem has $\nu=\ell=2$. Meanwhile $e^{2\pi ia_D}=1$ to the working accuracy. The radial spectrum of QNMs arises from quantisation of the dual period, $a_D\in\mathbb N$, the angular one on $a=\ell+\frac12$.

\subsubsection{RTS.}\label{sec:RTSQNM}
We solve \eqref{eq:Xi}, with the map of Table~\ref{tab:map}, for two smooth geometries with $({h_1},{h_2})=(1,2)$ and $r_b/r_s=1.2,\,1.7$, in units of $r_s=1$ (background data in Table~\ref{tab:RTSbg}). By \eqref{eq:RTSmq}, ${h_2}=2$ is the smallest value admitting non-axisymmetric KK-neutral modes with $\ell\le2$: $m=0$ for $\ell\le1$ and $m\in\{0,\pm2\}$ for $\ell=2$, odd $m$ belonging to the KK-charged sector $\hat{m}_\tau\neq0$. Both geometries lie in the window $1<r_b/r_s<2$, where static Top Stars are free of Gregory--Laflamme instabilities \cite{BDMR,Bah:2021irr,Bena:2024hoh,Dima:2024cok}, on the two sides of $r_b/r_s=3/2$, where the light ring of the axisymmetric modes reaches the cap. The accessory parameter is $\Lambda_{\rm eff}=\Lambda(\hat\omega)-g_{\rm RTS}\,\hat\omega^2$, cf.~\eqref{eq:RTSgeff}, with $\Lambda(\hat\omega)$ the prolate eigenvalue of \eqref{eq:rec} at $\Omega_1=0$. As for the {RKTB}, every frequency has also been determined independently from a Leaver continued fraction (Appendix~\ref{app:numerics-RTS}); the two agree with a median difference of $7\times10^{-14}$. We determined all modes with $\ell\le2$ in the region $0.02\le{\rm Re}\,\hat\omega\le3$, $|{\rm Im}\,\hat\omega|\le1.5$ ($r_b/r_s=1.2$) and $0.02\le{\rm Re}\,\hat\omega\le3$, $|{\rm Im}\,\hat\omega|\le2.75$ ($r_b/r_s=1.7$), in units $r_s=1$; in both geometries they are the first seven members, $n\le6$, of each tower. Since $\hat\omega$ is already the frequency measured by an asymptotic static observer, we quote ${\cal M}\hat\omega$  (for $G_4=1$), directly comparable with Table~\ref{tab:{RKTB}-fund}; the fundamental modes are listed in Table~\ref{tab:RTS-fund}, the complete list in Table~\ref{tab:RTS-full}.

\begin{table}[t]
\centering\scriptsize
\renewcommand{\arraystretch}{1.15}\setlength{\tabcolsep}{1.5pt}
\begin{tabular}{@{}cc|lll|lll@{}}
\toprule
 & & \multicolumn{3}{c|}{$r_b/r_s=1.2$} & \multicolumn{3}{c}{$r_b/r_s=1.7$}\\
$\ell$ & $m$ & $M\hat\omega_0$ & $\frac{|{\rm Im}\,\hat\omega_0|}{{\rm Re}\,\hat\omega_0}$ & $\nu$ & $M\hat\omega_0$ & $\frac{|{\rm Im}\,\hat\omega_0|}{{\rm Re}\,\hat\omega_0}$ & $\nu$\\[2pt]
\midrule
0 & $0$ & $0.229049-i\,0.05344$ & $0.233$ & $-0.2296+i\,0.1941$ & $0.246273-i\,0.1456$ & $0.591$ & $-0.0570+i\,0.3477$\\
\midrule
1 & $0$ & $0.496702-i\,0.02386$ & $0.0480$ & $0.4879+i\,0.1714$ & $0.559580-i\,0.1210$ & $0.216$ & $0.4810+i\,0.4621$\\
\midrule
2 & $-2$ & $0.763533-i\,0.02429$ & $0.0318$ & $1.0939+i\,0.2536$ & $0.855051-i\,0.1983$ & $0.232$ & $1.2736+i\,0.6704$\\
2 & $0$ & $0.781289-i\,0.01103$ & $0.0141$ & $1.0494+i\,0.1616$ & $0.897825-i\,0.1133$ & $0.126$ & $1.0145+i\,0.5239$\\
2 & $+2$ & $0.912980-i\,0.06730$ & $0.0737$ & $0.8413+i\,0.4655$ & $0.960589-i\,0.2744$ & $0.286$ & $1.1869+i\,0.8201$\\
\bottomrule
\end{tabular}
\caption{Fundamental QNM of each $(\ell,m)$ tower of the RTS with $({h_1},{h_2})=(1,2)$, and the value of $\nu=a-\frac12$ at the quasinormal frequency; in the angular problem the same quantity is fixed to $\nu=\ell$. Since ${\cal J}>0$, $m>0$ is co-rotating.}
\label{tab:RTS-fund}
\end{table}
\textbf{Structure of the spectrum:} as for the {RKTB}, all the modes have ${\rm Im}\,\hat\omega<0$ and, for fixed $(\ell,m)$, form a single tower along which the damping grows monotonically. At every quasinormal frequency $a=\nu+\frac12$ is not a half-integer (Table~\ref{tab:RTS-fund}), while $e^{2\pi ia_D}=1$ to the working accuracy, in agreement with the simultaneous quantisation \eqref{eq:RTScoupled}. In units of the ADM mass the real parts are comparable to those of the {RKTB}, e.g.\ ${\cal M}\,{\rm Re}\,\hat\omega_0=0.781,\ 0.898$ for $\ell=2$, $m=0$ ($r_b/r_s=1.2,\,1.7$), against $0.814,\ 0.745$ ($p=2,3$), but no mode is long-lived. At $r_b/r_s=1.2$ the least damped modes, axisymmetric and counter-rotating ($m<0$), have $|{\rm Im}\,\hat\omega_0|/{\rm Re}\,\hat\omega_0$ between $1.4\times10^{-2}$ and $5\times10^{-2}$ for $\ell\ge1$; the smallest value, for $\ell=2$, $m=0$, exceeds its $p=2$ and $p=3$ counterparts by a factor $3$ and by four orders of magnitude. At $r_b/r_s=1.7$ all the modes have $|{\rm Im}\,\hat\omega|/{\rm Re}\,\hat\omega\gtrsim0.12$. The ordering in $m$ is consistent with the eikonal limit, in which for $m\neq0$ the barrier at the cap, $\mathcal{\widetilde{N}}_4 (r_+)= \alpha^2m^2$, merges with the co-rotating light ring.

\textbf{Near-zone parameter:} the mechanism of the long-lived {RKTB} modes carries over with $p|m_\phi|\to {h_1}|m|/{h_2}$. By \eqref{eq:RTSA}--\eqref{eq:RTSB}, the mass proportional to the frequency is $m_2=-\kappa$ for $m\le0$, while for $m>0$ it is shifted, $m_2={h_1}m/{h_2}-\kappa$. The ratio
\begin{equation}\label{eq:epsRTS}
\varepsilon_{\rm RTS}\equiv\frac{|{q}|}{|\kappa|}=\frac{2\omega\sqrt{1-\xi^{2}}\,(r_+-r_-)}{\kappa}\simeq
\begin{cases}0.22 & r_b/r_s=1.2\\ 1.46 & r_b/r_s=1.7\end{cases},\qquad \kappa \equiv \frac{\omega \,r_s^{3/2}\sqrt{r_b-r_s}}{r_+-r_s},
\end{equation}
is independent of the frequency and larger than $\varepsilon_p$ in \eqref{eq:epsp}. At ${q}\to0$, ${\varrho}=1$ now requires a pole of $\Gamma(\frac12+m_2+a)$, $\frac12+m_2(\hat\omega^{(0)}_n)+a_0(\hat\omega^{(0)}_n)=-n$, with $a_0$ from \eqref{eq:a0}. For $r_b/r_s=1.2$ and $\ell=2$, $m\le0$, this reproduces ${\rm Re}\,\hat\omega_0$ within $0.4\%$, but the first-order width correction of \eqref{eq:ReQNM} misses ${\rm Im}\,\hat\omega_0$ by $30$--$90\%$; for $\ell\le1$, for $m>0$ and for all the modes at $r_b/r_s=1.7$ the condition has no solution close to the QNMs. The RTS geometries thus lie at the border ($r_b/r_s=1.2$) or outside ($r_b/r_s=1.7$) of the near-zone regime that produces the narrow resonances of the {RKTB}, which accounts for the absence of long-lived modes. As for the {RKTB}, the shift ${h_1}m/{h_2}$ moves the pole of co-rotating modes to higher frequency, where $|{q}|$ is larger, and these modes are the most damped.


\subsection{Reality of Love numbers}\label{sec:{RKTB}-Love}

Another observale property of non-singular horizonless micro-state geometries is their deformability that is measured by real Love numbers for small but non-zero real values of the frequency $\omega$. The Love numbers vanish in the static (zero-frequency) limit as for BHs in $D=4$, that however acquire purely imaginary Love numbers when $\omega\neq 0$. The `dissipative' nature of this effect is naturally associated to the presence of the horizon, wherein dynamical perturbations can be damped (see \cite{Rodriguez:2026iot} for a recent review and \cite{Hui:2020xxx,Charalambous:2021mea} for recent attempts to explain the vanishing of the static Love numbers for BHs  case). The tidal response of a BH or fuzzball can be extracted from the behaviour of the radial wave-function in the  near zone
$${\cal M} \sim r_+-r_-\ll r\ll\hat\omega^{-1}$$
Recall that the stationary scalar wave-function can be written as
\begin{equation}
\Phi_\omega(r,\theta, \phi) = \sum_{\ell, m}R_{\omega, \ell, m}(r) S_{\omega,\ell,m}(\theta)e^{i{m}\phi}
\end{equation}
where we made explicit the dependence of both $R(r)$ and $S(\theta)$ on $\omega$ that enters both via the terms inthe equations proportional to $\omega$ and $\omega^2$ and via the dependence of Carter constant $\Lambda$ on $\omega$.
Following \cite{CFMP, Bianchi:2022qph}, the near zone is captured by a double-scaling limit of the radial curve \eqref{eq:QSW}, in which both gauge couplings $z$ and $q/z$ are small. With $z\equiv z_{\rm SW}=(r_+-r_-)/(r-r_-)$, $ r_+-r_-=2\alpha/p$ and $q=2i\epsilon=-4i\hat\omega \alpha/p$, this is
\begin{equation}\label{eq:Love-region}
z\approx\frac{r_+-r_-}{r}\ll1,\qquad\Big|\frac qz\Big|\approx2\hat\omega\,r\ll1
\qquad\Longleftrightarrow\qquad 2\frac{\alpha}{p}\ll r\ll\hat\omega^{-1}.
\end{equation}
In this limit the solution regular at the bolt can be written as
\begin{equation}\label{eq:Love-def}
\Psi_{\omega,\ell,m} (z)\underset{z\to0}{\sim}z^{-\ell-\Delta\nu(\ell,m, \omega)}+z^{\ell+1+\Delta\nu(\ell,m, \omega)}\,\mathcal L\,\big[1+\dots\big],\qquad a=\ell+\tfrac12+\Delta\nu=\nu+\tfrac12 ,
\end{equation}
with $a$ the quantum period of the radial curve, so that $a\to\ell+\frac12$ as $\omega\to0$.
The dynamical Love and dissipation numbers are identified with the real and imaginary parts of \cite{LeTiec:2020spy,Charalambous:2021mea,Combaluzier--Szteinsznaider:2025eoc},
\begin{equation}\label{eq:Love-LTheta}
\mathcal L=\mathcal{L}_R+i\,\mathcal{L}_I .
\end{equation}
Equivalently, the radial solution behaves as
\begin{equation}
\label{eq:Love-k}
R_\ell(r)\simeq c_s\,r^{a(\ell)-\frac12}\big[1+\dots\big]+c_r\,r^{-a(\ell)-\frac12}\big[1+\dots\big],\qquad \widehat{\cal L}(\omega)\equiv\frac{c_r}{c_s}\equiv (r_+-r_-)^{2a}\,\mathcal L(\omega).
\end{equation}
In the double-scaling limit the irregular point at infinity, and with it $m_3$, decouples.
The two masses that survive are those associated to the behaviors at the regular points $r_+$ and $r_-$, i.e.\ $m_1=A-B$ and $m_2=A+B$ of \eqref{eq:diz}.
The radial equation then becomes hyper-geometric, with exponents $\pm A$ at $r_+$, $\pm B$ at $r_-$ and $-\frac12\pm a$ at infinity. Its solution regular at the bolt is
\begin{equation}\label{eq:Love-nz}
R(r)=(r-r_+)^{A}\,(r-r_-)^{B}\;{}_2F_1\Big(\tfrac12+m_2+a,\ \tfrac12+m_2-a;\ 1+2A;\ \frac{r_+-r}{r_+-r_-}\Big).
\end{equation}
In the notation of \cite{CFMP} this is the solution $\Psi_-\sim(1-z)^{\frac12+k_0}$, with $k_0=A$ selected by regularity rather than by an ingoing condition, $k_0\pm p_0=m_{1,2}$ and $p_2=a$.
The connection formula at infinity gives\footnote{This is eq.~(4.28) of \cite{CFMP} with $p_2\to a$ and $k_0\pm p_0\to m_{1,2}$; the overall sign of their eq.~(6.3) differs from the one obtained directly from \eqref{eq:Love-nz}. At $\omega=0$ we have checked that \eqref{eq:Love-Gamma} coincides with the exact ratio of the hyper-geometric connection coefficients.}
\begin{align}
\mathcal L&=\frac{F^{-1}_{-+}\big[\begin{smallmatrix}k_0&k_{12}\\ p_0&p_2\end{smallmatrix}\big]}{F^{-1}_{--}\big[\begin{smallmatrix}k_0&k_{12}\\ p_0&p_2\end{smallmatrix}\big]}
=\frac{\Gamma(-2p_2)\,\Gamma(\frac12+k_0+p_0+p_2)\,\Gamma(\frac12+k_0-p_0+p_2)}{\Gamma(2p_2)\,\Gamma(\frac12+k_0+p_0-p_2)\,\Gamma(\frac12+k_0-p_0-p_2)}\nonumber\\
&=\frac{\Gamma(-2a)\,\Gamma(\frac12+a+m_1)\,\Gamma(\frac12+a+m_2)}{\Gamma(2a)\,\Gamma(\frac12-a+m_1)\,\Gamma(\frac12-a+m_2)} .\label{eq:Love-Gamma}
\end{align}

\noindent\textbf{Reality: the masses.}
Since $\overline{\Gamma(z)}=\Gamma(\bar z)$, \eqref{eq:Love-Gamma} is real whenever $m_1,m_2\in\mathbb R$ and $a\in\mathbb R$.
The first condition is a direct consequence of smoothness.
By \eqref{eq:ansatz} and \eqref{eq:diz}, it is clear that $m_1$ and $m_2$ are real:\eqref{eq:branches}
\begin{equation}\label{eq:Love-masses}
m_\phi\ge0:\ \ m_1=\kappa,\ \ m_2=p m_\phi-\kappa;\qquad m_\phi<0:\ \ m_1=-p m_\phi+\kappa,\ \ m_2=-\kappa\,.
\end{equation}
Both are real for real $\omega$.
Moreover $2A=p|m_\phi|\in\mathbb N_{0}$ automatically, by the regularity condition \eqref{periodicity of phi}. Unlike for the RTS,  $m_\phi$ is unrestricted.

Two features of the geometry are responsible. The exponent at the bolt is $\omega$-independent, because the boundary condition there is regularity and not an ingoing condition. The point $r_-$ lies outside the spacetime, which ends at the bolt $r=r_+$.
For a (rotating) black hole the horizon exponent is instead $\propto i(\omega-m_\phi\Omega_H)/\kappa_H$, so that $m_{1,2}\notin\mathbb R$ and $\mathcal{L}_I\neq0$ already for real $a$; the purely imaginary Kerr result are given in \cite[eq.~(113)]{BILT-Kerr}, see also \cite{LeTiec:2020spy,Charalambous:2021mea}.

\noindent\textbf{Reality: the quantum period.}
By \eqref{eq:Matone}, the quantum Matone relation for the radial curve can be written at fixed $\Lambda_{\rm eff}$ as
\begin{equation}\label{eq:Love-Matone}
\Lambda_{\rm eff}+\tfrac14=a^2-q\,\partial_q{\cal F}^{\rm NS}_{\rm inst}(a;q,m_1,m_2,m_3),\qquad m_3=i\tau .
\end{equation}
We use $\Lambda_{\rm eff}$ rather than $u$, because $u=\Lambda_{\rm eff}+\frac14+\frac q2(1-m_1-m_2)$ has an imaginary part $-i\epsilon(2A-1)$ for real $\omega$. This term is an artefact of the dictionary, and cancels against the corresponding term of $q\partial_q{\cal F}^{\rm NS}_{\rm inst}$.
By contrast $\Lambda_{\rm eff}$ is real. The coefficients $\lambda_i$  of $N_4$ are real for real $\omega$, and so is the angular eigenvalue $\Lambda$, which is the eigenvalue of the real Sturm--Liouville problem \eqref{angular_equation}.

The recursion \eqref{eq:alpha}--\eqref{eq:beta}, from which ${\cal F}^{\rm NS}_{\rm inst}$ is obtained, depends on $(q,m_3)$ only through ${q}^2$, ${q}m_3=-2\epsilon\tau$ and $m_3^2$, cf.\ \eqref{product}.
It does so via rational functions with real coefficients of $(a,m_1,m_2)$. This is the invariance $s_q\to-s_q$ below \eqref{eq:dict-signed}.
Writing ${\cal F}^{\rm NS}_{\rm inst}=\sum_{n\ge1}q^n{\cal F}^{\rm NS}_n$, this implies ${\cal F}^{\rm NS}_n(a;m_1,m_2,-m_3)=(-1)^n{\cal F}^{\rm NS}_n(a;m_1,m_2,m_3)$.
At real $\omega$ the combinations ${q}^2$, ${q}\,m_3$ and $m_3^2$ are real, so every term ${q}^n{\cal F}^{\rm NS}_n$ is real for real $a$. The first two terms from \eqref{F_frk} are
\begin{equation}\label{eq:Love-F12}
q{\cal F}^{\rm NS}_1=-\epsilon\tau\Big(1+\frac{4m_1m_2}{4a^2-1}\Big),\qquad
{q}^2{\cal F}^{\rm NS}_2=-\frac{\epsilon^2}{2}\Big[\frac{T(a-\frac12)}{2a-1}-\frac{T(a+\frac12)}{2a+1}\Big],
\end{equation}
with $T\propto k^2-m_3^2=k^2+\tau^2$ as in \eqref{product}. Inverting \eqref{eq:Love-Matone} order by order gives
\begin{equation}\label{eq:Love-aexp}
a=\sum_{k\ge0}a_nq^n,\qquad a_0=\sqrt{\Lambda_{\rm eff}+\tfrac14},\qquad
a_1=\frac{m_3}{4a_0}\Big(1+\frac{4m_1m_2}{4a_0^2-1}\Big),\qquad a_n\in i^n\,\mathbb R .
\end{equation}
In other words $a_{2n+1}=-\bar a_{2n+1}$ and $a_{2n}=\bar a_{2n}$. Each term $a_nq^n$ is real, and so is $a$ at every order.
For $a_0=\ell+\frac12$, \eqref{eq:Love-aexp} reproduces the small-frequency expansion of $\nu$ below \eqref{eq:CF}, completed as in \eqref{eq:nusmall}.
At this point the expansion also meets the small denominator $\beta_{-2\ell-1}\simeq(2\ell+1)\delta$, with $\delta=a-\ell-\frac12$. It enters only at high order. Since \eqref{eq:CF} is linear in $\delta$ at leading order, with real coefficient $-(2\ell+1)$, its solution remains real order by order for $\ell\ge1$.

As shown in Section~\ref{sec:MST}, $a(\omega)$ contains odd powers of $\omega$, proportional to $m_\phi$, through $B$, and the only symmetry is $\nu(\omega,m_\phi)=\nu(-\omega,-m_\phi)$.
Unlike for static Top Stars \cite[eq.~(5.23)]{Bianchi-Bini-DiRusso}, reality therefore does not follow from the parity of the expansion in $\omega$, but from the reality structure of the coefficients.

Beyond perturbation theory, $\cos2\pi a$ is real for real $\omega$ (Appendix~\ref{app:monodromy}). Hence $a$ can leave the real axis only when $2a\in\mathbb N$, where $\Gamma(-2a)$ is singular and the split \eqref{eq:Love-def} into source and response breaks down. This is the threshold $\hat\omega_*$.
For the $p=2$ background of Table~\ref{tab:bg}, integrating the radial equation along the $A$-cycle, we find
\begin{equation}\label{eq:Love-threshold}
{\cal M}\hat\omega_*\simeq0.297,\ 0.481,\ 0.430,\ 0.717,\ 0.673\qquad\text{for}\qquad(\ell,m_\phi)=(0,0),\ (1,0),\ (1,1),\ (2,0),\ (2,1).
\end{equation}
In all these cases $\nu$ decreases from $\ell$ and reaches $\ell-\frac12$, i.e.\ $2a=2\ell$, at $\hat\omega_*$. Above the threshold $a$ acquires an imaginary part, and $\mathcal L$ is no longer real.

\noindent\textbf{Static limit:}
For $\omega\to0$ one has $a=\ell+\frac12+\delta$ with $\delta=O(\hat\omega^2)$, while $\{m_1,m_2\}\to\{0,p|m_\phi|\}$, up to $\pm\kappa=O(\omega)$.
Here $p|m_\phi|$ can exceed $\ell$, and two cases arise.
For $p|m_\phi|\le\ell$ the poles of $\Gamma(-2a)$ and $\Gamma(\frac12-a+m_{1,2})$ compensate, and
\begin{equation}\label{eq:Love-static1}
\lim_{\omega\to0}\mathcal L=(-1)^{p|m_\phi|+1}\,\frac{(\ell!)^2\,(\ell+p|m_\phi|)!\,(\ell-p|m_\phi|)!}{2\,(2\ell+1)!\,(2\ell)!}\;\lim_{\omega\to0}\frac{\omega^2C^2}{\delta},
\end{equation}
which is real and finite. For $p|m_\phi|>\ell$, the factor $\Gamma(\frac12-a+m_{1,2})$ carrying $p|m_\phi|$ tends to $(p|m_\phi|-\ell-1)!$, only one denominator pole survives, and
\begin{equation}\label{eq:Love-static2}
\mathcal L\simeq(-1)^{\ell}\,{\rm sgn}(m_\phi)\,\frac{(\ell!)^2\,(\ell+p|m_\phi|)!}{2\,(2\ell+1)!\,(2\ell)!\,(p|m_\phi|-\ell-1)!}\;\frac{\kappa}{\delta}\ \propto\ \frac1\omega .
\end{equation}
$\mathcal L$ stays real but seems to have no finite static limit. It changes sign with $\omega\,m_\phi$, as required by $\nu(\omega,m_\phi)=\nu(-\omega,-m_\phi)$. Actually, for $\omega =0 $, the radial equation becomes an associated Legendre equation and the solution can exist only if $-\ell \le p m_\phi \le +\ell$, so $p|m_\phi|>\ell$  has to be excluded.
\begin{figure}[!htbp]
    \centering
    \includegraphics[trim=0pt 15pt 0pt 0pt, clip,width=0.7\linewidth]{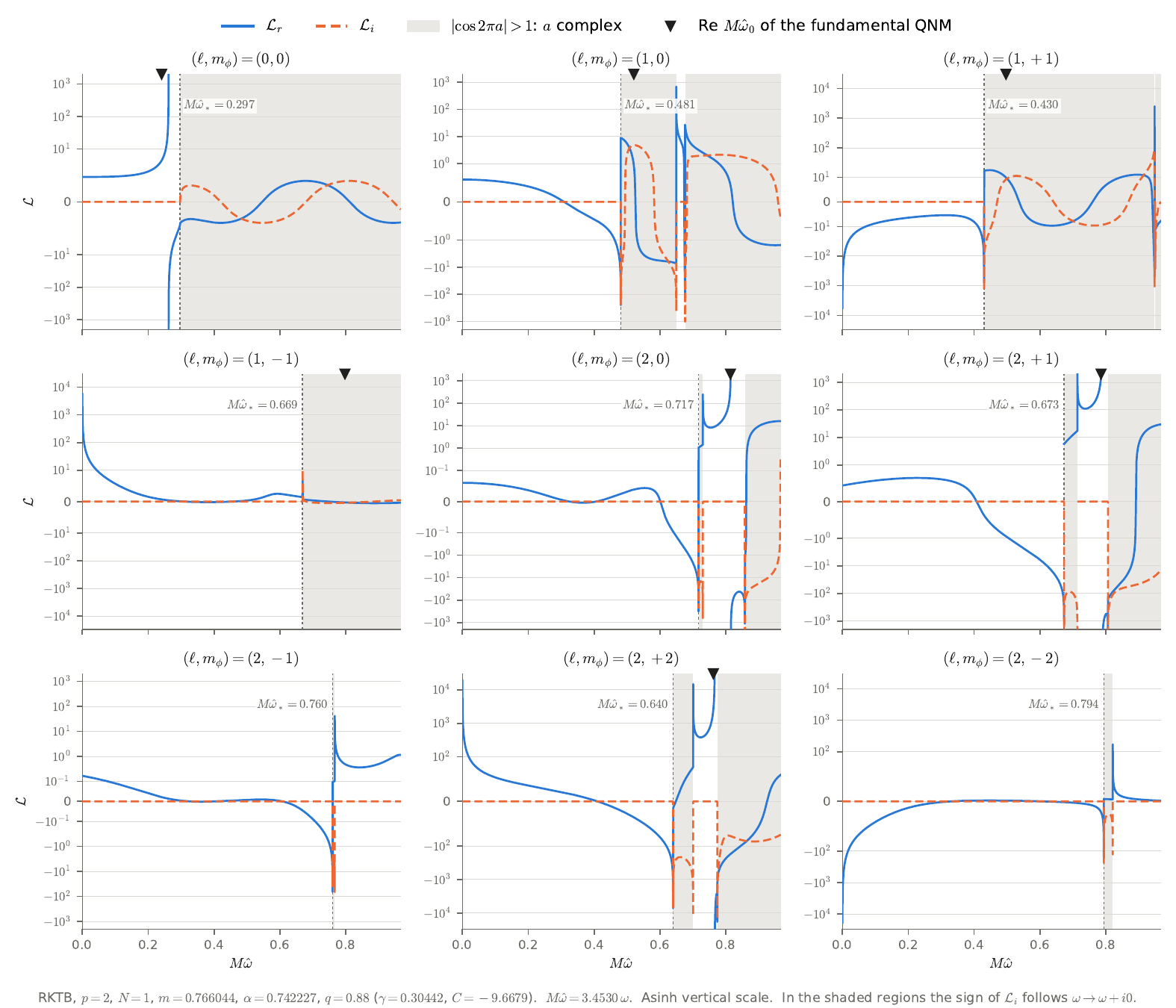}
    \caption{Plot of love numbers for several values of $(\ell,m_\phi)$ for the solution $(p,M,\alpha,Q)=(2,0.766,0.742,0.88)$. The red curves are $\mathcal{L}_I$ while the blue ones are $\mathcal{L}_R$. The shaded regions give $a$ complex because we cross an $\hat\omega_*$. In all cases $|\mathcal{L}_i|<10^{-11}$ in the unshaded regions.}
    \label{fig:Love_RKTB}
\end{figure}
%


At this order, however, the response exponent $r^{-a-\frac12}$ is degenerate, up to $r^{2\delta}$, with the $(2\ell+1)$-th term of the source series. That term has a real coefficient of the same order, as in \cite[eqs.~(102)--(104)]{BILT-Kerr}.
The static limit therefore depends on the order of the limits $\omega\to0$ and $\ell\to p|m_\phi|$ \cite{CFMP,Hui:2020xxx,Charalambous:2021mea,Charalambous:2021kcz,
Charalambous:2022rre}. Reality is not affected.
\subsubsection{RTS.}\label{sec:RTS-Love}
With the map of Table~\ref{tab:map}, the argument applies verbatim in the near zone $r_+-r_-\ll r\ll\hat\omega^{-1}$. The masses $m_{1,2}=\mathsf A\mp\mathsf B$ are real for real $\omega$, one of them being $\pm\kappa=O(\omega)$ and the other carrying the integer shift ${h_1}|m_\phi|/{h_2}$, and $\Lambda_{\rm eff}=\Lambda(\hat\omega)-g_{\rm RTS}\,\hat\omega^2$ is real, since the prolate eigenvalue is real for real $\Omega_2^2=+{\alpha}^2\hat\omega^2$. Hence $a$, and with it $\mathcal L$, is real below the thresholds
\begin{equation}\label{eq:RTSLove-threshold}
{\cal M}\hat\omega_*\simeq
\begin{cases}
0.241,\ 0.430,\ 0.766,\ 0.814,\ 0.616 & \quad r_b/r_s=1.2,\\
0.240,\ 0.468,\ 0.892,\ 0.992,\ 0.787 & \quad r_b/r_s=1.7,
\end{cases}
\end{equation}
for $(\ell,m)=(0,0),\ (1,0),\ (2,0),\ (2,2),\ (2,-2)$, respectively, with $({h_1},{h_2})=(1,2)$.
Since ${h_1}|m|/{h_2}\le|m|\le\ell$, only the finite static limit \eqref{eq:Love-static1} occurs, with $p|m_\phi|\to {h_1}|m|/{h_2}$ (Table~\ref{tab:RTS-Love}). For $m=0$ the response and the degenerate $(2\ell+1)$-th term of the source series cancel up to $O(\omega^2)$, reproducing the vanishing static value of \cite[eq.~(5.21)]{Bianchi-Bini-DiRusso}; for ${h_1}\to0$, $(r_+-r_-)^{2a}\mathcal L$ reduces to \cite[eq.~(5.22)]{Bianchi-Bini-DiRusso} once the $O(\omega)$ masses $\pm\kappa$ are dropped.
\begin{table}[h]
\centering\small
\renewcommand{\arraystretch}{1.15}
\begin{tabular}{lccc}
\toprule
$(\ell,m_\phi)$ & ${h_1}|m_\phi|/{h_2}$ & $r_b/r_s=1.2$ & $r_b/r_s=1.7$\\
\midrule
$(1,0)$    & $0$ & $0.16297$  & $0.036298$\\
$(2,0)$    & $0$ & $0.018112$ & $0.0040420$\\
$(2,\pm2)$ & $1$ & $-0.02717$ & $-0.006063$\\
\bottomrule
\end{tabular}
\caption{Static limit of $\mathcal L$ for the RTS with $({h_1},{h_2})=(1,2)$, eq.~\eqref{eq:Love-static1} with $p|m_\phi|\to {h_1}|m_\phi|/{h_2}$, in units $r_s=1$. The values at $\hat\omega=0.0025$ differ from these by less than $5\times10^{-4}$ in relative terms, and $|\mathcal L_I|<10^{-35}$.}
\label{tab:RTS-Love}
\end{table}

\section{Conclusions and outlook}

Let summarize our results, draw some conclusions and outline directions for future investigation.

Aim of the present investigation was to study static and dynamical properties of RKTB \cite{BL} and RTS \cite{BDMR}, that represent two extremely simple classes of rotating, non-singular and horizonless solutions of five-dimensional gravity coupled to Maxwell fields. Their reduction to four dimensions produces compact objects that can mimic BH micro-states and yet enjoy properties that allow discriminating them from putative BHs \cite{Bianchi:2020bxa,Bianchi:2020miz,Ikeda:2021uvc,Bena:2020see}.

In particular, differently from Kerr BHs in $D=4$ \cite{Bianchi:2024shc} and Myers-Perry BHs in higher dimensions \cite{Bianchi:2025xol},  we have shown that RTS have always positive mass quadrupole moment, while RKTB may have positive but small mass quadrupole moment for some choice of the parameters. RKTB break equatorial symmetry and have even spin multipoles and odd mass multipoles. 

We have then focused on geodesics and shown their separability even for massive probes and for non-zero KK momentum. For massless probes with zero KK momentum we have studied critical (null) geodesics and found that for RKTB only in the counter-rotating case the critical radius $r_c$ is larger than the radial position of the bolt $r_+$, while for RTS one can have co-rotating null geodesics too, at least for small rotation and for $r_b<3r_s/2$.

We finally studied massless scalar perturbations of the background and solved the combined polar angular and radial dynamics -- both leading to confluent Heun equations (CHE) -- relying on the MST approach \cite{MST1,MST2,MST3,MST4} and on quantum Seiberg-Witten curves \cite{Aminov:2020yma, BCGM, BILT-Kerr, Bonelli-Iossa-Lichtig-Tanzini, CFMP,Bianchi-Bini-DiRusso}. The resulting spheroidal harmonics turned out to enjoy equatorial symmetry in the RTS case but not in the RKTB case. To the best of our knowledge the closed-form expressions that we found in the latter case have not yet appeared in the literature, since BHs typically enjoy equatorial symmetry \cite{BertiCardosoCasals}.

We found the eigenvalue $\Lambda =\ell(\ell+1) + ...$ of the real Sturm-Liouville problem associated to the polar angular dynamics from the quantization of the $a$-period of the qSW curve $a=\ell +{1\over 2}$ or equivalently of the `UN-renormalized angular momentum' $\nu= a -{1\over 2}=\ell$ (exactly) in the MST approach \cite{Cipriani:2025ikx}. 

We plugged the resulting $\Lambda$ into the radial equation and found the QNMs from the quantization of the dual $a_D$-period of the qSW curve \cite{Aminov:2020yma,BILT-Kerr,Bonelli-Iossa-Lichtig-Tanzini, Bianchi:2021xpr,DiRusso:2024hmd,HPoghosyan}. In the radial case the relevant `renormalized angular momentum' $\nu$ in the MST aproach is still related to the $a$-period of the qSW curve but it is not quantized\footnote{Please note that $\nu$ and $a$  in the radial problem are NOT the same ones as in the angular problem, yet they are exactly related by $a=\nu+{1\over 2}$ in both cases.}. All the QNMs have negative imaginary part\footnote{We double checked this property by Leaver method of continued fraction.} offering circumstantial evidence for the stability of both RKTB and RTS under linear scalar perturbations. We conjecture that they remain stable -- like their non-rotating cousins \cite{Bah:2021irr,Bena:2024hoh,Dima:2024cok}-- even under linearized metric tensor and gauge vector field perturbations that tend to mix with each other, thus making the problem of disentangling the two trickier \cite{Bena:2024hoh}. We leave the problem to future work.

We identified two classes of QNMs, those associated to the unstable light-ring, which are expected to dominate the `prompt ring-down', with $|Im\omega/Re\omega| \sim {\cal O}(n/\ell)$ with $n$ the overtone number and those associated to the meta-stable light ring,   with $|Im\omega/Re\omega| << 1$ that are expected to show up in possible Echoes \cite{Bianchi:2021xpr,CFMP,Ikeda:2021uvc}. 

Finally we studied the tidal response of RKTBs and RTSs and found that the Love `numbers' are real for small $\omega$, contary to BHs that have purely imaginary `dissipative' dynamical Love `numbers', and vanish in the static limit as for BHs. We also discussed the appearance of imaginary parts when the $a$-period gets shifted by 1/2 and $2a$ becomes an integer. Similar tidal resonances have been found even in the non-rotating case \cite{DiRusso:2024hmd}.

We did not compute other observables such as the absorption cross section, which is simply zero because of the regularity at the cap or the bolt, as for circular fuzzballs \cite{Bianchi:2022qph}
or the super-radiant modes associated to ergo-region instability \cite{Cardoso:2005gj, JMaRTinstMB} since neither RKTB not RTS have an ergoregion \cite{BL, BDMR}. As for static Top Stars we expect no charge instability \cite{Cipriani:2024ygw} either, which would require to study the behavior of charged probes. We leave the problem to future work, too. We limited our analysis to linear order. Given the solutions we found for RKTB and RTS, we can study self-force effects as for non-rotating Top Stars \cite{Bianchi:2025aei,
DiRusso:2025lip,Bianchi:2024rod,Bini:2026oie} and for W-solitons \cite{Bianchi:2025ydq}, originally proposed in \cite{Dima:2025tjz}.

Last but not least we did not address similar problems in the so-called Asymptotically Flat RTS (AFRTS) \cite{pHpPjS} since, notwithstanding separability of geodesic motion and of linearized scalar wave equation, the explicit form of the metric and gauge field are quite involved and harder to manipulate. It would be very interesting to work their properties at the level that we worked the ones of RKTBs and RTSs. 

\section*{Acknowledgements}

We would like to thank I. Bena, F. Fucito, F. Morales, G. Di Russo, G. Dibitetto, A. Cipriani, D. Bini, A. Grassi, C. Iossa, D. Mauri, M. Giorgi, ...
M.B. would like to thank Universit\'e de Geneve and CERN for hospitality during completion of this project.
A. L. would like to thank University of Roma Tor Vergata and INFN for hospitality during completion of this project.

\appendix
\section{Supergravity background}
\label{Appendix Supergravity}

In this appendix we collect our supergravity conventions and the brane set-up from which the solutions of Section~\ref{sec:RKTB solution} are built. They coincide with those of~\cite{BL}; we repeat them here so that the present paper is self-contained.

We start from eleven-dimensional supergravity, whose bosonic action reads~\cite{Chakraborty:2025ger}
\begin{equation}
\label{action sugra}
	S_{11} = \frac{1}{16 \pi G_{11}} \int  R *_{11} 1 -\frac{1}{2} F_4 \wedge *_{11} F_4 - \frac{1}{6} F_4 \wedge F_4 \wedge A_3 \, ,
\end{equation}
with $F_4 = dA_3$; the subscript of $*_{11}$ indicates that the Hodge dual is taken with respect to the eleven-dimensional metric. The final term is the Chern--Simons coupling. The variation of~\eqref{action sugra} with respect to $A_3$ yields the equation of motion
\begin{equation}
	d  *_{11} F_4 = -\frac{1}{2} F_4 \wedge F_4  \, .
\end{equation}
The eleven-dimensional spacetime is decomposed as $\mathcal{M}_5\times T^2_1\times T^2_2\times T^2_3$. The five-dimensional factor $\mathcal{M}_5$ is made of the time direction together with a four-dimensional Euclidean base; the latter includes the circle $S^1$, with coordinate $\tau$, on which the reduction from eleven to ten dimensions is performed. In the present case, the base is the Euclidean Kerr--Taub--bolt space~\eqref{EKTB base space}.

Table~\ref{tab:brane setup} displays the branes. The configuration contains three M5 branes, each of which wraps $S^1$ together with two of the tori, and three M2 branes, each wrapping only one torus. Each brane is smeared over the compact directions that it does not wrap. Although this brane configuration is supersymmetric (BPS), the solutions studied here are non-supersymmetric, since the Euclidean Kerr--Taub--bolt base is Ricci-flat but not hyper-K\"ahler.

\begin{table}[h!]
    \centering
    \begin{tabular}{|c|c|c|c|c|c|c|}
        \hline
         & $0$ & $1$ & $2\hspace{1mm}3\hspace{1mm}4$ & $5\hspace{1mm}6$ & $7\hspace{1mm}8$ & $9 \hspace{1mm} 10$ \\
        \hline
         & $\mathbb{R}_{t}$ & $S^1_{\tau}$ & $\mathcal{M}_3$ & $T^2_1$ & $T^2_2$ & $T^2_3$ \\
        \hline
        $M5_1$ & $\times$ & $\times$ & $\perp$ & -- & $\times$ & $\times$ \\
        \hline
        $M5_2$ & $\times$ & $\times$ & $\perp$ & $\times$ & -- & $\times$ \\
        \hline
        $M5_3$ & $\times$ & $\times$ & $\perp$ & $\times$ & $\times$ & -- \\
        \hline
        $M2_1$ & $\times$ & -- & $\perp$ & $\times$ & -- & -- \\
        \hline
        $M2_2$ & $\times$ & -- & $\perp$ & -- & $\times$ & -- \\
        \hline
        $M2_3$ & $\times$ & -- & $\perp$ & -- & -- & $\times$ \\
        \hline
    \end{tabular}
    \caption{Directions occupied by each brane: $\times$ denotes a worldvolume direction, $\perp$ a transverse one, and -- a direction over which the brane is smeared. The labels are chosen such that M2$_I$ and M5$_I$ are, respectively, the electric and magnetic sources of the same gauge field $A^I$.}
    \label{tab:brane setup}
\end{table}

Within the floating-brane ansatz~\cite{FloatingBrane}, the three-form potential is written as
\begin{equation}
\label{eq:A3 ansatz}
	A_3 = \sum_{I=1}^{3} A^{I} \wedge dz^{2I+3} \wedge dz^{2I+4} \,, \qquad
	A^{I} = - \varepsilon\, Z^{-1}_I \left(dt+W\right) + B^{I} \, ,
\end{equation}
where $(z^{2I+3},z^{2I+4})$ denote the coordinates of $T^2_I$, namely directions $5,\dots,10$ in Table~\ref{tab:brane setup}, while $Z_I$ and $W$ are the warp factors and the rotation one-form of the metric~\eqref{eq:11d metric}. The one-forms $B^I$ are defined on the base, and their field strengths $\Theta^I \equiv dB^I$ describe the magnetic fluxes through the bolt. The sign $\varepsilon = \pm 1$ chooses between self-dual and anti-self-dual fluxes, $\Theta^I = \varepsilon *_4 \Theta^I$, with $*_4$ the Hodge dual on the base. The ansatz owes its name to the fact that M2-brane probes parallel to the M2$_I$ experience no force, and the equations of motion reduce to a linear system on a Ricci-flat base.

Upon compactification on the three tori one obtains five-dimensional $\mathcal{N}=2$ supergravity coupled to two vector multiplets, also called the STU model or $U(1)^3$ supergravity~\cite{Chakraborty:2025ger}. In our setting this reduction is straightforward: since each brane is either wrapped on a given torus or smeared over it, none of the fields depends on the torus coordinates. The resulting five-dimensional action reads
\begin{equation}
	S_5 = \frac{1}{16 \pi G_5} \int R_5 *_5 1 - \frac{1}{2}Q_{IJ}\, d X^{I} \wedge *_5 d X^{J}-  \frac{1}{2} Q_{IJ}\, F^{I} \wedge *_5 F^{J}- \frac{C_{IJK}}{6}\, F^{I} \wedge F^{J} \wedge A^K \, ,
\end{equation}
with $F^I\equiv dA^I$, $C_{IJK} = |\epsilon_{IJK}|$, $G_5 = G_{11}/V_{T^6}$, where $V_{T^6}$ is the volume of the three tori, and $Q_{IJ}=\mathrm{diag}\big((X^1)^{-2}, (X^2)^{-2},(X^3)^{-2}\big)$. The scalars satisfy the constraint $X^1 X^2 X^3 = 1$, hence only two of them are independent.
They encode the sizes of the tori; by comparison with~\eqref{eq:11d metric},
\begin{equation}
	X^I = \frac{Z}{Z_I}\,, \qquad Z \equiv \left(Z_1 Z_2 Z_3\right)^{1/3} .
\end{equation}
Since $X^1X^2X^3=1$, the overall volume of $T^6$ does not vary, and the five-dimensional metric is simply given by the first two terms of~\eqref{eq:11d metric},
\begin{equation}
	ds_5^2 = -Z^{-2}\left(dt+W\right)^2 + Z\, ds^2_{\mathrm{EKTB}} \,.
\end{equation}

The choice $X^1=X^2=X^3=1$, $A^1=A^2=A^3\equiv A$ defines a consistent truncation. For the solutions considered in this paper, it amounts to $Z_1=Z_2=Z_3$, that is, to equal fluxes $q_1=q_2=q_3\equiv q$. The action then reduces to
\begin{equation}
	S_5^{\mathrm{min}} = \frac{1}{16 \pi G_5} \int  R_5 *_5 1 -  \frac{3}{2} F\wedge *_5 F- F \wedge F \wedge A  \, ,
\end{equation}
where $F\equiv dA$. This is the bosonic part of minimal five-dimensional supergravity, namely Einstein--Maxwell theory supplemented by a Chern--Simons term with a coefficient determined by supersymmetry. Another normalisation is also frequently used in the literature,
\begin{equation}
	S_5^{\mathrm{min}} = \frac{1}{16 \pi G_5} \int R_5 *_5 1 -  \frac{1}{2} F\wedge *_5 F- \frac{1}{3 \sqrt{3}}F \wedge F \wedge A  \, .
\end{equation}
The two forms are connected by the rescaling $A_{\mathrm{here}}= A_{\mathrm{lit}}/\sqrt{3}$, which brings the Maxwell term to its canonical normalisation. All numerical results presented in this paper are derived within minimal supergravity, as this lowers the number of free parameters.

\section{Massive geodesics}\label{app:massive}
For massive probes, Hamilton-Jacobi equations are still separable and we can do a similar analysis starting with \eqref{eq:radial_angular_ode}. The analysis needs then to be done for new radial and angular functions:
\begin{equation}
    \begin{split}
    \mathcal{R}_\mu(r) &= \mathcal{R}(r)+\mu^2 \mathcal{R}_Z(r)  \\
    \Theta_\mu(\theta) &=\Theta(\theta)+\mu^2 \Theta_Z(\theta)
    \end{split}
\end{equation}
with:
\begin{equation}
    \mathcal{R}_Z(r) \equiv r^2-N^2+\frac{Q^2}{N-M}\big(r+N\big)\, ,\qquad \Theta_Z(\theta) \equiv -\alpha^2\cos^2\theta+\alpha \left(\frac{Q^2}{N-M}-2N\right) \cos\theta\,.
\end{equation}
This will not change the behavior of $\mathcal{N}_4$ at infinity and near $r_+$, and we note that $\Theta_Z$ is also asymmetric.
\section{Numerical evaluation of the RKTB spectrum}\label{app:numerics}
 
\begin{table}[t]
\centering\small
\renewcommand{\arraystretch}{1.12}
\begin{tabular}{@{}lrr@{}}
\toprule
 & $p=2$ & $p=3$\\
\midrule
$M$ & $0.7660444431$ & $0.8388522721$\\
$\alpha$ & $0.7422271990$ & $0.5773800515$\\
$\gamma$ & $0.3044241225$ & $0.6422337617$\\
$r_+$ & $1.137158043$ & $1.031312289$\\
$r_-$ & $0.3949308436$ & $0.6463922549$\\
$G_4M$ & $3.625010880$ & $8.404234752$\\
$G_4J$ & $-3.766667585$ & $-8.811318687$\\
$\chi\equiv J/M^2$ & $-0.286641$ & $-0.124751$\\
$g_{\rm eff}$ & $-51.177808$ & $-102.30379$\\
$G_4M\sqrt{1-\gamma^2}$ & $3.452955499$ & $6.441920195$\\
$\varepsilon_p$ & $0.146$ & $0.0252$\\
\bottomrule
\end{tabular}
\caption{Background data of the two RKTB geometries at $q=0.88$, in units $N=1$. $M$ and $\alpha$ are fixed by regularity at the bolt, $\alpha=p\,(r_+-r_-)/2$ and $|U_r(r_+)|=2N(r_+-r_-)$; for $p=2$, $M=\cos(2\pi/9)$. $G_4\cM$ and $G_4\cJ$ are given by~\eqref{M self dual}, $g_{\rm eff}$ is the coefficient of $\omega^2$ in $\Lambda_{\rm eff}$, and $\varepsilon_p$ is defined in~\eqref{eq:epsp}.}
\label{tab:bg}
\end{table}

\subsection{Accuracy and completeness.} For all the 94 modes the two determination agree, with a median difference of $4\times10^{-12}$; the largest, $5\times10^{-7}$, occurs for the most damped modes and is dominated by the quadrature error on $\partial_a\F$. Completeness within the search region of Section~\ref{sec:QNMspectrum} was checked, for every $(\ell,m_\phi)$, by evaluating the Leaver approach on a $70\times40$ grid and refining every local minimum. Resonances too narrow to be resolved by the grid are long-lived, and are located independently near the solutions of the first of eqs. \eqref{eq:ReQNM}.
 
\subsection{Complete list} Table \ref{tab:RRB-full} lists all the modes with $\ell\le2$ in the search region, in units of the ADM mass, $M\hat\omega=G_4M\sqrt{1-\gamma^2}\,\omega$. All the digits shown are common to the two determinations (nine decimals in $M\,{\rm Re}\,\hat\omega$ for 88 of the 94 modes).
 
{\small
\renewcommand{\arraystretch}{1.06}
\begin{longtable}{@{}cccllrl@{}}
\caption{Scalar QNMs of the RKTB geometries at $Q=0.88$, $N=1$, $m_\tau=0$ (background in Table~\ref{tab:bg}). $n$ orders the members of each $(\ell,m_\phi)$ tower by increasing ${\rm Re}\,\hat\omega$; $m_\phi<0$ is co-rotating.}\label{tab:RRB-full}\\
\toprule
$\ell$ & $m_\phi$ & $n$ & $M\,{\rm Re}\,\hat\omega$ & $M\,{\rm Im}\,\hat\omega$ & $|{\rm Im}\,\hat\omega|/{\rm Re}\,\hat\omega$ & $\nu$\\
\endfirsthead
\toprule
$\ell$ & $m_\phi$ & $n$ & $M\,{\rm Re}\,\hat\omega$ & $M\,{\rm Im}\,\hat\omega$ & $|{\rm Im}\,\hat\omega|/{\rm Re}\,\hat\omega$ & $\nu$\\
\endhead
\bottomrule
\endfoot
\midrule
\multicolumn{7}{c}{$p=2$}\\
\midrule
0 & $0$ & 0 & $0.240948769$ & $-0.04432$ & $0.184$ & $-0.1862+i\,0.0956$\\
0 & $0$ & 1 & $0.453669713$ & $-0.2123$ & $0.468$ & $-0.0835+i\,0.6198$\\
0 & $0$ & 2 & $0.704861357$ & $-0.3895$ & $0.553$ & $0.1551+i\,1.0725$\\
0 & $0$ & 3 & $0.963851820$ & $-0.5566$ & $0.578$ & $0.3991+i\,1.4849$\\
0 & $0$ & 4 & $1.217872491$ & $-0.7153$ & $0.587$ & $0.6677+i\,1.8674$\\
0 & $0$ & 5 & $1.464015272$ & $-0.8748$ & $0.598$ & $0.9740+i\,2.2611$\\
\addlinespace[2pt]
1 & $0$ & 0 & $0.520565097$ & $-0.01375$ & $0.0264$ & $0.5256+i\,0.1697$\\
1 & $0$ & 1 & $0.678440732$ & $-0.1288$ & $0.190$ & $0.4709+i\,0.5155$\\
1 & $0$ & 2 & $0.875602303$ & $-0.3020$ & $0.345$ & $0.6574+i\,1.0350$\\
1 & $0$ & 3 & $1.10379073$ & $-0.4927$ & $0.446$ & $0.9313+i\,1.5079$\\
1 & $0$ & 4 & $1.35779087$ & $-0.6870$ & $0.506$ & $0.2116+i\,1.9894$\\
\addlinespace[2pt]
1 & $+1$ & 0 & $0.496718302$ & $-0.04046$ & $0.0815$ & $0.5882+i\,0.2955$\\
1 & $+1$ & 1 & $0.679538108$ & $-0.1865$ & $0.274$ & $0.7335+i\,0.6304$\\
1 & $+1$ & 2 & $0.899470394$ & $-0.3595$ & $0.400$ & $0.8594+i\,1.0239$\\
1 & $+1$ & 3 & $1.138715618$ & $-0.5360$ & $0.471$ & $1.0503+i\,1.4822$\\
1 & $+1$ & 4 & $1.388148790$ & $-0.7100$ & $0.511$ & $1.2820+i\,1.9378$\\
\addlinespace[2pt]
1 & $-1$ & 0 & $0.797354276$ & $-0.2098$ & $0.263$ & $0.4345+i\,0.7118$\\
1 & $-1$ & 1 & $1.001514744$ & $-0.4108$ & $0.410$ & $0.6165+i\,1.1886$\\
1 & $-1$ & 2 & $1.233193086$ & $-0.6059$ & $0.491$ & $0.8201+i\,1.6791$\\
1 & $-1$ & 3 & $1.478583541$ & $-0.7914$ & $0.535$ & $1.0546+i\,2.1570$\\
\addlinespace[2pt]
2 & $0$ & 0 & $0.813673290$ & $-3.493\times10^{-3}$ & $4.29\times10^{-3}$ & $1.2851+i\,0.0115$\\
2 & $0$ & 1 & $0.958826504$ & $-0.07407$ & $0.0772$ & $1.1130+i\,0.4490$\\
2 & $0$ & 2 & $1.123600968$ & $-0.2153$ & $0.192$ & $1.1472+i\,0.9165$\\
2 & $0$ & 3 & $1.313303343$ & $-0.3836$ & $0.292$ & $1.3428+i\,1.3985$\\
2 & $0$ & 4 & $1.523073712$ & $-0.5666$ & $0.372$ & $1.5990+i\,1.8333$\\
\addlinespace[2pt]
2 & $+1$ & 0 & $0.785246354$ & $-0.01782$ & $0.0227$ & $1.2403+i\,0.0879$\\
2 & $+1$ & 1 & $0.941241380$ & $-0.1258$ & $0.134$ & $1.2547+i\,0.6300$\\
2 & $+1$ & 2 & $1.127495813$ & $-0.2790$ & $0.247$ & $1.3721+i\,1.0380$\\
2 & $+1$ & 3 & $1.335212210$ & $-0.4501$ & $0.337$ & $1.5337+i\,1.4948$\\
2 & $+1$ & 4 & $1.559605895$ & $-0.6298$ & $0.404$ & $1.7664+i\,1.9587$\\
\addlinespace[2pt]
2 & $+2$ & 0 & $0.762024841$ & $-0.03518$ & $0.0462$ & $1.2611+i\,0.1861$\\
2 & $+2$ & 1 & $0.928262839$ & $-0.1660$ & $0.179$ & $1.3635+i\,0.7410$\\
2 & $+2$ & 2 & $1.127247786$ & $-0.3285$ & $0.291$ & $1.5608+i\,1.1502$\\
2 & $+2$ & 3 & $1.346623230$ & $-0.5017$ & $0.373$ & $1.7627+i\,1.5412$\\
2 & $+2$ & 4 & $1.579860778$ & $-0.6779$ & $0.429$ & $1.9663+i\,1.9500$\\
\addlinespace[2pt]
2 & $-1$ & 0 & $1.107439334$ & $-0.1213$ & $0.110$ & $0.8807+i\,0.6449$\\
2 & $-1$ & 1 & $1.268692975$ & $-0.2941$ & $0.232$ & $1.0745+i\,1.1395$\\
2 & $-1$ & 2 & $1.459928165$ & $-0.4830$ & $0.331$ & $1.2673+i\,1.6199$\\
\addlinespace[2pt]
2 & $-2$ & 0 & $1.331531299$ & $-0.3967$ & $0.298$ & $1.0136+i\,1.2500$\\
2 & $-2$ & 1 & $1.537722732$ & $-0.6052$ & $0.394$ & $1.2497+i\,1.7924$\\
\midrule
\multicolumn{7}{c}{$p=3$}\\
\midrule
0 & $0$ & 0 & $0.229586969$ & $-0.01193$ & $0.0520$ & $-0.1037+i\,0.0121$\\
0 & $0$ & 1 & $0.426678430$ & $-0.09312$ & $0.218$ & $-0.2853+i\,0.3138$\\
0 & $0$ & 2 & $0.655132728$ & $-0.1970$ & $0.301$ & $-0.2026+i\,0.7118$\\
0 & $0$ & 3 & $0.894300869$ & $-0.2971$ & $0.332$ & $-0.0940+i\,1.0515$\\
0 & $0$ & 4 & $1.135281609$ & $-0.3937$ & $0.347$ & $0.0224+i\,1.3704$\\
0 & $0$ & 5 & $1.375008347$ & $-0.4882$ & $0.355$ & $0.1496+i\,1.6799$\\
0 & $0$ & 6 & $1.612578264$ & $-0.5824$ & $0.361$ & $0.2898+i\,1.9894$\\
0 & $0$ & 7 & $1.848392356$ & $-0.6771$ & $0.366$ & $0.4400+i\,2.3063$\\
\addlinespace[2pt]
1 & $0$ & 0 & $0.489363208$ & $-1.686\times10^{-4}$ & $3.45\times10^{-4}$ & $0.7774+i\,1.9\times10^{-4}$\\
1 & $0$ & 1 & $0.6732067$ & $-0.01642$ & $0.0244$ & $0.5213+i\,0.1865$\\
1 & $0$ & 2 & $0.845456937$ & $-0.09267$ & $0.110$ & $0.4001+i\,0.3851$\\
1 & $0$ & 3 & $1.046880733$ & $-0.1943$ & $0.186$ & $0.3866+i\,0.8435$\\
1 & $0$ & 4 & $1.266987306$ & $-0.3009$ & $0.237$ & $0.4889+i\,1.2338$\\
1 & $0$ & 5 & $1.499512697$ & $-0.4075$ & $0.272$ & $0.6057+i\,1.5963$\\
1 & $0$ & 6 & $1.74026$ & $-0.5115$ & $0.294$ & $0.7223+i\,1.9470$\\
\addlinespace[2pt]
1 & $+1$ & 0 & $0.480514303$ & $-2.104\times10^{-3}$ & $4.38\times10^{-3}$ & $0.7454+i\,0.0033$\\
1 & $+1$ & 1 & $0.657604757$ & $-0.04313$ & $0.0656$ & $0.5663+i\,0.3172$\\
1 & $+1$ & 2 & $0.84995563$ & $-0.1341$ & $0.158$ & $0.6152+i\,0.5610$\\
1 & $+1$ & 3 & $1.064367438$ & $-0.2348$ & $0.221$ & $0.5868+i\,0.8358$\\
1 & $+1$ & 4 & $1.290081255$ & $-0.3365$ & $0.261$ & $0.6083+i\,1.2152$\\
1 & $+1$ & 5 & $1.522143141$ & $-0.4374$ & $0.287$ & $0.6962+i\,1.5891$\\
1 & $+1$ & 6 & $1.757675572$ & $-0.5367$ & $0.305$ & $0.8066+i\,1.9459$\\
\addlinespace[2pt]
1 & $-1$ & 0 & $0.986881905$ & $-0.1216$ & $0.123$ & $0.2425+i\,0.5928$\\
1 & $-1$ & 1 & $1.187840384$ & $-0.2471$ & $0.208$ & $0.3387+i\,0.9512$\\
1 & $-1$ & 2 & $1.407173015$ & $-0.3652$ & $0.260$ & $0.3925+i\,1.2798$\\
1 & $-1$ & 3 & $1.635597887$ & $-0.4776$ & $0.292$ & $0.4351+i\,1.6343$\\
1 & $-1$ & 4 & $1.868561197$ & $-0.5853$ & $0.313$ & $0.5065+i\,2.0058$\\
\addlinespace[2pt]
2 & $0$ & 0 & $0.744509790$ & $-7.140\times10^{-7}$ & $9.59\times10^{-7}$ & $1.7038+i\,6.5\times10^{-7}$\\
2 & $0$ & 1 & $0.950514616$ & $-5.162\times10^{-4}$ & $5.43\times10^{-4}$ & $1.4619+i\,0.0009$\\
2 & $0$ & 2 & $1.117259948$ & $-0.02121$ & $0.0190$ & $1.0733+i\,0.2040$\\
2 & $0$ & 3 & $1.280809456$ & $-0.09453$ & $0.0738$ & $1.0680+i\,0.5198$\\
2 & $0$ & 4 & $1.467478424$ & $-0.1890$ & $0.129$ & $0.9837+i\,0.8936$\\
2 & $0$ & 5 & $1.669448913$ & $-0.2911$ & $0.174$ & $1.0405+i\,1.3038$\\
2 & $0$ & 6 & $1.882984248$ & $-0.3970$ & $0.211$ & $1.1437+i\,1.6705$\\
\addlinespace[2pt]
2 & $+1$ & 0 & $0.739479500$ & $-2.956\times10^{-5}$ & $4.00\times10^{-5}$ & $1.6855+i\,3.1\times10^{-5}$\\
2 & $+1$ & 1 & $0.933484926$ & $-4.793\times10^{-3}$ & $5.14\times10^{-3}$ & $1.4399+i\,0.0127$\\
2 & $+1$ & 2 & $1.096836252$ & $-0.05202$ & $0.0474$ & $1.1341+i\,0.3663$\\
2 & $+1$ & 3 & $1.278546280$ & $-0.1381$ & $0.108$ & $1.1815+i\,0.7113$\\
2 & $+1$ & 4 & $1.478357673$ & $-0.2347$ & $0.159$ & $1.1806+i\,1.0061$\\
2 & $+1$ & 5 & $1.690091500$ & $-0.3359$ & $0.199$ & $1.1852+i\,1.3740$\\
2 & $+1$ & 6 & $1.910661934$ & $-0.4391$ & $0.230$ & $1.2573+i\,1.7563$\\
\addlinespace[2pt]
2 & $+2$ & 0 & $0.733898228$ & $-1.889\times10^{-4}$ & $2.57\times10^{-4}$ & $1.6648+i\,2.4\times10^{-4}$\\
2 & $+2$ & 1 & $0.917058035$ & $-0.01354$ & $0.0148$ & $1.4348+i\,0.0497$\\
2 & $+2$ & 2 & $1.085611240$ & $-0.07948$ & $0.0732$ & $1.1903+i\,0.4749$\\
2 & $+2$ & 3 & $1.277725037$ & $-0.1713$ & $0.134$ & $1.2629+i\,0.8418$\\
2 & $+2$ & 4 & $1.484988794$ & $-0.2699$ & $0.182$ & $1.3358+i\,1.1455$\\
2 & $+2$ & 5 & $1.702160423$ & $-0.3710$ & $0.218$ & $1.3883+i\,1.4409$\\
2 & $+2$ & 6 & $1.926173958$ & $-0.4726$ & $0.245$ & $1.4300+i\,1.7670$\\
\addlinespace[2pt]
2 & $-1$ & 0 & $1.297159863$ & $-0.02814$ & $0.0217$ & $0.8698+i\,0.1517$\\
2 & $-1$ & 1 & $1.449503827$ & $-0.1194$ & $0.0824$ & $0.7910+i\,0.6954$\\
2 & $-1$ & 2 & $1.631917953$ & $-0.2274$ & $0.139$ & $0.8695+i\,1.0819$\\
2 & $-1$ & 3 & $1.832411121$ & $-0.3398$ & $0.185$ & $0.9248+i\,1.4265$\\
\addlinespace[2pt]
2 & $-2$ & 0 & $1.71578852$ & $-0.2696$ & $0.157$ & $0.6290+i\,1.1124$\\
2 & $-2$ & 1 & $1.927352078$ & $-0.4018$ & $0.208$ & $0.7493+i\,1.5719$\\
\end{longtable}
}

\subsection{High regime}
\paragraph*{Setup.} Massless scalar, KK-neutral ($m_\tau=0$), axisymmetric ($m_\phi=0$) modes of the RKTB geometries with $p=2,3$ at $q=0.88$, $N=1$ (background data in the appendix of the draft). Frequencies are measured by the asymptotic static observer, in units of the ADM mass, $M\hat\omega=G_4M\sqrt{1-\gamma^2}\,\omega$ ($G_4=1$).

\begin{itemize}
\item \textbf{Content.} Tables~\ref{tab:p2} and~\ref{tab:p3} list all the modes with $-M\,{\rm Im}\,\hat\omega\le0.8$: 80 for $p=2$ and 151 for $p=3$. The search extends in ${\rm Re}\,\hat\omega$ beyond the last mode of each $\ell$, and every tower was followed until its next member leaves the strip, so that within the strip every list is complete. For $\ell\le2$ the lists coincide with the entries of the appendix of the draft that fall in the strip.
\item \textbf{Labelling.} $n$ orders the modes of a given $\ell$ by increasing ${\rm Re}\,\hat\omega$; for every $\ell$ this coincides with the ordering by increasing $|{\rm Im}\,\hat\omega|$.
\item \textbf{Search.} For each $(p,\ell)$ the Leaver continued fraction (appendix of the draft) was scanned on a dense grid covering the strip, and every local minimum was refined. Each tower was then continued to the edge of the strip.
\item \textbf{Determination and accuracy.} All frequencies are roots of the Leaver continued fraction evaluated in 40-digit arithmetic, converged in depth. This is necessary: for $p=3$ the widths of the trapped modes reach $10^{-27}$, far below double precision. Unmarked entries (205) are also roots of $\Xi=e^{2\pi ia_D}=1$, with a certified instanton-branch representative of $a$ and $|1-\Xi|<10^{-5}$ at the Leaver frequency. Entries marked $^\dagger$ (24) are trapped modes with $-M\,{\rm Im}\,\hat\omega<10^{-8}$: the root of $\Xi=1$ reproduces ${\rm Re}\,\hat\omega$ to $10^{-14}$, while ${\rm Im}\,\hat\omega$ lies below the resolution of our double-precision implementation of $\Xi$ and is determined by Leaver alone. The two entries marked $^\ddagger$ ($p=3$: $\ell=2$, $n=9$; $\ell=6$, $n=5$) are confirmed by Leaver only: for the first the quadrature of $\partial_a\mathfrak F$ does not converge, for the second no representative of $a$ on the instanton branch was found.
\end{itemize}

\section*{Comments}
\begin{itemize}
\item \textbf{Stability.} All 231 modes have ${\rm Im}\,\hat\omega<0$: no instability in the $m_\phi=0$ sector up to $\ell=10$.
\item \textbf{Eikonal structure.} At real frequency the radial equation reads $\frac{d}{dr}(\Delta\frac{dR}{dr})-\frac{\cN_4^{\rm w}}{\Delta}R=0$, with $\cN_4^{\rm w}(r_+)=0$ for $m_\phi=0$. Since $\cN_4^{\rm w}=\cN_4(r;\, E=\omega\,,\,m_\tau=0,\,J=m_\phi)$, the light ring (double root of $\cN_4^{\rm w}$) is exactly the geodesic light ring at $J=0$, and together with the cap condition $\partial_r\cN_4^{\rm w}(r_+)=0$ it fixes two values of $\Lambda$, independent of $\ell$. Translated into frequency with the exact $\Lambda(\omega)$, they define, for every $\ell$, a cap frequency $\omega_{\rm cap}$ and a light-ring frequency $\omega_{\rm LR}$, with Lyapunov exponent $\lambda$:
\begin{center}\small
\begin{tabular}{@{}lcccccc@{}}\toprule
 & $r_{\rm LR}$ & $\Lambda_{\rm LR}$ & $\Lambda_{\rm cap}$ & $M\omega_{\rm cap}/(\ell+\frac12)$ & $M\omega_{\rm LR}/(\ell+\frac12)$ & $M\lambda$\\\midrule
$p=2$ & $2.462$ & $104.66$ & $144.64$ & $0.287$ & $0.338$ & $0.163$\\
$p=3$ & $3.668$ & $228.78$ & $649.6$ & $0.252$ & $0.426$ & $0.229$\\\bottomrule
\end{tabular}
\end{center}
($r_{\rm LR}$ and $\Lambda$ in units $N=1$; frequency ratios at $\ell=10$, reached within $2\%$ already at $\ell=2$). $r_{\rm LR}$ and $\Lambda_{\rm LR}$ are geodesic quantities and can be compared directly with the light rings of the geodesic analysis. The frequencies, instead, are obtained with the exact $\Lambda(\omega)$. Unlike the RTS, where the angular problem is spherical and $\Lambda\simeq(\ell+\frac12)^2$, here $\Omega_1\chi$ and $\Omega_2^2\chi^2$ grow as $\omega^2$, like $\Lambda$; a closed eikonal formula for $\Omega_c$ in ${\rm Re}\,\hat\omega\simeq(\ell+\frac12)\Omega_c$ would require the angular Bohr--Sommerfeld condition $\oint p_\chi\,d\chi=2\pi(\ell+\frac12)$, $p_\chi^2=(\Lambda-\Omega_1\chi-\Omega_2^2\chi^2)/(1-\chi^2)^2$ ($m_\phi=0$), which we have not worked out. Both geometries are in the situation of the RTS with $r_b/r_s=1.2$: the cap lies below the light ring, and for $\omega_{\rm cap}<{\rm Re}\,\hat\omega<\omega_{\rm LR}$ there is a cavity between the cap and the light-ring barrier. The window is much wider for $p=3$. Just above $\omega_{\rm LR}$ the maximum and the minimum of $\cN_4^{\rm w}$ merge, so the barrier is shallow: the light ring is close to coalescence with the bottom of the cavity.
\item \textbf{Trapped modes.} The lowest members of each tower fall in the cavity window \cite{Heidmann:2023ojf,Bianchi:2022qph,DiRusso:2024hmd,Dima:2024cok}. Their number grows with $\ell$: for $p=2$ one mode for $\ell=2$--$5$, two for $\ell=6$--$8$, three for $\ell=9,10$; for $p=3$ it grows almost linearly, from one at $\ell=1$ to nine at $\ell=10$. They are exponentially long-lived: $-M\,{\rm Im}\,\hat\omega_0$ decreases from $3.5\times10^{-3}$ ($\ell=2$) to $1.7\times10^{-10}$ ($\ell=10$) for $p=2$, and from $7.1\times10^{-7}$ to $1.5\times10^{-27}$ for $p=3$. The decay in $\ell$ is the WKB tunnelling through the light-ring barrier: the slope $-\Delta\ln|{\rm Im}\,\hat\omega_0|$ per unit $\ell$ is compared below with $\Delta T$, $T=2\int\sqrt{\cN_4^{\rm w}}\,dr/\Delta$ over the barrier at $\omega={\rm Re}\,\hat\omega_0$:
\begin{center}\small
\begin{tabular}{@{}lccccccc@{}}\toprule
$\ell\to\ell+1$ & $3\to4$ & $4\to5$ & $5\to6$ & $6\to7$ & $7\to8$ & $8\to9$ & $9\to10$\\\midrule
$p=2$: data & $1.949$ & $2.087$ & $2.158$ & $2.200$ & $2.227$ & $2.248$ & $2.263$\\
\phantom{$p=2$:} WKB & $2.139$ & $2.196$ & $2.230$ & $2.252$ & $2.268$ & $2.279$ & $2.289$\\
$p=3$: data & $5.869$ & $5.933$ & $5.974$ & $6.002$ & $6.022$ & $6.038$ & $6.051$\\
\phantom{$p=3$:} WKB & $6.007$ & $6.023$ & $6.038$ & $6.050$ & $6.060$ & $6.068$ & $6.075$\\\bottomrule
\end{tabular}
\end{center}
The difference decreases as $1/\ell$. The two descriptions are compatible: the near-zone law $|{\rm Im}\,\omega|\propto|q|^{2{\rm Re}\,a}$ of Sec.~4.1.4 is itself exponential in $\ell$, and at large $\ell$, where $|q|$ is no longer small, the barrier integral gives the correct rate.
\item \textbf{Modes above the barrier.} The modes with ${\rm Re}\,\hat\omega>\omega_{\rm LR}$ do not form a light-ring tower ${\rm Im}\,\hat\omega=-(n+\frac12)\lambda$: their spacing in ${\rm Im}\,\hat\omega$ grows along the tower and stays well below $\lambda$ for $p=3$ ($0.05$--$0.10$ against $M\lambda=0.229$). As for the RTS with $r_b/r_s=1.2$, they are resonances of the region bounded by the cap, accumulating above $\omega_{\rm LR}$. Their number in the strip grows with $\ell$, from 5 to 7 for $p=2$ and from 9 to 10 for $p=3$.
\end{itemize}

\begin{table}[p]
\centering\scriptsize\renewcommand{\arraystretch}{0.98}
\begin{minipage}[t]{0.48\textwidth}\centering
\begin{tabular}{@{}rrll@{}}
\toprule
$\ell$ & $n$ & $M\,{\rm Re}\,\hat\omega$ & $-M\,{\rm Im}\,\hat\omega$\\
\midrule
0 & 0 & $0.240949$ & $0.04432$\\
 & 1 & $0.453670$ & $0.2123$\\
 & 2 & $0.704861$ & $0.3895$\\
 & 3 & $0.963852$ & $0.5566$\\
 & 4 & $1.217872$ & $0.7153$\\
\midrule
1 & 0 & $0.520565$ & $0.01375$\\
 & 1 & $0.678441$ & $0.1288$\\
 & 2 & $0.875602$ & $0.3020$\\
 & 3 & $1.103791$ & $0.4927$\\
 & 4 & $1.357791$ & $0.6870$\\
\midrule
2 & 0 & $0.813673$ & $3.493\times10^{-3}$\\
 & 1 & $0.958827$ & $0.07407$\\
 & 2 & $1.123601$ & $0.2153$\\
 & 3 & $1.313303$ & $0.3836$\\
 & 4 & $1.523074$ & $0.5666$\\
 & 5 & $1.748168$ & $0.7580$\\
\midrule
3 & 0 & $1.105977$ & $6.426\times10^{-4}$\\
 & 1 & $1.254052$ & $0.03973$\\
 & 2 & $1.401218$ & $0.1547$\\
 & 3 & $1.570416$ & $0.3015$\\
 & 4 & $1.756970$ & $0.4674$\\
 & 5 & $1.958832$ & $0.6453$\\
\midrule
4 & 0 & $1.396180$ & $9.149\times10^{-5}$\\
 & 1 & $1.554377$ & $0.01857$\\
 & 2 & $1.690163$ & $0.1094$\\
 & 3 & $1.846479$ & $0.2381$\\
 & 4 & $2.017737$ & $0.3877$\\
 & 5 & $2.202499$ & $0.5521$\\
 & 6 & $2.399566$ & $0.7267$\\
\midrule
5 & 0 & $1.685157$ & $1.135\times10^{-5}$\\
 & 1 & $1.855231$ & $7.019\times10^{-3}$\\
 & 2 & $1.985274$ & $0.07397$\\
 & 3 & $2.132073$ & $0.1873$\\
 & 4 & $2.292728$ & $0.3226$\\
 & 5 & $2.465280$ & $0.4740$\\
 & 6 & $2.648881$ & $0.6373$\\
\midrule
6 & 0 & $1.973555$ & $1.312\times10^{-6}$\\
 & 1 & $2.153937$ & $1.989\times10^{-3}$\\
 & 2 & $2.284831$ & $0.04603$\\
 & 3 & $2.423271$ & $0.1453$\\
 & 4 & $2.575987$ & $0.2681$\\
 & 5 & $2.739501$ & $0.4077$\\
 & 6 & $2.912942$ & $0.5602$\\
 & 7 & $3.095731$ & $0.7228$\\
\\[-8pt]\bottomrule
\end{tabular}
\end{minipage}\hfill
\begin{minipage}[t]{0.48\textwidth}\centering
\begin{tabular}{@{}rrll@{}}
\toprule
$\ell$ & $n$ & $M\,{\rm Re}\,\hat\omega$ & $-M\,{\rm Im}\,\hat\omega$\\
\midrule
7 & 0 & $2.261657$ & $1.454\times10^{-7}$\\
 & 1 & $2.449479$ & $4.202\times10^{-4}$\\
 & 2 & $2.587534$ & $0.02529$\\
 & 3 & $2.718258$ & $0.1099$\\
 & 4 & $2.864540$ & $0.2216$\\
 & 5 & $3.021024$ & $0.3506$\\
 & 6 & $3.186587$ & $0.4931$\\
 & 7 & $3.360673$ & $0.6464$\\
\midrule
8 & 0 & $2.549588$ & $1.568\times10^{-8}$\\
 & 1 & $2.742415$ & $7.183\times10^{-5}$\\
 & 2 & $2.891299$ & $0.01170$\\
 & 3 & $3.016217$ & $0.07973$\\
 & 4 & $3.156783$ & $0.1813$\\
 & 5 & $3.307522$ & $0.3007$\\
 & 6 & $3.466766$ & $0.4341$\\
 & 7 & $3.633872$ & $0.5787$\\
 & 8 & $3.808450$ & $0.7326$\\
\midrule
9 & 0 & $2.837411$ & $1.656\times10^{-9}$\\
 & 1 & $3.033727$ & $1.082\times10^{-5}$\\
 & 2 & $3.193873$ & $4.287\times10^{-3}$\\
 & 3 & $3.316925$ & $0.05403$\\
 & 4 & $3.451802$ & $0.1459$\\
 & 5 & $3.597617$ & $0.2566$\\
 & 6 & $3.751614$ & $0.3816$\\
 & 7 & $3.912980$ & $0.5182$\\
 & 8 & $4.081293$ & $0.6644$\\
\midrule
10 & 0 & $3.125160$ & $1.723\times10^{-10}$$^\dagger$\\
 & 1 & $3.324064$ & $1.509\times10^{-6}$\\
 & 2 & $3.493677$ & $1.192\times10^{-3}$\\
 & 3 & $3.620232$ & $0.03307$\\
 & 4 & $3.749067$ & $0.1148$\\
 & 5 & $3.890458$ & $0.2173$\\
 & 6 & $4.039948$ & $0.3346$\\
 & 7 & $4.196473$ & $0.4637$\\
 & 8 & $4.359528$ & $0.6027$\\
 & 9 & $4.528810$ & $0.7503$\\
\\[-8pt]\bottomrule
\end{tabular}
\end{minipage}
\caption{$p=2$: all modes with $m_\phi=0$, $\ell\le10$ and $-M\,{\rm Im}\,\hat\omega\le0.8$. Marks as described in the text.}\label{tab:p2}
\end{table}

\begin{table}[p]
\centering\scriptsize\renewcommand{\arraystretch}{0.98}
\begin{minipage}[t]{0.32\textwidth}\centering
\begin{tabular}{@{}rrll@{}}
\toprule
$\ell$ & $n$ & $M\,{\rm Re}\,\hat\omega$ & $-M\,{\rm Im}\,\hat\omega$\\
\midrule
0 & 0 & $0.229587$ & $0.01193$\\
 & 1 & $0.426678$ & $0.09312$\\
 & 2 & $0.655133$ & $0.1970$\\
 & 3 & $0.894301$ & $0.2971$\\
 & 4 & $1.135282$ & $0.3937$\\
 & 5 & $1.375008$ & $0.4882$\\
 & 6 & $1.612578$ & $0.5824$\\
 & 7 & $1.848392$ & $0.6771$\\
 & 8 & $2.083242$ & $0.7728$\\
\midrule
1 & 0 & $0.489363$ & $1.686\times10^{-4}$\\
 & 1 & $0.673207$ & $0.01642$\\
 & 2 & $0.845457$ & $0.09267$\\
 & 3 & $1.046881$ & $0.1943$\\
 & 4 & $1.266987$ & $0.3009$\\
 & 5 & $1.499513$ & $0.4075$\\
 & 6 & $1.740262$ & $0.5115$\\
 & 7 & $1.985686$ & $0.6117$\\
 & 8 & $2.232815$ & $0.7077$\\
 & 9 & $2.479270$ & $0.8000$\\
\midrule
2 & 0 & $0.744510$ & $7.140\times10^{-7}$\\
 & 1 & $0.950515$ & $5.162\times10^{-4}$\\
 & 2 & $1.117260$ & $0.02121$\\
 & 3 & $1.280809$ & $0.09453$\\
 & 4 & $1.467478$ & $0.1890$\\
 & 5 & $1.669449$ & $0.2911$\\
 & 6 & $1.882984$ & $0.3970$\\
 & 7 & $2.105629$ & $0.5047$\\
 & 8 & $2.335714$ & $0.6130$\\
 & 9 & $2.572373$ & $0.7216$$^\ddagger$\\
\midrule
3 & 0 & $0.998133$ & $2.266\times10^{-9}$\\
 & 1 & $1.212993$ & $4.158\times10^{-6}$\\
 & 2 & $1.404752$ & $1.097\times10^{-3}$\\
 & 3 & $1.561242$ & $0.02654$\\
 & 4 & $1.720921$ & $0.09851$\\
 & 5 & $1.899947$ & $0.1878$\\
 & 6 & $2.091798$ & $0.2852$\\
 & 7 & $2.293857$ & $0.3875$\\
 & 8 & $2.504361$ & $0.4926$\\
 & 9 & $2.721921$ & $0.5993$\\
 & 10 & $2.945406$ & $0.7070$\\
\midrule
4 & 0 & $1.251359$ & $6.406\times10^{-12}$$^\dagger$\\
 & 1 & $1.470177$ & $2.147\times10^{-8}$\\
 & 2 & $1.674484$ & $1.343\times10^{-5}$\\
 & 3 & $1.855441$ & $2.010\times10^{-3}$\\
 & 4 & $2.005117$ & $0.03236$\\
 & 5 & $2.162781$ & $0.1035$\\
 & 6 & $2.337029$ & $0.1891$\\
 & 7 & $2.522362$ & $0.2827$\\
 & 8 & $2.716746$ & $0.3814$\\
 & 9 & $2.918852$ & $0.4837$\\
 & 10 & $3.127632$ & $0.5884$\\
 & 11 & $3.342209$ & $0.6945$\\
\\[-8pt]\bottomrule
\end{tabular}
\end{minipage}\hfill
\begin{minipage}[t]{0.32\textwidth}\centering
\begin{tabular}{@{}rrll@{}}
\toprule
$\ell$ & $n$ & $M\,{\rm Re}\,\hat\omega$ & $-M\,{\rm Im}\,\hat\omega$\\
\midrule
5 & 0 & $1.504419$ & $1.698\times10^{-14}$$^\dagger$\\
 & 1 & $1.725558$ & $9.012\times10^{-11}$$^\dagger$\\
 & 2 & $1.935830$ & $9.639\times10^{-8}$\\
 & 3 & $2.131893$ & $3.438\times10^{-5}$\\
 & 4 & $2.303847$ & $3.359\times10^{-3}$\\
 & 5 & $2.448985$ & $0.03856$\\
 & 6 & $2.605502$ & $0.1091$\\
 & 7 & $2.776437$ & $0.1920$\\
 & 8 & $2.957179$ & $0.2824$\\
 & 9 & $3.146046$ & $0.3781$\\
 & 10 & $3.341962$ & $0.4776$\\
 & 11 & $3.544093$ & $0.5800$\\
 & 12 & $3.751745$ & $0.6843$\\
 & 13 & $3.964314$ & $0.7899$\\
\midrule
6 & 0 & $1.757392$ & $4.322\times10^{-17}$$^\dagger$\\
 & 1 & $1.980073$ & $3.332\times10^{-13}$$^\dagger$\\
 & 2 & $2.193992$ & $5.385\times10^{-10}$\\
 & 3 & $2.397391$ & $3.166\times10^{-7}$\\
 & 4 & $2.586534$ & $7.735\times10^{-5}$\\
 & 5 & $2.750633$ & $5.239\times10^{-3}$$^\ddagger$\\
 & 6 & $2.892918$ & $0.04503$\\
 & 7 & $3.048721$ & $0.1150$\\
 & 8 & $3.217200$ & $0.1957$\\
 & 9 & $3.394511$ & $0.2836$\\
 & 10 & $3.579208$ & $0.3767$\\
 & 11 & $3.770374$ & $0.4738$\\
 & 12 & $3.967317$ & $0.5739$\\
 & 13 & $4.169464$ & $0.6762$\\
 & 14 & $4.376317$ & $0.7802$\\
\midrule
7 & 0 & $2.010315$ & $1.069\times10^{-19}$$^\dagger$\\
 & 1 & $2.234097$ & $1.130\times10^{-15}$$^\dagger$\\
 & 2 & $2.450516$ & $2.565\times10^{-12}$$^\dagger$\\
 & 3 & $2.658474$ & $2.206\times10^{-9}$\\
 & 4 & $2.856066$ & $8.748\times10^{-7}$\\
 & 5 & $3.039105$ & $1.599\times10^{-4}$\\
 & 6 & $3.196237$ & $7.722\times10^{-3}$\\
 & 7 & $3.336945$ & $0.05169$\\
 & 8 & $3.492254$ & $0.1211$\\
 & 9 & $3.658829$ & $0.2002$\\
 & 10 & $3.833465$ & $0.2859$\\
 & 11 & $4.014886$ & $0.3768$\\
 & 12 & $4.202289$ & $0.4717$\\
 & 13 & $4.395074$ & $0.5696$\\
 & 14 & $4.592755$ & $0.6701$\\
 & 15 & $4.794911$ & $0.7724$\\
\\[-8pt]\bottomrule
\end{tabular}
\end{minipage}\hfill
\begin{minipage}[t]{0.32\textwidth}\centering
\begin{tabular}{@{}rrll@{}}
\toprule
$\ell$ & $n$ & $M\,{\rm Re}\,\hat\omega$ & $-M\,{\rm Im}\,\hat\omega$\\
\midrule
8 & 0 & $2.263206$ & $2.592\times10^{-22}$$^\dagger$\\
 & 1 & $2.487814$ & $3.594\times10^{-18}$$^\dagger$\\
 & 2 & $2.706065$ & $1.091\times10^{-14}$$^\dagger$\\
 & 3 & $2.917204$ & $1.285\times10^{-11}$$^\dagger$\\
 & 4 & $3.120055$ & $7.304\times10^{-9}$\\
 & 5 & $3.312566$ & $2.164\times10^{-6}$\\
 & 6 & $3.490016$ & $3.098\times10^{-4}$\\
 & 7 & $3.640989$ & $0.01085$\\
 & 8 & $3.781073$ & $0.05847$\\
 & 9 & $3.935999$ & $0.1274$\\
 & 10 & $4.101049$ & $0.2051$\\
 & 11 & $4.273531$ & $0.2891$\\
 & 12 & $4.452302$ & $0.3780$\\
 & 13 & $4.636642$ & $0.4709$\\
 & 14 & $4.826020$ & $0.5670$\\
 & 15 & $5.020009$ & $0.6656$\\
 & 16 & $5.218246$ & $0.7663$\\
\midrule
9 & 0 & $2.516074$ & $6.183\times10^{-25}$$^\dagger$\\
 & 1 & $2.741327$ & $1.090\times10^{-20}$$^\dagger$\\
 & 2 & $2.960981$ & $4.261\times10^{-17}$$^\dagger$\\
 & 3 & $3.174487$ & $6.579\times10^{-14}$$^\dagger$\\
 & 4 & $3.381042$ & $5.027\times10^{-11}$$^\dagger$\\
 & 5 & $3.579398$ & $2.107\times10^{-8}$\\
 & 6 & $3.767343$ & $4.955\times10^{-6}$\\
 & 7 & $3.939523$ & $5.692\times10^{-4}$\\
 & 8 & $4.085154$ & $0.01463$\\
 & 9 & $4.225289$ & $0.06532$\\
 & 10 & $4.379896$ & $0.1338$\\
 & 11 & $4.543693$ & $0.2103$\\
 & 12 & $4.714397$ & $0.2928$\\
 & 13 & $4.890974$ & $0.3801$\\
 & 14 & $5.072767$ & $0.4712$\\
 & 15 & $5.259298$ & $0.5655$\\
 & 16 & $5.450184$ & $0.6625$\\
 & 17 & $5.645103$ & $0.7616$\\
\midrule
10 & 0 & $2.768928$ & $1.457\times10^{-27}$$^\dagger$\\
 & 1 & $2.994697$ & $3.181\times10^{-23}$$^\dagger$\\
 & 2 & $3.215462$ & $1.557\times10^{-19}$$^\dagger$\\
 & 3 & $3.430805$ & $3.049\times10^{-16}$$^\dagger$\\
 & 4 & $3.640143$ & $3.006\times10^{-13}$$^\dagger$\\
 & 5 & $3.842632$ & $1.672\times10^{-10}$$^\dagger$\\
 & 6 & $4.036949$ & $5.520\times10^{-8}$\\
 & 7 & $4.220701$ & $1.070\times10^{-5}$\\
 & 8 & $4.387799$ & $9.965\times10^{-4}$\\
 & 9 & $4.528954$ & $0.01902$\\
 & 10 & $4.669580$ & $0.07220$\\
 & 11 & $4.823904$ & $0.1403$\\
 & 12 & $4.986651$ & $0.2158$\\
 & 13 & $5.155858$ & $0.2970$\\
 & 14 & $5.330584$ & $0.3828$\\
 & 15 & $5.510222$ & $0.4724$\\
 & 16 & $5.694334$ & $0.5652$\\
 & 17 & $5.882573$ & $0.6605$\\
 & 18 & $6.074648$ & $0.7581$\\
\\[-8pt]\bottomrule
\end{tabular}
\end{minipage}
\caption{$p=3$: all modes with $m_\phi=0$, $\ell\le10$ and $-M\,{\rm Im}\,\hat\omega\le0.8$. Marks as described in the text.}\label{tab:p3}
\end{table}
\section{Numerical evaluation of the RTS spectrum}\label{app:numerics-RTS}
\begin{table}[t]
\centering\small
\renewcommand{\arraystretch}{1.12}
\begin{tabular}{@{}lrr@{}}
\toprule
 & $r_b/r_s=1.2$ & $r_b/r_s=1.7$\\
\midrule
$-a^2$ & $0.003333333333$ & $0.04083333333$\\
$r_+$ & $1.215470054$ & $1.754145188$\\
$r_-$ & $0.9845299462$ & $0.9458548116$\\
$r_+-r_-$ & $0.2309401077$ & $0.8082903769$\\
$\xi^t$ & $0.05461765324$ & $0.03239153452$\\
$R_{S^1}$ & $10.95356983$ & $9.881979682$\\
$G_4M$ & $0.7998503993$ & $0.9248161951$\\
$G_4J$ & $0.06000482757$ & $0.1121500392$\\
$\chi\equiv J/M^2$ & $0.093793$ & $0.131126$\\
$C_{\rm RTS}$ & $2.075525520$ & $1.109415056$\\
$g_{\rm RTS}$ & $3.690793679$ & $5.827895756$\\
$\varepsilon_{\rm RTS}$ & $0.222$ & $1.46$\\
\bottomrule
\end{tabular}
\caption{Background data of the two RTS geometries with $({h_1},{h_2})=(1,2)$, in units $r_s=1$. $a^2$ is fixed by the quantisation (3.41) and $R_{S^1}$ by (3.42) of \cite{BDMR}, $r_\pm$ by (3.7) of \cite{BDMR}, $\xi^t$ by (3.23) of \cite{BDMR}; $G_4M$ is given by \eqref{eq:RTStau}, $G_4J$ by (3.67) of~\cite{BDMR}, $g_{\rm RTS}$ by \eqref{eq:RTSgeff}, and $C_{\rm RTS}$, $\varepsilon_{\rm RTS}$ by \eqref{eq:epsRTS}.}
\label{tab:RTSbg}
\end{table}

\subsection{Accuracy and completeness.} For all the 70 modes the two determinations agree, with a median difference of $7\times10^{-14}$ and a largest difference of $7\times10^{-11}$; $|\Xi-1|<2\times10^{-13}$ at the Leaver frequencies, and the check of the instanton branch is satisfied to $10^{-24}$. The modes span $0.14\le|q|\le1.34$ for $r_b/r_s=1.2$ and $0.50\le|q|\le5.4$ for $r_b/r_s=1.7$: within the explored range, the integral representation~\eqref{eq:Matone} remains reliable well beyond the small-$q$ regime. Completeness within the search region of Section \ref{sec:RTSQNM} was checked, for every $(\ell,m)$, by evaluating the Leaver condition with the first seven inversions on a $300\times210$ grid covering $0.02\le{\rm Re}\,\hat\omega\le3$ and $|{\rm Im}\,\hat\omega|\le2$ ($r_b/r_s=1.2$) or $|{\rm Im}\,\hat\omega|\le3$ ($r_b/r_s=1.7$), logarithmic in ${\rm Im}\,\hat\omega$ close to the real axis, and refining every local minimum; a root is retained if its position is stable under an increase of the depth. In both geometries the search region contains exactly the members $n\le6$ of each tower, the next overtones lying at $|{\rm Im}\,\hat\omega|\ge1.53$ ($r_b/r_s=1.2$) and $|{\rm Im}\,\hat\omega|\ge2.84$ ($r_b/r_s=1.7$).

\subsection{Complete list.} Table~\ref{tab:RTS-full} lists all the modes with $\ell\le2$ in the search region, i.e.\ the first seven members of every tower, in units of the ADM mass, $M\hat\omega=G_4M\,\hat\omega$. All the digits shown are common to the two determinations.

{\small
\renewcommand{\arraystretch}{1.06}
\begin{longtable}{@{}cccllrl@{}}
\caption{Scalar QNMs of the RTS geometries with $({h_1},{h_2})=(1,2)$, $r_s=1$, $\hat p=0$ (background in Table~\ref{tab:RTSbg}). $n$ orders the members of each $(\ell,m)$ tower by increasing ${\rm Re}\,\hat\omega$; $m>0$ is co-rotating.}\label{tab:RTS-full}\\
\toprule
$\ell$ & $m$ & $n$ & $M\,{\rm Re}\,\hat\omega$ & $M\,{\rm Im}\,\hat\omega$ & $|{\rm Im}\,\hat\omega|/{\rm Re}\,\hat\omega$ & $\nu$\\
\endfirsthead
\toprule
$\ell$ & $m$ & $n$ & $M\,{\rm Re}\,\hat\omega$ & $M\,{\rm Im}\,\hat\omega$ & $|{\rm Im}\,\hat\omega|/{\rm Re}\,\hat\omega$ & $\nu$\\
\endhead
\bottomrule
\endfoot
\midrule
\multicolumn{7}{c}{$r_b/r_s=1.2$}\\
\midrule
0 & $0$ & 0 & $0.229048591$ & $-0.05344$ & $0.233$ & $-0.2296+i\,0.1941$\\
0 & $0$ & 1 & $0.422316982$ & $-0.2351$ & $0.557$ & $0.0470+i\,0.7649$\\
0 & $0$ & 2 & $0.648657477$ & $-0.4275$ & $0.659$ & $0.4077+i\,1.2695$\\
0 & $0$ & 3 & $0.883456264$ & $-0.6182$ & $0.700$ & $0.7760+i\,1.7628$\\
0 & $0$ & 4 & $1.121407023$ & $-0.8077$ & $0.720$ & $1.1462+i\,2.2520$\\
0 & $0$ & 5 & $1.360871077$ & $-0.9964$ & $0.732$ & $1.5170+i\,2.7394$\\
0 & $0$ & 6 & $1.601186361$ & $-1.185$ & $0.740$ & $1.8882+i\,3.2257$\\
\addlinespace[2pt]
1 & $0$ & 0 & $0.496701901$ & $-0.02386$ & $0.0480$ & $0.4879+i\,0.1714$\\
1 & $0$ & 1 & $0.636567778$ & $-0.1587$ & $0.249$ & $0.4877+i\,0.7049$\\
1 & $0$ & 2 & $0.812389592$ & $-0.3407$ & $0.419$ & $0.7736+i\,1.1945$\\
1 & $0$ & 3 & $1.012828580$ & $-0.5349$ & $0.528$ & $1.0888+i\,1.6505$\\
1 & $0$ & 4 & $1.227995589$ & $-0.7310$ & $0.595$ & $1.4094+i\,2.1074$\\
1 & $0$ & 5 & $1.451683212$ & $-0.9263$ & $0.638$ & $1.7323+i\,2.5752$\\
1 & $0$ & 6 & $1.680517937$ & $-1.120$ & $0.667$ & $2.0609+i\,3.0548$\\
\addlinespace[2pt]
2 & $-2$ & 0 & $0.763532500$ & $-0.02429$ & $0.0318$ & $1.0939+i\,0.2536$\\
2 & $-2$ & 1 & $0.899754498$ & $-0.1432$ & $0.159$ & $1.1344+i\,0.7104$\\
2 & $-2$ & 2 & $1.061718483$ & $-0.3047$ & $0.287$ & $1.3082+i\,1.2388$\\
2 & $-2$ & 3 & $1.242856507$ & $-0.4852$ & $0.390$ & $1.5656+i\,1.7155$\\
2 & $-2$ & 4 & $1.439276754$ & $-0.6746$ & $0.469$ & $1.8463+i\,2.1807$\\
2 & $-2$ & 5 & $1.647035170$ & $-0.8678$ & $0.527$ & $2.1443+i\,2.6536$\\
2 & $-2$ & 6 & $1.862915444$ & $-1.062$ & $0.570$ & $2.4607+i\,3.1346$\\
\addlinespace[2pt]
2 & $0$ & 0 & $0.781288622$ & $-0.01103$ & $0.0141$ & $1.0494+i\,0.1616$\\
2 & $0$ & 1 & $0.905805633$ & $-0.1099$ & $0.121$ & $1.0070+i\,0.6145$\\
2 & $0$ & 2 & $1.054304413$ & $-0.2653$ & $0.252$ & $1.1849+i\,1.1526$\\
2 & $0$ & 3 & $1.224066434$ & $-0.4444$ & $0.363$ & $1.4233+i\,1.6324$\\
2 & $0$ & 4 & $1.411519562$ & $-0.6349$ & $0.450$ & $1.6940+i\,2.1178$\\
2 & $0$ & 5 & $1.612413829$ & $-0.8301$ & $0.515$ & $1.9991+i\,2.6084$\\
2 & $0$ & 6 & $1.823017461$ & $-1.027$ & $0.563$ & $2.3278+i\,3.0964$\\
\addlinespace[2pt]
2 & $+2$ & 0 & $0.912979587$ & $-0.06730$ & $0.0737$ & $0.8413+i\,0.4655$\\
2 & $+2$ & 1 & $1.042043586$ & $-0.2192$ & $0.210$ & $1.0361+i\,1.0183$\\
2 & $+2$ & 2 & $1.198330174$ & $-0.3992$ & $0.333$ & $1.2608+i\,1.5312$\\
2 & $+2$ & 3 & $1.375817357$ & $-0.5924$ & $0.431$ & $1.5450+i\,2.0417$\\
2 & $+2$ & 4 & $1.569379700$ & $-0.7910$ & $0.504$ & $1.8634+i\,2.5380$\\
2 & $+2$ & 5 & $1.774489572$ & $-0.9908$ & $0.558$ & $2.1969+i\,3.0253$\\
2 & $+2$ & 6 & $1.987756084$ & $-1.190$ & $0.599$ & $2.5368+i\,3.5088$\\
\midrule
\multicolumn{7}{c}{$r_b/r_s=1.7$}\\
\midrule
0 & $0$ & 0 & $0.246272651$ & $-0.1456$ & $0.591$ & $-0.0570+i\,0.3477$\\
0 & $0$ & 1 & $0.405291140$ & $-0.5135$ & $1.27$ & $0.6109+i\,0.7930$\\
0 & $0$ & 2 & $0.606771662$ & $-0.8932$ & $1.47$ & $1.3365+i\,1.2289$\\
0 & $0$ & 3 & $0.818248091$ & $-1.271$ & $1.55$ & $2.0696+i\,1.6668$\\
0 & $0$ & 4 & $1.033392541$ & $-1.647$ & $1.59$ & $2.8044+i\,2.1056$\\
0 & $0$ & 5 & $1.250345219$ & $-2.023$ & $1.62$ & $3.5396+i\,2.5451$\\
0 & $0$ & 6 & $1.468362194$ & $-2.398$ & $1.63$ & $4.2748+i\,2.9851$\\
\addlinespace[2pt]
1 & $0$ & 0 & $0.559580255$ & $-0.1210$ & $0.216$ & $0.4810+i\,0.4621$\\
1 & $0$ & 1 & $0.648846371$ & $-0.4318$ & $0.665$ & $1.0161+i\,0.9059$\\
1 & $0$ & 2 & $0.784055910$ & $-0.8012$ & $1.02$ & $1.6294+i\,1.2795$\\
1 & $0$ & 3 & $0.954299474$ & $-1.186$ & $1.24$ & $2.2881+i\,1.6671$\\
1 & $0$ & 4 & $1.143746252$ & $-1.572$ & $1.37$ & $2.9698+i\,2.0753$\\
1 & $0$ & 5 & $1.343471416$ & $-1.956$ & $1.46$ & $3.6655+i\,2.4975$\\
1 & $0$ & 6 & $1.549163342$ & $-2.338$ & $1.51$ & $4.3706+i\,2.9281$\\
\addlinespace[2pt]
2 & $-2$ & 0 & $0.855051009$ & $-0.1983$ & $0.232$ & $1.2736+i\,0.6704$\\
2 & $-2$ & 1 & $0.957975894$ & $-0.4908$ & $0.512$ & $1.7349+i\,1.1640$\\
2 & $-2$ & 2 & $1.081163924$ & $-0.8326$ & $0.770$ & $2.2832+i\,1.5653$\\
2 & $-2$ & 3 & $1.229224519$ & $-1.201$ & $0.977$ & $2.8953+i\,1.9523$\\
2 & $-2$ & 4 & $1.398199502$ & $-1.580$ & $1.13$ & $3.5523+i\,2.3453$\\
2 & $-2$ & 5 & $1.581807248$ & $-1.962$ & $1.24$ & $4.2385+i\,2.7465$\\
2 & $-2$ & 6 & $1.775161603$ & $-2.345$ & $1.32$ & $4.9432+i\,3.1546$\\
\addlinespace[2pt]
2 & $0$ & 0 & $0.897824707$ & $-0.1133$ & $0.126$ & $1.0145+i\,0.5239$\\
2 & $0$ & 1 & $0.968926543$ & $-0.3884$ & $0.401$ & $1.4791+i\,1.0494$\\
2 & $0$ & 2 & $1.067584910$ & $-0.7197$ & $0.674$ & $2.0066+i\,1.4858$\\
2 & $0$ & 3 & $1.194089724$ & $-1.085$ & $0.909$ & $2.6145+i\,1.8889$\\
2 & $0$ & 4 & $1.346120615$ & $-1.466$ & $1.09$ & $3.2736+i\,2.2827$\\
2 & $0$ & 5 & $1.516991311$ & $-1.851$ & $1.22$ & $3.9613+i\,2.6773$\\
2 & $0$ & 6 & $1.700638892$ & $-2.238$ & $1.32$ & $4.6646+i\,3.0747$\\
\addlinespace[2pt]
2 & $+2$ & 0 & $0.960589282$ & $-0.2744$ & $0.286$ & $1.1869+i\,0.8201$\\
2 & $+2$ & 1 & $1.032954583$ & $-0.5972$ & $0.578$ & $1.7245+i\,1.3396$\\
2 & $+2$ & 2 & $1.136608852$ & $-0.9589$ & $0.844$ & $2.3275+i\,1.7665$\\
2 & $+2$ & 3 & $1.270576024$ & $-1.340$ & $1.05$ & $2.9773+i\,2.1751$\\
2 & $+2$ & 4 & $1.427946618$ & $-1.728$ & $1.21$ & $3.6567+i\,2.5872$\\
2 & $+2$ & 5 & $1.601259445$ & $-2.117$ & $1.32$ & $4.3546+i\,3.0062$\\
2 & $+2$ & 6 & $1.785220981$ & $-2.505$ & $1.40$ & $5.0643+i\,3.4311$\\
\end{longtable}
}

\subsection{High regime}

\paragraph*{Setup.} Massless scalar, KK-neutral ($\hat p=0$), axisymmetric ($m=0$) modes of the RTS with $({h_1},{h_2})=(1,2)$, $r_s=1$, $r_b/r_s=1.2$ and $1.7$ (backgrounds of Section~4.2.4). Frequencies are measured by the asymptotic static observer, in units of the ADM mass, $M\hat\omega$ ($G_4=1$).
\begin{itemize}\setlength{\itemsep}{1pt}
\item \emph{Content.} Tables~\ref{tab:all12} and~\ref{tab:all17} list all the modes found with $-M\,{\rm Im}\,\hat\omega\le0.800$ ($r_b/r_s=1.2$) and $\le1.480$ ($r_b/r_s=1.7$), i.e.\ $|{\rm Im}\,\hat\omega|\le1$ and $1.6$ in units $r_s=1$: 73 and 56 modes respectively. The search in ${\rm Re}\,\hat\omega$ extends beyond the last mode listed for each $\ell$, so that within this strip every list is complete. For $\ell\le2$ the towers continued to larger $|{\rm Im}\,\hat\omega|$ are those of the appendix of the draft.
\item \emph{Labelling.} $n$ orders the modes of a given $\ell$ by increasing ${\rm Re}\,\hat\omega$; for every $\ell$ this coincides with the ordering by increasing $|{\rm Im}\,\hat\omega|$.
\item \emph{Determination and accuracy.} All frequencies are computed with the Leaver continued fraction, converged in depth. Unmarked entries (68) are also roots of $\Xi(\hat\omega)=1$, i.e.\ $a_D\in\mathbb Z$: the two determinations agree with a median difference of $3\times10^{-14}$ (largest $3\times10^{-12}$). Entries marked $^\dagger$ (61) are confirmed by direct numerical integration of the radial equation instead: 7 of them have $n\le2$ ($r_b/r_s=1.7$, $\ell=9,10$; $r_b/r_s=1.2$, $\ell=6$, $n=0$), where the instanton-branch continuation needed by $\Xi$ fails (agreement $\le3\times10^{-10}$); the remaining 54 are the overtones $n\ge3$ for $\ell\ge3$, for which $\Xi$ has not been evaluated (median $5\times10^{-9}$, largest $2\times10^{-7}$, limited by the integration at large $|{\rm Im}\,\hat\omega|$). All digits quoted are therefore significant.
\end{itemize}

\begin{table}[p]\centering\footnotesize\renewcommand{\arraystretch}{1.02}
\begin{tabular}[t]{@{}cc|ll@{}}\toprule
$\ell$ & $n$ & $M\,{\rm Re}\,\hat\omega$ & $-M\,{\rm Im}\,\hat\omega$\\\midrule
0 & 0 & $0.229049$ & $0.0534$\\
 & 1 & $0.422317$ & $0.2351$\\
 & 2 & $0.648657$ & $0.4275$\\
 & 3 & $0.883456$ & $0.6182$\\
\midrule
1 & 0 & $0.496702$ & $0.0239$\\
 & 1 & $0.636568$ & $0.1587$\\
 & 2 & $0.812390$ & $0.3407$\\
 & 3 & $1.012829$ & $0.5349$\\
 & 4 & $1.227996$ & $0.7310$\\
\midrule
2 & 0 & $0.781289$ & $0.0110$\\
 & 1 & $0.905806$ & $0.1099$\\
 & 2 & $1.054304$ & $0.2653$\\
 & 3 & $1.224066$ & $0.4444$\\
 & 4 & $1.411520$ & $0.6349$\\
\midrule
3 & 0 & $1.067748$ & $4.702\times10^{-3}$\\
 & 1 & $1.187045$ & $0.0774$\\
 & 2 & $1.321729$ & $0.2111$\\
 & 3 & $1.473410$ & $0.3717$$^\dagger$\\
 & 4 & $1.640389$ & $0.5491$$^\dagger$\\
 & 5 & $1.820891$ & $0.7363$$^\dagger$\\
\midrule
4 & 0 & $1.353918$ & $1.759\times10^{-3}$\\
 & 1 & $1.473077$ & $0.0538$\\
 & 2 & $1.598848$ & $0.1703$\\
 & 3 & $1.739541$ & $0.3148$$^\dagger$\\
 & 4 & $1.892997$ & $0.4781$$^\dagger$\\
 & 5 & $2.058380$ & $0.6543$$^\dagger$\\
\midrule
5 & 0 & $1.639213$ & $5.740\times10^{-4}$\\
 & 1 & $1.761722$ & $0.0360$\\
 & 2 & $1.880677$ & $0.1380$\\
 & 3 & $2.013965$ & $0.2690$$^\dagger$\\
 & 4 & $2.158223$ & $0.4196$$^\dagger$\\
 & 5 & $2.312804$ & $0.5845$$^\dagger$\\
 & 6 & $2.477155$ & $0.7598$$^\dagger$\\
\bottomrule\end{tabular}
\hspace{1.5cm}
\begin{tabular}[t]{@{}cc|ll@{}}\toprule
$\ell$ & $n$ & $M\,{\rm Re}\,\hat\omega$ & $-M\,{\rm Im}\,\hat\omega$\\\midrule
6 & 0 & $1.923663$ & $1.677\times10^{-4}$$^\dagger$\\
 & 1 & $2.051768$ & $0.0229$\\
 & 2 & $2.165262$ & $0.1114$\\
 & 3 & $2.293070$ & $0.2311$$^\dagger$\\
 & 4 & $2.430686$ & $0.3705$$^\dagger$\\
 & 5 & $2.577373$ & $0.5250$$^\dagger$\\
 & 6 & $2.732729$ & $0.6908$$^\dagger$\\
\midrule
7 & 0 & $2.207515$ & $4.552\times10^{-5}$\\
 & 1 & $2.342306$ & $0.0136$\\
 & 2 & $2.451718$ & $0.0889$\\
 & 3 & $2.575121$ & $0.1987$$^\dagger$\\
 & 4 & $2.707659$ & $0.3285$$^\dagger$\\
 & 5 & $2.848352$ & $0.4735$$^\dagger$\\
 & 6 & $2.996803$ & $0.6306$$^\dagger$\\
 & 7 & $3.152726$ & $0.7972$$^\dagger$\\
\midrule
8 & 0 & $2.490993$ & $1.178\times10^{-5}$\\
 & 1 & $2.632642$ & $7.370\times10^{-3}$\\
 & 2 & $2.739635$ & $0.0696$\\
 & 3 & $2.859192$ & $0.1708$$^\dagger$\\
 & 4 & $2.987651$ & $0.2920$$^\dagger$\\
 & 5 & $3.123617$ & $0.4285$$^\dagger$\\
 & 6 & $3.266624$ & $0.5774$$^\dagger$\\
 & 7 & $3.416413$ & $0.7364$$^\dagger$\\
\midrule
9 & 0 & $2.774243$ & $2.958\times10^{-6}$\\
 & 1 & $2.922259$ & $3.561\times10^{-3}$\\
 & 2 & $3.028832$ & $0.0530$\\
 & 3 & $3.144746$ & $0.1461$$^\dagger$\\
 & 4 & $3.269790$ & $0.2597$$^\dagger$\\
 & 5 & $3.401894$ & $0.3887$$^\dagger$\\
 & 6 & $3.540492$ & $0.5301$$^\dagger$\\
 & 7 & $3.685306$ & $0.6819$$^\dagger$\\
\midrule
10 & 0 & $3.057344$ & $7.270\times10^{-7}$\\
 & 1 & $3.210816$ & $1.514\times10^{-3}$\\
 & 2 & $3.319183$ & $0.0387$\\
 & 3 & $3.431455$ & $0.1243$$^\dagger$\\
 & 4 & $3.553533$ & $0.2308$$^\dagger$\\
 & 5 & $3.682387$ & $0.3529$$^\dagger$\\
 & 6 & $3.817321$ & $0.4876$$^\dagger$\\
 & 7 & $3.958016$ & $0.6328$$^\dagger$\\
 & 8 & $4.104280$ & $0.7868$$^\dagger$\\
\bottomrule\end{tabular}
\caption{All axisymmetric scalar QNMs of the RTS with $r_b/r_s=1.2$ in the strip $-M\,{\rm Im}\,\hat\omega\le0.800$. Unmarked: $\Xi=1$ and Leaver; $^\dagger$: Leaver and direct integration.}\label{tab:all12}
\end{table}

\begin{table}[p]\centering\footnotesize\renewcommand{\arraystretch}{1.02}
\begin{tabular}[t]{@{}cc|ll@{}}\toprule
$\ell$ & $n$ & $M\,{\rm Re}\,\hat\omega$ & $-M\,{\rm Im}\,\hat\omega$\\\midrule
0 & 0 & $0.246273$ & $0.1456$\\
 & 1 & $0.405291$ & $0.5135$\\
 & 2 & $0.606772$ & $0.8932$\\
 & 3 & $0.818248$ & $1.2706$\\
\midrule
1 & 0 & $0.559580$ & $0.1210$\\
 & 1 & $0.648846$ & $0.4318$\\
 & 2 & $0.784056$ & $0.8012$\\
 & 3 & $0.954299$ & $1.1860$\\
\midrule
2 & 0 & $0.897825$ & $0.1133$\\
 & 1 & $0.968927$ & $0.3884$\\
 & 2 & $1.067585$ & $0.7197$\\
 & 3 & $1.194090$ & $1.0848$\\
 & 4 & $1.346121$ & $1.4656$\\
\midrule
3 & 0 & $1.240900$ & $0.1096$\\
 & 1 & $1.302848$ & $0.3658$\\
 & 2 & $1.386908$ & $0.6691$\\
 & 3 & $1.489241$ & $1.0076$$^\dagger$\\
 & 4 & $1.611721$ & $1.3701$$^\dagger$\\
\midrule
4 & 0 & $1.585926$ & $0.1075$\\
 & 1 & $1.641371$ & $0.3521$\\
 & 2 & $1.717542$ & $0.6368$\\
 & 3 & $1.807451$ & $0.9535$$^\dagger$\\
 & 4 & $1.911884$ & $1.2956$$^\dagger$\\
\midrule
5 & 0 & $1.931986$ & $0.1062$\\
 & 1 & $1.982289$ & $0.3428$\\
 & 2 & $2.052962$ & $0.6144$\\
 & 3 & $2.135508$ & $0.9147$$^\dagger$\\
 & 4 & $2.229184$ & $1.2392$$^\dagger$\\
\bottomrule\end{tabular}
\hspace{1.5cm}
\begin{tabular}[t]{@{}cc|ll@{}}\toprule
$\ell$ & $n$ & $M\,{\rm Re}\,\hat\omega$ & $-M\,{\rm Im}\,\hat\omega$\\\midrule
6 & 0 & $2.278675$ & $0.1053$\\
 & 1 & $2.324729$ & $0.3362$\\
 & 2 & $2.391023$ & $0.5980$\\
 & 3 & $2.468467$ & $0.8858$$^\dagger$\\
 & 4 & $2.555220$ & $1.1961$$^\dagger$\\
\midrule
7 & 0 & $2.625778$ & $0.1046$\\
 & 1 & $2.668237$ & $0.3313$\\
 & 2 & $2.730809$ & $0.5855$\\
 & 3 & $2.804290$ & $0.8635$$^\dagger$\\
 & 4 & $2.886135$ & $1.1623$$^\dagger$\\
\midrule
8 & 0 & $2.973172$ & $0.1042$\\
 & 1 & $3.012538$ & $0.3275$\\
 & 2 & $3.071840$ & $0.5756$\\
 & 3 & $3.142013$ & $0.8457$$^\dagger$\\
 & 4 & $3.220073$ & $1.1352$$^\dagger$\\
 & 5 & $3.305195$ & $1.4418$$^\dagger$\\
\midrule
9 & 0 & $3.320777$ & $0.1038$$^\dagger$\\
 & 1 & $3.357452$ & $0.3245$$^\dagger$\\
 & 2 & $3.413827$ & $0.5677$$^\dagger$\\
 & 3 & $3.481113$ & $0.8313$$^\dagger$\\
 & 4 & $3.556060$ & $1.1128$$^\dagger$\\
 & 5 & $3.637500$ & $1.4107$$^\dagger$\\
\midrule
10 & 0 & $3.668541$ & $0.1035$$^\dagger$\\
 & 1 & $3.702852$ & $0.3221$$^\dagger$\\
 & 2 & $3.756575$ & $0.5611$$^\dagger$\\
 & 3 & $3.821276$ & $0.8192$$^\dagger$\\
 & 4 & $3.893544$ & $1.0942$$^\dagger$\\
 & 5 & $3.971971$ & $1.3846$$^\dagger$\\
\bottomrule\end{tabular}
\caption{All axisymmetric scalar QNMs of the RTS with $r_b/r_s=1.7$ in the strip $-M\,{\rm Im}\,\hat\omega\le1.480$. Unmarked: $\Xi=1$ and Leaver; $^\dagger$: Leaver and direct integration.}\label{tab:all17}
\end{table}

\section*{Comments}
\begin{itemize}\setlength{\itemsep}{2pt}
\item \textbf{Stability.} All 129 modes have ${\rm Im}\,\hat\omega<0$: no instability in the $m=0$ sector up to $\ell=10$.

\item \textbf{Eikonal structure.} For $m=0$, at large $\ell$ the radial equation reads $d^2R/dr_*^2+Q\,R=0$, $Q=\hat\omega^2P(r)-(\ell+\frac12)^2\Delta(r)$, $dr_*=dr/\Delta$. The light ring is the maximum of $\Delta/P$ on $r\ge r_+$, $\Omega_c^2=\max(\Delta/P)$, and the cap value is $\Omega_{\rm cap}^2=(r_+-r_-)/P'(r_+)$. The two geometries realise the two possible situations :\\
$r_b/r_s=1.2$: light ring at $r_c=1.486\,r_s$, $M\Omega_c=0.3095>M\Omega_{\rm cap}=0.2826$, with a cavity between the cap and the light-ring barrier;\\
$r_b/r_s=1.7$: light ring on the cap, $M\Omega_c=M\Omega_{\rm cap}=0.3486$.

\item \textbf{$r_b/r_s=1.2$: trapped modes.} The fundamental mode falls in the cavity window $\Omega_{\rm cap}<{\rm Re}\,\hat\omega/(\ell+\frac12)<\Omega_c$ from $\ell=3$ on, and the first overtone from $\ell=9$. These modes are exponentially long-lived: $-M\,{\rm Im}\,\hat\omega_0$ decreases from $4.7\times10^{-3}$ at $\ell=3$ to $7.3\times10^{-7}$ at $\ell=10$. The modes above the barrier do not form a light-ring tower ${\rm Im}\,\hat\omega=-(n+\frac12)\lambda$, $M\lambda=0.1317$: they are cavity resonances accumulating to the right of $\Omega_c$, with spacing in ${\rm Im}\,\hat\omega$ decreasing with $\ell$. This is why the number of modes in the strip grows with $\ell$, from 4 at $\ell=0$ to 9 at $\ell=10$.

\item \textbf{$r_b/r_s=1.7$: light ring on the cap.} Near the cap the radial equation can be solved exactly, giving
\[
\hat\omega_n\simeq\big(\ell+\tfrac12\big)\Omega_c-i\big(n+\tfrac12\big)\lambda_{\rm cap},\qquad
\lambda_{\rm cap}=\frac{(r_+-r_-)\sqrt{\Omega_c^2P''(r_+)/2-1}}{\Omega_c\,P'(r_+)},\qquad M\lambda_{\rm cap}=0.2040 ,
\]
with damping independent of $\ell$ at leading order. The modes indeed line up at ${\rm Re}\,\hat\omega/(\ell+\frac12)\to\Omega_c$ and ${\rm Im}\,\hat\omega\to-(n+\frac12)\lambda_{\rm cap}$. At $\ell=10$ the deviations are $0.2\%$ in ${\rm Re}\,\hat\omega_0/(\ell+\frac12)$ and $1.5,\,5,\,10\%$ in ${\rm Im}\,\hat\omega_{0,1,2}$, decreasing as $1/\ell$ and growing with $n$ (about $23\%$ for $n=5$). The number of modes in the strip stays between 4 and 6.

\item \textbf{Computational limit of $\Xi$.} The quantisation on $a_D$ works wherever the instanton-branch continuation of $a$ (from $L\to tL$ at fixed masses) succeeds. It fails for $|L|\gtrsim5.8$ at $r_b/r_s=1.7$ ($\ell=9,10$), and for the trapped fundamental mode at $\ell=6$, $r_b/r_s=1.2$. That the failure is due to a branch point of $a(u)$ on the path is a conjecture, which we have not verified. The frequencies themselves are unaffected (Leaver and direct integration agree).
\end{itemize}

\section{Monodromy and reality of the quantum period}\label{app:monodromy}

Fix a base point $r_0>r_+$ and a basis $(R_1,R_2)$ of solutions of the radial equation: eq.~\eqref{eq:RTSradial} for the RTS, or its canonical form \eqref{eq:QSW} with the dictionary \eqref{eq:diz} for the RKTB.
Let $\gamma_A$ be the $A$-cycle, a closed path in the complex $r$-plane encircling once both regular singular points $r_+$ and $r_-$.
Analytic continuation along $\gamma_A$ defines the monodromy matrix $M$,
\begin{equation}
R_i\ \xrightarrow{\ \gamma_A\ }\ \sum_jM_{ij}R_j,\qquad\det M=1,
\end{equation}
where $\det M=1$ because the Wronskian $\Delta\,(R_1R_2'-R_2R_1')$ is constant (in the canonical form, $\Psi_1\Psi_2'-\Psi_2\Psi_1'$ is constant).
In the annulus $0<|z_{\rm SW}|<1$ there are two Floquet solutions, $R_\pm\sim r^{-\frac12\pm a}$ times a single-valued series, as in the MST representation $\sum_nC_nF_{\nu+n}$ with $a=\nu+\frac12$.
Along $\gamma_A$ they transform as $R_\pm\to-e^{\pm2\pi ia}R_\pm$, so that
\begin{equation}\label{eq:app-trM}
{\rm tr}\,M=-2\cos2\pi a .
\end{equation}
For real $\omega$ the coefficients of the radial equation are real. For the RTS these are $\Delta(r)$, $\mathcal C(r)$ and $A$. For the RKTB they are the coefficients of \eqref{eq:QSW}, which involve only $x^2$, $x\,m_3$, $(m_1\pm m_2)^2$ and $\Lambda_{\rm eff}$.
If $R(r)$ is a solution, so is $\overline{R(\bar r)}$, and complex conjugation maps $\gamma_A$ into $\gamma_A^{-1}$. Therefore $\bar M$ is conjugate to $M^{-1}$, and since ${\rm tr}\,M^{-1}={\rm tr}\,M$ for $M\in SL(2,\mathbb C)$,
\begin{equation}
\cos2\pi a\in\mathbb R\quad\Longrightarrow\quad a\in\mathbb R\quad\text{or}\quad{\rm Re}\,a\in\tfrac12\mathbb Z .
\end{equation}
Starting from $a=\ell+\frac12$ at $\omega=0$, where the expansion \eqref{eq:Love-aexp} is real, $a$ can thus become complex only after reaching a point with $2a\in\mathbb Z$. This is the threshold $\hat\omega_*$.
Numerically, $\cos2\pi a$ is obtained by integrating the canonical form along the circle $|z_{\rm SW}|=\frac12$, which does not involve the MST recursion. Where both exist, it agrees with $\cos2\pi(\nu+\frac12)$ from the root of \eqref{eq:CF} to ten digits.

\bibliographystyle{utphys}
\bibliography{biblio-2}
\end{document}